\documentclass{article}
\usepackage{arxiv}

\usepackage[utf8]{inputenc} 
\usepackage[T1]{fontenc}    
\usepackage[hidelinks]{hyperref}     
\usepackage{url}            
\usepackage{booktabs}       
\usepackage{amsfonts}       
\usepackage{nicefrac}       
\usepackage{microtype}      
\usepackage{lipsum}
\usepackage{bm}
\usepackage{graphicx}
\usepackage{subcaption}
\usepackage{amsmath}
\usepackage{placeins}
\graphicspath{ {./images/} }

\title{Willis-type thermal homogenization for particulate composites under harmonic heat transfer}

\author{
  Chunlin Wu \\
  Shanghai Institute of Applied Mathematics \\and Mechanics
  ,Shanghai University\\
  Shanghai, 200044 \\
  \texttt{chunlinwu@shu.edu.cn} \\
\And 
  Ning Ding \\ 
  Shanghai Institute of Applied Mathematics \\and Mechanics
  ,Shanghai University\\
  Shanghai, 200044 \\
  \texttt{24723988@shu.edu.cn}
  \And
  Huiming Yin  \\
  Department of Civil Engineering and \\Engineering Mechanics,
  Columbia University\\
  New York, NY, 10027 \\
  \texttt{yin@civil.columbia.edu} \\
}

\begin{document}
\maketitle
\begin{abstract}
This paper develops a Willis-type thermal homogenization framework for two-dimensional particulate composites under harmonic heat transfer and establishes a unique effective constitutive relation without introducing residual fields. The local thermal fields are obtained using the dual-reciprocity inclusion-based boundary element method, in which two eigen-fields are introduced to simulate the mismatches in thermal conductivity and heat capacity. Instead of retrieving effective kernels from several independent macroscopic solutions generated by residual fields, the complete constitutive operator is constructed upon relations between eigen-fields and generalized macroscopic responses. The effective constitutive relations are validated against a one-dimensional analytical solution and an extended Fourier-series homogenization scheme. The Fourier formulae further establish the opposite-wavenumber reciprocity relations between two coupling tensors and reflection-symmetry parity relations, which explain when transverse coupling components vanish and the sign reversal of off-diagonal conductivity components under reflection. Numerical results further demonstrate that the de-phased average is required to extract physically consistent effective thermal properties under the Bloch conditions, and reveal their nonlocal frequency/wavenumber-dependent behavior. The framework provides a general computational method for Willis-type homogenization of particulate composites and is extendable to three-dimensional, multiphysical problems. 
\end{abstract}

\keywords{Willis-type thermal coupling \and Bloch-form boundary condition \and Dual-Reciprocity Inclusion-based Boundary Element Method \and Eshelby's Tensor \and Green's function technique}

\section{Introduction}\label{sec:1 intro} 
Metamaterials have been created to meet various engineering needs which natural materials may not provide, such as electromagnetic metamaterials in nondestructive testing \cite{Grimberg2013}, acoustic metamaterials in noise reduction \cite{muhlestein2017experimental}, and mechanical metamaterials in impact protection \cite{Diallo2026}. For heating control, thermal metamaterials are designed to handle precise control on heat conduction, convection, and  radiation as well \cite{Li2021}. Xu and Huang \cite{Xu2019} proposed an effective-medium theory to design radiative metamaterials with anisotropic geometries, which can tailor thermal radiation for thermal protection. Su et al. \cite{Su2023} demonstrated the passive thermal rectification with a linearly graded microstructure and realized macroscopic asymmetric heat transfer with a tailored environment. These interesting designs were built upon effective constitutive models, which bridge microstructure features to macroscopic performances.  

To predict effective constitutive behavior from the heterogeneous microstructure, pioneers have proposed several homogenization methods \cite{nemat2013micromechanics}, which can be classified into two categories: (a) the micromechanical-based effective medium theories \cite{Mura1987} and (b) numerical realization-based multiscale methods \cite{PinhodaCruz2009}. The former approximates the overall response through volume average of the microscopic fields, such as Mori-Tanaka and self-consistent models \cite{yin2016introduction}; while the latter seeks macroscopic constitutive relations from the governing equation of the composites, i.e., the asymptotic expansion method implemented with finite element method (FEM) or boundary element method (BEM) \cite{PinhodaCruz2009}. For unidirectional periodic composites, Kaminski \cite{Kaminski2003a} developed the asymptotic expansion method and implemented it with FEM. Specifically, the author developed the closed-form effective thermal conductivity with the effective heat capacity calculated by simple spatial averaging. Kaminski \cite{Kaminski2003} extended the same framework to two-dimensional (2D) and three-dimensional (3D) composites, and demonstrated the homogenized properties for both deterministic and stochastic microstructures. Subsequently, Alzina et al. \cite{Alzina2006} applied the two-scale asymptotic expansion method to evaluate the effective thermal conductivity for cryogenic structures. Moreover, Andrianov et al. \cite{Andrianov2013} proposed the closed-form effective thermal conductivity for two-phase composites with square inclusions. Xu \cite{Xu2014} developed the stochastic asymptotic homogenization method for both steady-state and transient heat transfer with random thermal conductivity. Other related work can be found in a comprehensive review \cite{Matine2015}. 

The primary limitation of the asymptotic expansion method lies in that it generally requires a proper scale separation between the microscale and macroscale. For instance, the expansion variable $\varepsilon = l / L \ll 1$ is required \cite{Auriault2012}, where $l$ and $L$ refer to the characteristic length of the representative volume element (RVE) and the overall composite, respectively. This primary limitation has been discussed and studied in elastodynamics \cite{Nassar2015}. The authors systematically compared homogenized properties using the asymptotic expansion method and Willis' coupling theory, and the results indicate that Willis' coupling theory is exact. As Willis' coupling theory does not require any assumed scale separations, Nassar et al. \cite{Nassar2015} showed that the homogenized properties by Willis' coupling theory can reproduce the dispersion relation. However, Shmuel and Willis \cite{Gal2025a} demonstrated that, in harmonic heat transfer, the effective constitutive relations not only depend on the microstructure but also on the frequency.

To understand and address such interesting coupling phenomena of effective properties, Willis \cite{Willis1980} first proposed a generalized effective constitutive framework for elastodynamics, where both the averaged stress and momentum depend on the averaged strain and velocity. Subsequently, Willis \cite{Willis1997} has proposed that the effective constitutive relations are spatio-temporal convolution, which reduces to the bi-anisotropic tensor in the frequency domain. Since Willis' theory can explain frequency-dependent macroscopic behaviors of composites, recent progress has been made in piezoelectricity \cite{PernasSalomon2020, PernasSalomon2021, Muhafra2022} and heat transfer \cite{Gal2025a, Shmuel2025}. Particularly, Shmuel and Willis \cite{Shmuel2025} pointed out the following inconsistencies: (a) the coupling terms vanish in the coupling model in the absence of time modulation \cite{Torrent2018}; and (b) the uniqueness of the constitutive relation becomes questionable with the one-way coupling \cite{Xu2019}. In analogy with Willis' elastodynamic constitutive relations, Shmuel and Willis \cite{Shmuel2025} proved that both the averaged sensible heat (heat storage) and heat flux depend on the averaged temperature variation and temperature gradients, and the coupling terms are correlated with the factor $i \omega$, which is an admissible representative of Willis constitutive relation \cite{Wu2026_rspa}. The uniqueness issue is particularly relevant under Bloch-form excitation because the averaged temperature and temperature gradient are kinematically related. Willis \cite{Willis2011, Willis2012} resolves the uniqueness issue by introducing residual fields, such as inelastic strain in elastodynamics. In the present work, the unique constitutive relations are constructed through connecting eigen-fields with the macroscopic responses, which avoids introducing residual fields. 

Moreover, Shmuel and Willis \cite{Gal2025a} discussed that the effective constitutive relations can be established among averaged sensible heat, heat flux, temperature changing rate, and temperature gradients. However, such a change in constitutive relations can lead to variations in coupling tensors. As Srivastava \cite{Srivastava2015} summarized in the comprehensive review, the primary concept of Willis' constitutive relations is to satisfy the governing equation on average, which is fundamentally different from the above works. Specifically, for harmonic heat transfer, the effective thermal conductivity cannot be derived from the quotient between the averaged heat flux and (minus) temperature gradients, and a similar rule applies to the effective heat capacity, which will be confirmed in the present paper. 

Although Willis' effective constitutive relation provides a general framework to describe macroscopic behaviors of composite, many efforts have been devoted to obtain explicit effective properties. Milton and Willis \cite{Milton2007} utilized the polarization method \cite{Willis1980}, and pointed out that the effective properties (in the frequency domain) may be solved by the iterative method for weakly heterogeneous composites. In elastodynamics, the polarization method replaces the entire composites, including the matrix and inhomogeneities, with the comparison medium and polarization fields, which are determined by consistent conditions. For instance, Willis \cite{Willis1980} employed eigenstress and eigenvelocity to simulate material mismatches in stiffness tensor and mass density, respectively. For a 1D periodic microstructure, Willis \cite{Willis2009} derived the elastodynamic Green's function for the specific microstructure, i.e., two-phase composites containing a concentrated mass, and proved that the coupling tensors are even function of the macroscopic wavenumber. Moreover, specific examples on weighted average indicates that the effective properties significantly rely on the way of observation \cite{Willis2009}. 

The above works primarily rely on the microstructure-specific Green's function, which becomes the major obstacle to evaluating effective properties. To address such limitations, Nemat-Nasser and Srivastava \cite{NematNasser2011} proposed a Fourier-series-based method, in which the elastic fields can be divided into the de-phased average and the fluctuation parts. The authors have shown that the fluctuation part can be attributed to disturbances by polarization fields, and they provided the explicit formulae for 1D periodic composites. Subsequently, Srivastava and Nemat-Nasser \cite{Srivastava2011} extended the framework to 3D periodic composites under Bloch-form boundary conditions, which exhibits good agreement with \cite{Minagawa1976} on the dispersion relation. In contrast to Willis' previous works \cite{Willis1980, Milton2007}, the authors employed eigenstrain and eigenmomentum to simulate material mismatches. Subsequently, Nemat-Nasser and Srivastava \cite{NematNasser2013} further discussed bounds on effective dynamic properties. However, some issues may arise for the Fourier-series-based method as follows: a) for inhomogeneities with other geometric shapes, such as circular voids, it is challenging to approximate inhomogeneities with cuboids or squares, which leads to higher-order series; and b) when the number of inhomogeneities significantly increase, i.e., 100 inhomogeneities, the particle interactions require higher-order series to achieve convergent predictions. Recently, Lee et al. \cite{Lee2023} developed a Green's function-based dynamic homogenization framework for piezoelectric composites using the Hashin-Shtrikman variational approach, which presents a 1D verification with homogenization limit.

In contrast, the present paper performs Willis-type thermal homogenization of composites under harmonic heat transfer. Wu et al. \cite{Wu2025a} derive Eshelby's tensor for ellipsoidal inclusions with polynomial-form eigen-fields, while the subsequent work \cite{Wu2026} demonstrates the difficulty of obtaining a closed-form Eshelby's tensor for circular and polygonal inclusions. Since Willis-type homogenization requires the heat equation to be satisfied both locally and globally, the accuracy of both local and global (averaged) fields is important. Consequently, the lack of closed-form Eshelby's tensors has become a major obstacle preventing the application of Eshelby's equivalent inclusion method in unsteady-state heat transfer. Alternatively, our recent work \cite{Wu_2025_EABE} interpreted the heat equation as a quasi-static form, in which the heat generation/storage rate and eigen-fields are considered as source fields. In contrast to the above cases with microstructure-specific and unsteady-state Green's function, the proposed dual-reciprocity inclusion-based boundary element method (DR-iBEM) uses the well-established infinite-domain steady-state Green's function, while the disturbances by eigen-fields are evaluated through closed-form steady-state Eshelby's tensors, such as circular \cite{Mura1987} and polygonal shapes \cite{Wu2021, Trotta2017}. Therefore, the present implementation provides an efficient and simple way to evaluate effective thermal properties. Although the present paper is restricted to 2D harmonic heat transfer for illustration, the framework is general, which can be extended to 3D and elastodynamic problems. 

This paper is organized as follows. Section 2 briefly reviews the Willis-type effective constitutive relations and introduces assumptions and weighted averages in the homogenization procedure for particulate composites. Section 3 presents the DR-iBEM formulae for 2D composites, and the DR-iBEM algorithm is verified against the finite element analysis by comparing local thermal fields. Section 4 derives the homogenization scheme to extract effective thermal properties of particulate composites, thereby providing the unique constitutive relations without the introduction of residual fields. Section 5 develops a Fourier-series-based homogenization scheme, establishes the opposite-wavenumber reciprocity relation, and derives the additional relations associated with microstructural reflection symmetry. Subsequently, Section 6 cross-validates the effective properties for composites containing square inhomogeneities. The frequency-wavenumber dependence of the coupling terms for circular particulate composites is investigated, and a benchmark is provided that Willis' residual-field-based unique constitutive relations can be recovered by the present formulation for a 1D laminate. Finally, some conclusive remarks are provided in Section 7. 

\section{Willis-type effective constitutive law of particulate composites}

\subsection{Boundary value problem of the local solution of a composite specimen}
A composite specimen can be considered as a 2D bounded domain $\mathcal{D}$ containing multiple $N_I$ subdomains $\Omega^I$ ($I = 1, 2, 3, \ldots, N_I$) subjected to prescribed temperature $u^{BC}$ and heat flux $q^{BC}$. Each subdomain is assumed to be perfectly bonded with the matrix, and therefore the temperature and heat flux are continuous along the interface $\partial \Omega^I$. Note that in the numerical implementation, the non-circular inhomogeneities can be divided into small sub-squares. After the discretization, the sub-squares are treated as subdomains, and therefore they are indexed from $I = 1, 2, \ldots N_I$ ($N_I$ is updated), which inherit the same material as the original physical phase. In general, each subdomain $\Omega^I$ and the matrix $\Omega^0 = \mathcal{D} - \sum_{I=1}^{N_I}$ exhibits dissimilar thermal properties, such as the thermal conductivity $K^I$ and heat capacity $C_p^I$. Following our recent definition \cite{Wu2025-ijss}, the thermal properties can be defined as, 

\begin{equation}
    K(\textbf{x}) = K^c + \sum_{I=0}^{N_I} (K^I - K^c) \Theta^I(\textbf{x}), \quad C_p(x) = C_p^c + \sum_{I=0}^{N_I} (C_p^I - C_p^c) \Theta^I(\textbf{x})
    \label{eq:properties}
\end{equation}
where $K^c$ and $C_p^c$ are the thermal conductivity and heat capacity of the comparison medium; when the superscript $I = 0$, it refers to thermal properties of the matrix phase; and $\Theta^I(\textbf{x})$ is the characteristic function of the subdomain (including the matrix). In this paper, the algorithm is developed for the circular inhomogeneities and extended to square ones in the similar fashion, and the latter will be utilized in cross-validation against the Fourier-series-based homogenization scheme in Section 6. The specific characteristic function for circular case can be expressed as, 

\begin{equation}
 \Theta^I(\textbf{x}) = \begin{cases} 1 & |\textbf{x} - \textbf{x}^{IC}| \leq a^I \\ 0 & \text{otherwise} \end{cases}
    \label{eq:char_func}
\end{equation}
where $\textbf{x}^{IC}$ is the center of the $I^{th}$ subdomain; $a^{I}$ refers to the radius of the circular subdomain. 

When square inhomogeneities are considered, the characteristic function can be written as:
\begin{equation}
\Theta^I(\textbf{x}) = \begin{cases} 1 & |x_1 - x_1^{IC}| \leq \dfrac{b^I}{2} \thinspace \thinspace \text{and} \thinspace \thinspace |x_2 - x_2^{IC}| \leq \dfrac{b^I}{2} \\ 0 & \text{otherwise} \end{cases} 
    \label{eq:char_func_square}
\end{equation}
where $b^{I}$ refers to the edge length of the square. This paper will demonstrate the details with the circular case, and the square case will be reproduced at the end of Section 3. Note that square inhomogeneities are common in Fast-Fourier-Transform (FFT) related methods \cite{Mura1987,zhou2015numerical}, which discretize the subdomain into small grids. In general, singularity exists at the vertices of a square for Eshelby's tensor \cite{Rodin1996, Wu2021}, which cannot be captured by the FFT exactly, but exhibit negligible influence on the effective properties with the domain integral in homogenization.  This paper focuses on the general framework for homogenization, rather than rigorous treatment of local singular fields. Our recent work \cite{Yang2026} has proved the singular eigen-fields for angular inhomogeneities in heat transfer, and the treatment of such singularities can be extended to square inhomogeneities as well. 

The unsteady-state heat equation with spatially varying thermal properties can be written as, 

\begin{equation}
     \frac{\partial \left[ K(\textbf{x}) \overline{u}_{,i}(\textbf{x}, t) \right]}{\partial x_i} = C_p(\textbf{x}) \frac{\partial \overline{u}(\textbf{x}, t)}{\partial t} - \overline{Q}(\textbf{x}, t)
    \label{eq:heat_eqn}
\end{equation}
where $\overline{u}(\textbf{x}, t)$ refers to the temperature; and $\overline{Q}(\textbf{x}, t)$ is the volumetric heat source fields. The homogenization is conducted under harmonic heat transfer (in the frequency domain) following the Bloch-form boundary conditions. For harmonic heat transfer, the thermal quantities can be defined by the temperature field in the following form: 
\begin{equation}
    \overline{u}(\textbf{x}, t) = \text{Re} \left[ u(\textbf{x}) \thinspace \exp[-i \omega t] \right]
    \label{eq:harm_def}
\end{equation}
where $i = \sqrt{-1}$ is the imaginary unit which is different from the subscript of index $i$; $\omega$ is the circular frequency of the system; $\text{Re}[.]$ refers to the real part of the quantity; and $u(\textbf{x})$ is the spatial part of the temperature field. The same form can be applied to heat flux and other variables by decoupling space and time with a sinusoidal function. Hence, the heat equation in the frequency domain can be rewritten as, 

\begin{equation}
     \frac{\partial \left[ K(\textbf{x}) u_{,i}(\textbf{x}) \right]}{\partial x_i} = -i \omega C_p(x) u(x) - Q(\textbf{x})
    \label{eq:heat_eqn_har}
\end{equation}

The local constitutive relations can be expressed as, 

\begin{equation}
    F(\textbf{x}) = C_p(\textbf{x}) u(\textbf{x}), \quad q_i(\textbf{x}) = -K(\textbf{x}) u_{,i}(\textbf{x})
    \label{eq:local_cons}
\end{equation}
where $F(\textbf{x})$ and $q_i(\textbf{x})$ refer to the sensible heat and heat flux, respectively. 

Under the harmonic loading condition, both temperature and heat flux change with time and space. When the particles are relatively small, the specimen can be homogenized into a statistically homogeneous material, and the effective constitutive law can be different from Eq. \eqref{eq:local_cons}. Instead, the Willis-type constitutive law has been proposed for 1D laminates \cite{Gal2025a, Shmuel2025} and will be used for 2D particulate composites in the following.  

\subsection{Weighted averages for Willis-type effective constitutive law}

As Srivastava commented in the review paper \cite{Srivastava2015}, the effective properties of particulate composites after the homogenization are required to satisfy the governing equation on average. Before introducing the effective constitutive relations, the weighted spatial average is defined by using temperature for example: 

\begin{equation}
    \langle u \rangle = \frac{1}{A_{\mathcal{D}}} \int_{\mathcal{D}} f(\textbf{x}) \thinspace u(\textbf{x}) \thinspace d\textbf{x}
    \label{eq:def_avg}
\end{equation}
where $A_{\mathcal{D}}$ refers to the area of the 2D specimen domain; and $f(\textbf{x})$ is the weight function for average. Specifically, when $f(\textbf{x}) \equiv 1$, the weighted average reduces to the conventional spatial average, which is denoted as $\langle . \rangle^{f=1}$; and when $f(\textbf{x}) \equiv \exp [-i \bm{\zeta} \cdot \textbf{x}]$, the weighted average becomes the de-phased average at $\bm{\zeta}$, which stands for the macroscopic thermal wave-vector (wavenumber) \cite{Srivastava2011}. In the present paper, the de-phased average is applied for periodic composites under the Bloch-form boundary conditions, and effective properties are compared with the conventional domain averages that can be canceled within a whole wavelength. 

Following the recent work \cite{Gal2025a, Shmuel2025} on Willis-type constitutive relations for harmonic heat transfer, the effective constitutive relations can be expressed as, 

\begin{equation}
    \begin{bmatrix} \langle F \rangle \\ \langle \textbf{q} \rangle \end{bmatrix} = \begin{bmatrix} C^\text{eff} & \bm{\xi}^\text{eff} \\ 
    \bm{\chi}^\text{eff} & -\textbf{K}^\text{eff}
    \end{bmatrix} \begin{bmatrix} \langle u \rangle \\ \langle u_{,i} \rangle \end{bmatrix}
    \label{eq:cons_law}
\end{equation}
where $C^\text{eff}$ is the effective heat capacity;  $\textbf{K}^\text{eff}$ refers to the effective thermal conductivity, which should be interpreted as a $2 \times 2$ matrix for 2D problems; $\bm{\xi}^\text{eff}$ and $\bm{\chi}^\text{eff}$ are two nonlocal Willis coupling vectors. Although the Willis-type constitutive law has been proposed and studied for decades \cite{Willis1997,Srivastava2015,Shmuel2025,Wu2026_rspa}, the case studies mainly focused on 1D cases for bianisotropic couplings. This paper advances the state of the art for Willis-type homogenization from 1D \cite{Wu2026_rspa} to 2D particulate composites based on the local solution with anisotropic microstructure. For reciprocal constituent materials and consistent source and averaging weights, the two coupling vectors form an adjoint pair \cite{Gal2025a}. The explicit reciprocity relation under the Bloch-form boundary condition will be provided in the homogenization formulation in Section 5, which will be further confirmed with two numerical homogenization methods. The main objective of the present work is to provide an efficient, robust framework to determine such effective material constants for the homogenized composite specimen from the local solution.

\section{Formulation and implementation of the DR-iBEM for the local solution}

The dual-reciprocity inclusion-based boundary element method (DR-iBEM) combines the dual-reciprocity boundary element method (DRBEM) and Eshelby's equivalent inclusion method (EIM), which serve as a boundary-only framework for solving dynamic problems in particular composites \cite{Wu_2025_EABE}. In this framework, the inhomogeneities are replaced with the comparison medium and two eigen-fields, the eigen-temperature-gradient (ETG) and eigen-heat-source (EHS), which handle material mismatches in thermal conductivity and heat capacity, respectively. Hence, the harmonic heat equation is recast into an equivalent quasi-static problem, in which the heat storage effects and contribution of eigen-fields are evaluated through domain integrals of steady-state thermal Green's function. Through the dual-reciprocity approximation, these domain integrals are converted into boundary integrals and classic steady-state thermal Eshelby's tensors. 

The proposed DR-iBEM using the steady-state Green's function exhibits two primary advantages over the iBEM \cite{yin2022inclusion} using the frequency-dependent Green's function: (1) The iBEM involves domain integrals of the frequency-dependent Green's functions (kernels) over spatially varying eigen-fields, which cannot be evaluated analytically but require internal cells with quadrature. Note that analytical formulae can be obtained in rare cases, such as Michelitsch et al. \cite{Michelitsch2003} and our recent work \cite{Wu2025a}, which studied spherical inclusions; while closed-form solutions are generally unavailable even for circular or square inclusions \cite{Wu2026}. Instead, the DR-iBEM is based on the steady-state thermal Green's function, and closed-form formulae exist for inclusions with arbitrary shapes \cite{Rodin1996, Wu2021}, which avoids numerical quadrature. (2) The Willis-type constitutive relations are nonlocal, which depend on frequency and wavenumber. In the conventional iBEM, the Green's function and its associated boundary/domain integrals are frequency-dependent. Therefore, the global coefficient matrices need to be re-evaluated at each excitation frequency, which significantly increases the computational burden for frequency sweeping. However, DR-iBEM treats the frequency variable $\omega$ as an external factor multiplying the coefficient matrix; once the global coefficient matrices are evaluated, they can be reused throughout the frequency sweeping to extract the Willis-type effective properties. 

\subsection{Formulation of the DR-iBEM}

For harmonic heat transfer, the inhomogeneities can be replaced by the comparison medium with two eigen-fields \cite{Wu2025a}. Specifically, the material mismatch in thermal conductivity and heat capacity can be simulated by the eigen-temperature-gradient (ETG) ($u_{i}^*$) and eigen-heat-source (EHS) ($Q^*$), respectively, which are expressed in terms of characteristic functions in Eq. (\ref{eq:char_func}), 

\begin{equation}
    u_i^*(\textbf{x}) = \sum_{I=0}^{N_I} u_{i}^{I*}(\textbf{x}) \Theta^I(\textbf{x}), \quad Q^*(\textbf{x}) = \sum_{I=0}^{N_I} Q^{I*}(\textbf{x}) \Theta^I(\textbf{x})
    \label{eq:eigen_fields}
\end{equation}
where $u^{I*}$ and $Q^{I*}$ are ETG and EHS distributed within the $I^{th}$ subdomain, respectively. 

According to Willis' theorem \cite{Willis1980}, the comparison medium can be selected as an arbitrary material different from the matrix and all inhomogeneities in the composites, and therefore, the eigen-fields depend on the comparison medium. When the matrix is selected as the comparison medium,  Eshelby's EIM is recovered, and eigen-fields in the matrix must be zero. However, when the comparison medium exhibits different thermal properties, it can be interpreted as a generalized Eshelby's EIM, and the eigen-fields in the matrix are nonzero. Nevertheless, the physical solution remains independent from the comparison medium's selection, provided that the equivalent conditions are satisfied throughout the entire domain. In the present formulation, this paper focuses on the comparison medium of the matrix phase, but a case study of different comparison media is provided in \ref{sec:verify_square}. 

The equivalent conditions for heat flux and sensible heat storage are, 

\begin{equation}
\begin{aligned}
    -K^c \left[ u_{,i}(\textbf{x}) - u_{i}^*(\textbf{x}) \right] = -K(\textbf{x}) u_{,i}(\textbf{x}) \\ 
    C_p^c u(\textbf{x}) - Q^*(\textbf{x}) =  C_p(\textbf{x}) u(\textbf{x})
\end{aligned}
    \label{eq:equiv_cond}
\end{equation}

With two equivalent conditions on heat flux and sensible heat storage, the constitutive relations in Eq. (\ref{eq:local_cons}) can be substituted into the heat equation as, 

\begin{equation}
    K^c \left[ u_{,ii}(\textbf{x}) - u^{*}_{i,i}(\textbf{x}) \right] = -i \omega \left[ C_p^c u(\textbf{x}) - Q^*(\textbf{x}) \right]
    \label{eq:heat_eqn_equiv}
\end{equation}

The DR-iBEM interprets Eq. (\ref{eq:heat_eqn_equiv}) as the quasi-static equation, 

\begin{equation}
    -K^c u_{,ii}(\textbf{x}) = \underbrace{i \omega C_p^c u(\textbf{x}) - i \omega Q^*(\textbf{x}) - K^c u_{i,i}^*(\textbf{x})}_{\text{``equivalent source field''}}
    \label{eq:heat_eqn_static}
\end{equation}
where the right-hand side is regarded as an equivalent source field. Let $\textbf{x}$ and $\textbf{x}'$ denote the field and source points, respectively. The infinite-domain Green's function of the comparison medium is \cite{yin2022inclusion}, 

\begin{equation}
    G(\textbf{x}, \textbf{x}') = \frac{1}{4 \pi K^c} \phi, \quad \phi = -2 \ln |\textbf{x} - \textbf{x}'|
    \label{eq:Gfunc_steady}
\end{equation}
where $\phi$ is the Newton's potential in 2D problems. Applying the Green's identity to Eq. (\ref{eq:heat_eqn_static}) yields: 
\begin{equation}
    \begin{split}
    u(\textbf{x}) & = -\int_{\partial \mathcal{D}} K^c G_{,i'}(\textbf{x}, \textbf{x}') n_i(\textbf{x}') u(\textbf{x}') \thinspace d\textbf{x}' + \int_{\partial \mathcal{D}} G(\textbf{x}, \textbf{x}') q_n(\textbf{x}') \thinspace d\textbf{x}' \\ 
    & +  i \omega  C_p^c\int_{\mathcal{D}} G(\textbf{x}, \textbf{x}') u(\textbf{x}')\thinspace d\textbf{x}' - \int_{\mathcal{D}} G(\textbf{x}, \textbf{x}') \left[ i \omega Q^*(\textbf{x}') + K^c u_{i,i'}^*(\textbf{x}') \right] \thinspace d\textbf{x}'  
    \end{split}
    \label{eq:temperature}
\end{equation}
where $q_n(\textbf{x}') = K^c n_i(\textbf{x}') u_{,i'}(\textbf{x}')$; $(.)_{,i'}$ refers to partial derivative with respect to $x_i'$; $\partial \mathcal{D}$ is the boundary of the domain $\mathcal{D}$; and $\textbf{n}$ stands for the outward unit normal vector on the boundary. The temperature can be expressed as two boundary integrals and two domain integrals, of which the domain integrals are converted into boundary integrals through the dual-reciprocity method and  Eshelby's formulation, respectively. 

\subsubsection*{(i) Dual-reciprocity treatment of the heat-storage integral}
The dual-reciprocity method approximates the temperature with a global interpolation functions, and then converts domain integrals into boundary integrals using the reciprocal property of the Green's function. Following the original work \cite{Partridge1991}, the temperature is approximated by the global radial basis function as,  

\begin{equation}
    u(\textbf{x}) = \sum_{m = 1}^{NN + NS} \beta(\textbf{x}, \textbf{x}^m) \alpha(\textbf{x}^m), \qquad \beta(\textbf{x}, \textbf{x}^m) = 1 + |\textbf{x} - \textbf{x}^m|
    \label{eq:temp_approx}
\end{equation}
where $NN$ and $NS$ refer to the number of boundary nodes and interior interpolation nodes, respectively; $\beta(\textbf{x}, \textbf{x}^m)$ is the global interpolation function; and $\alpha(\textbf{x}^m)$ is the virtual source density distributed at all boundary/interior nodes. The corresponding particular solution $\Gamma(\textbf{x}, \textbf{x}^m)$ can be defined as, 

\begin{equation}
    \Gamma(\textbf{x}, \textbf{x}^m) = \frac{|\textbf{x} - \textbf{x}^m|^2}{4} + \frac{|\textbf{x} - \textbf{x}^m|^3}{9}, \quad \nabla^2 \Gamma(\textbf{x}, \textbf{x}^m) = \beta(\textbf{x}, \textbf{x}^m)
    \label{eq:Gamma}
\end{equation}

Subsequently, using the reciprocal property of Green's function again, the first area integral in Eq. (\ref{eq:temperature}) can be converted to the following boundary integrals: 

\begin{equation}
\begin{split}
    i \omega  C_p^c\int_{\mathcal{D}} G(\textbf{x}, \textbf{x}') u(\textbf{x}')\thinspace d\textbf{x}' &= i \omega C_p^c \sum_{m=1}^{NN + NS} \alpha(\textbf{x}^m) \Big[ \int_{\partial \mathcal{D}} K^c G_{,i'}(\textbf{x}, \textbf{x}') n_i(\textbf{x}') \Gamma(\textbf{x}', \textbf{x}^m) \thinspace d\textbf{x}' \\ & \quad - \int_{\partial \mathcal{D}} G(\textbf{x}, \textbf{x}') \Gamma_{,i'}(\textbf{x}', \textbf{x}^m) n_i(\textbf{x}') \thinspace d\textbf{x}' \Big]
\end{split}
    \label{eq:temperature_bie}
\end{equation}

\subsubsection*{(ii) Eshelby's formulation of the domain integrals of eigen-fields}
The second area integral in Eq. (\ref{eq:temperature}) involves disturbances caused by eigen-fields, which can be evaluated through superposition of contributions from the matrix and all subdomains. Substituting Eq. (\ref{eq:eigen_fields}) into this area integral provides, 

\begin{equation}
\small
\begin{split}
    \int_{\mathcal{D}} G(\textbf{x}, \textbf{x}') \left[ i \omega Q^*(\textbf{x}') + K^c u_{i,i'}^*(\textbf{x}') \right] \thinspace d\textbf{x}' & = \sum_{I=0} \Big[ i \omega \int_{\Omega^I} G(\textbf{x}, \textbf{x}') Q^{I*}(\textbf{x}') \thinspace d\textbf{x}' + K^c \int_{\Omega^I} G_{,i}(\textbf{x}, \textbf{x}') u_{i}^{I*}(\textbf{x}') \thinspace d\textbf{x}' \Big] \\ & \quad + K^c \int_{\partial \mathcal{D}} G(\textbf{x}, \textbf{x}') u_{i}^*(\textbf{x}') n_i(\textbf{x}') \thinspace d\textbf{x}' 
\end{split}
    \label{eq:eigen_integrals}
\end{equation}
where the second item uses the divergence theorem, which leads to an additional boundary integral term. Note that when the matrix is selected as the comparison medium, the boundary integral term vanishes due to zero ETG on the external boundary. However, for a general comparison medium, this boundary integral term is retained and combined with the original boundary flux integral by defining the auxiliary flux: $q_{n}(\textbf{x}) = K^c n_{i}(\textbf{x}) \left[ u_{,i}(\textbf{x}) - u_i^*(\textbf{x}) \right],\textbf{x} \in \partial \mathcal{D} $. Therefore, the same boundary integral formulae remain applicable, provided that the auxiliary flux is used.

To illustrate the DR-iBEM algorithm, this paper assumes the eigen-fields in an inhomogeneity are uniformly distributed, i.e., $Q^{I*}(\textbf{x}) = Q^{I*}, u_{i}^*(\textbf{x}) = u_{i}^{I*}$ in Eq. \eqref{eq:eigen_fields}. Thanks to the piecewise continuity approximation of circular and square inhomogeneities, the domain integrals in Eq. (\ref{eq:eigen_integrals}) can be evaluated using the steady-state thermal Eshelby's tensors. After incorporating the external ETG contribution into the boundary flux variables, the disturbed temperature caused by eigen-fields can be written in terms of Eshelby's tensors as follows: 

\begin{equation}
    \begin{split}
    - \int_{\mathcal{D}} G(\textbf{x}, \textbf{x}') \left[ i \omega Q^*(\textbf{x}') + K^c u_{i,i'}^*(\textbf{x}') \right] \thinspace d\textbf{x}' = \sum_{I=0}^{N_I} \left[ -i \omega L^I(\textbf{x}) Q^{I*} + D^{I}_{i}(\textbf{x}) u_{i}^{I*} \right]
    \end{split}
    \label{eq:eigen_disturb}
\end{equation}
where $D_{i}^{I}(\textbf{x})$ and $L^{I}(\textbf{x})$ are Eshelby's tensors of the $I^{th}$ subdomain relating ETG and EHS to the disturbed temperature, respectively. The closed-form circular Eshelby's tensors have been reported in \cite{Mura1987, Wu2023_IJSS}, and those for squares can be found in \cite{Rodin1996, Wu2021}. Note that the uniform assumption is introduced here only to present the DR-iBEM algorithm in a compact notation. In the numerical examples, higher-order polynomial-form eigen-fields and smaller subdivisions of square inhomogeneities are adopted to improve description of the spatially varying eigen-fields, as shown in Section \ref{sec:verify_circular} and \ref{sec:verify_square}, respectively. 

The substitution of Eqs. (\ref{eq:temperature_bie}) and  (\ref{eq:eigen_disturb}) into (\ref{eq:temperature}) yields the temperature at any interior points $\textbf{x}$ as: 

\begin{equation}
\begin{split}
    u(\textbf{x}) = & -\int_{\partial \mathcal{D}} K^c G_{,i'}(\textbf{x}, \textbf{x}') n_i(\textbf{x}') u(\textbf{x}') \thinspace d\textbf{x}' + \int_{\partial \mathcal{D}} G(\textbf{x}, \textbf{x}') q_n(\textbf{x}') \thinspace d\textbf{x}' + \sum_{I=0}^{N_I} \left[ -i \omega L^I(\textbf{x}) Q^{I*} + D^{I}_{i}(\textbf{x}) u_{i}^{I*} \right] \\ 
    & +  i \omega C_p^c \sum_{m=1}^{NN + NS} \alpha(\textbf{x}^m) \Big[ \int_{\partial \mathcal{D}} K^c G_{,i'}(\textbf{x}, \textbf{x}') n_i(\textbf{x}') \Gamma(\textbf{x}', \textbf{x}^m) \thinspace d\textbf{x}' - \int_{\partial \mathcal{D}} G(\textbf{x}, \textbf{x}') \Gamma_{,i'}(\textbf{x}', \textbf{x}^m) n_i(\textbf{x}') \thinspace d\textbf{x}' \Big] 
\end{split}
    \label{eq:temp_final}
\end{equation}
where the temperature field is written in terms of boundary responses and eigen-fields. 

\subsection{Implementation of the DR-iBEM}
In the previous subsection, Eq. (\ref{eq:temp_final}) presents the integral formulation of temperature, which contains contribution from both boundary responses and eigen-fields. To further solve the integral formulation numerically, the DR-iBEM discretizes the boundary into $NN$ boundary nodes, $NE$ boundary elements, and some interior interpolation nodes $NS$, which improves the integral accuracy with the global interpolation function. 

Based on Eq. (\ref{eq:temp_final}), in addition to Eshelby's tensors $\bm{D}^I$ and $L^I$, four types of boundary integrals can be defined as, 

\begin{equation}
    \begin{aligned} 
    & H^e(\textbf{x}) = -\int_{\Omega^e} K^c G_{,i'}(\textbf{x}, \textbf{x}') n_i(\textbf{x}')  N^e(\textbf{x}')\thinspace d\textbf{x}',  \quad &S^e &= \int_{\Omega^e} G(\textbf{x}, \textbf{x}') N^e(\textbf{x}') \thinspace d\textbf{x}' \\ 
    & \Xi^e(\textbf{x}, \textbf{x}^m) = \int_{\Omega^e} K^c G_{,i'}(\textbf{x}, \textbf{x}') n_i(\textbf{x}')  N^e(\textbf{x}') \Gamma(\textbf{x}', \textbf{x}^m) \thinspace d\textbf{x}', \quad &\Pi^e &= \int_{\Omega^e} G(\textbf{x}, \textbf{x}') N^e(\textbf{x}') \Gamma_{,i'}(\textbf{x}, \textbf{x}^m) n_i(\textbf{x}') \thinspace d\textbf{x}'
    \end{aligned}
    \label{eq:BIE_def}
\end{equation}
where $N^e(\textbf{x})$ refers to the shape function within the $e^{th}$ boundary element $\Omega^e$. The integrals of $H^e$ and $S^e$ are the same as the boundary integrals for steady-state heat conduction \cite{yin2022inclusion}, which handles the boundary responses. In contrast, the integrals of $\Xi^e$ and $\Pi^e$ involve the potential function $\Gamma(\textbf{x}, \textbf{x}^m)$, which takes into account the heat generation rate effect. Although $H^e$, $S^e$, $\Xi^e$, and $\Pi^e$ are different boundary integrals in the continuous form, $\Xi^e$ and $\Pi^e$ in the discretization form, can be obtained by multiplying $H^e$ and $S^e$ with the potential function only, which does not require evaluating the boundary integral twice. 

Formally, the global matrix of the DR-iBEM can be established as, 

\begin{equation}
    \textbf{H} \textbf{U} + \textbf{S} \textbf{Q} + \textbf{M} \textbf{X} = 0
    \label{eq:form_global}
\end{equation}
where matrix components are listed as, 
\begin{equation}
\begin{aligned}
    & \textbf{H} = \begin{bmatrix} -c -H & \ldots & -i\omega L^I & D^I_{i} & \ldots \\ 
    \vdots & \vdots & \vdots & \vdots & \vdots \\
    -\Delta C_p^I H^I & \ldots & -i \omega \Delta C_p^I L^I &  \Delta C_p^I D_{i}^{I} & \ldots \\
    -\Delta K^I & \ldots & -i \omega \Delta K^I L^{I \prime} & \Delta K^I D_{i}^{I\prime} & \ldots \\
    \vdots & \vdots & \vdots & \vdots & \vdots
    \end{bmatrix} ,\quad
    \textbf{S} = \begin{bmatrix} S & \ldots & \Delta C_p^I S^I & \Delta K^I S^{I \prime} & \ldots  \end{bmatrix}^T \\
    & \textbf{E} = C_p^c \begin{bmatrix} \Xi & \ldots & \Delta C_p^I \Xi^I & \Delta K^I \Xi^{I \prime} & \ldots  \end{bmatrix}^T ,\quad
    \textbf{P} = C_p^c \begin{bmatrix} \Pi & \ldots & \Delta C_p^I \Pi^I & \Delta K^I \Pi^{I \prime} & \ldots  \end{bmatrix}^T \\
    & \textbf{M} = -i \omega \left( \textbf{E} - \textbf{P} \right) \begin{bmatrix} \beta \end{bmatrix}^{-1}, \quad \textbf{Q} = \begin{bmatrix} q_n \end{bmatrix}, \quad \textbf{X} =  \begin{bmatrix} u \end{bmatrix} ,\quad
    \textbf{U} = \begin{bmatrix} u & \ldots & Q^{I*} & u_{i}^{I*} & \ldots \end{bmatrix}
\end{aligned}
    \label{eq:global_mat}
\end{equation}
where $c$ in $\textbf{H}$ is the free term when the field point coincides with the source point with $c = 1, 0.5$ for interior points and boundary nodes, respectively; $\Delta C_p^I = C_p^c - C_p^I$ and $\Delta K^I = K^c - K^{I}$ represent the material mismatch in heat capacity and thermal conductivity, respectively; the superscript of $(.)^{\prime}$ represents the partial derivative with respect to the point, on which the boundary integral equation is established. 

Solving the global system of linear equations in Eq. (\ref{eq:form_global}), all boundary responses, interior temperature, and eigen-fields in each subdomain can be obtained. According to Eq. (\ref{eq:temp_final}), the temperature at any interior field point can be expressed formally, 

\begin{equation}
    u(\textbf{x}) = \textbf{H}(\textbf{x}) \textbf{U} + \textbf{S}(\textbf{x}) \textbf{Q} + \textbf{M}(\textbf{x}) \textbf{X}
    \label{eq:post_temp}
\end{equation}
where $\textbf{H}(\textbf{x}), \textbf{S}(\textbf{x}), \textbf{M}(\textbf{x})$ are three post-process coefficient matrices, which are related to the interior field point $\textbf{x}$. Because the free-term coefficient satisfies $c(\textbf{x}) = 1$ for interior field points, the $c$ term in Eq. (\ref{eq:global_mat}) has been moved to the left-hand side. Subsequently, the corresponding temperature gradient can be obtained by further differentiating Eq. (\ref{eq:post_temp}) with respect to $\textbf{x}$, $u_{,i}(\textbf{x}) = \textbf{H}_{,i}(\textbf{x}) \textbf{U} + \textbf{S}_{,i}(\textbf{x}) \textbf{Q} + \textbf{M}_{,i}(\textbf{x}) \textbf{X}$.

\subsection{Verification of the DR-iBEM} \label{sec:verify_circular}
This subsection provides numerical verification of the dual-reciprocity inclusion-based boundary method on local thermal fields. This section presents two case studies, (i) a single circular inhomogeneity; and (ii) two top-down circular inhomogeneities with intensive interactions. These two cases are solved through classic circular Eshelby's tensors. Two additional case studies using square sub-elements are presented in \ref{sec:verify_square}. The first case shows DR-iBEM convergence when refined square sub-elements approximate a circular inhomogeneity. The second case confirms that, for a square inhomogeneity, the local thermal fields are independent of the choice of the comparison medium when the eigen-fields are consistently solved.

\begin{figure}
    \centering
    \includegraphics[width=0.7\linewidth]{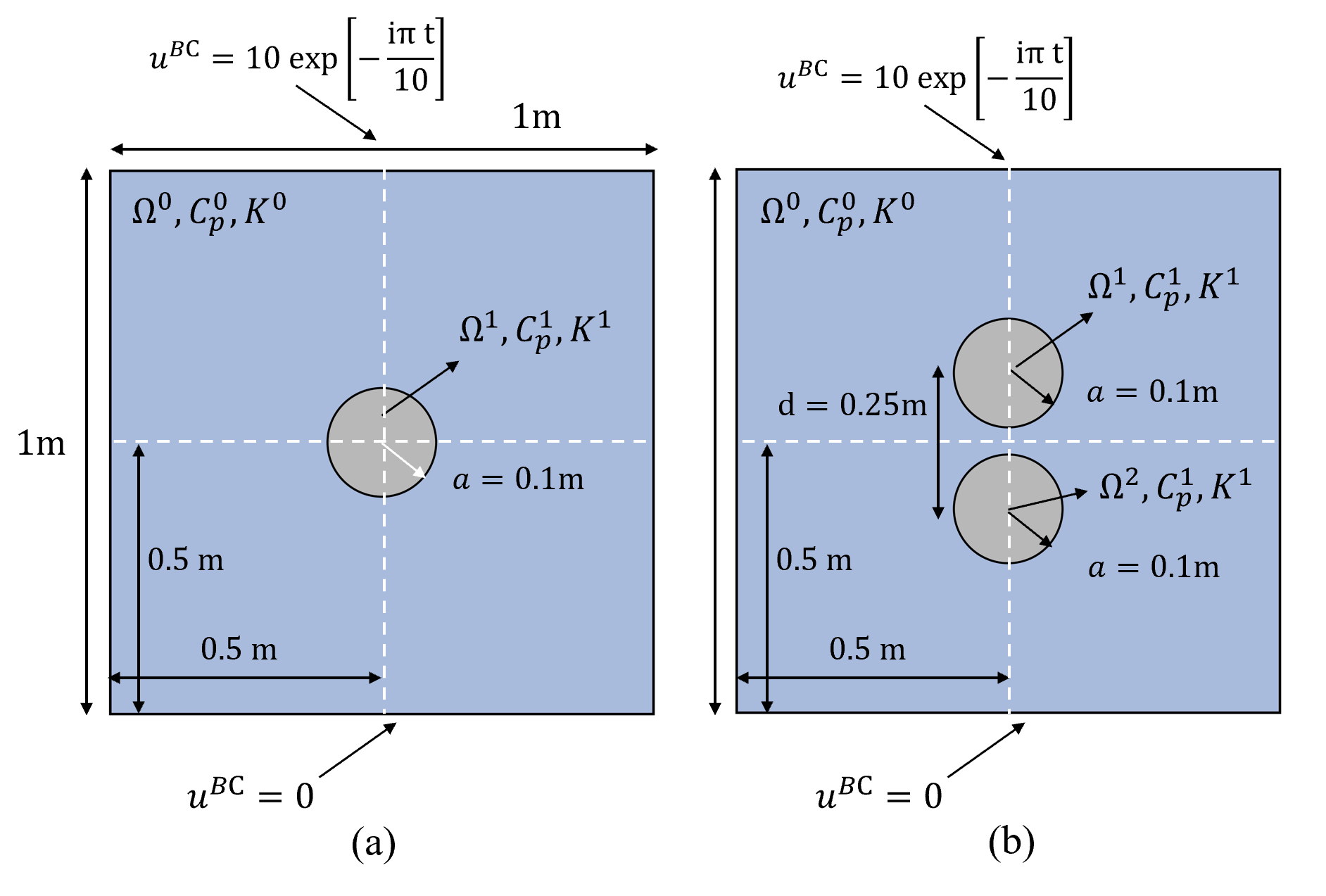}
    \caption{Geometry and boundary settings for numerical verification of the DR-iBEM algorithm on harmonic heat transfer. A square sample with length $1$ m contains (a) a single circular inhomogeneity placed at the center with radius $a = 0.1 m$; (b) two top-down circular inhomogeneities placed along the vertical centerline, and the distance between two centers is $d = 0.25$ m. The top line is subjected to harmonic thermal load $u^{BC}(t) = 10 \exp[-\frac{i \pi t}{10}]$, while the temperature of the bottom line is zero.}
    \label{fig:case}
\end{figure}

As Figs. \ref{fig:case} (a-b) indicate, the length of the square sample is $1$ m, which is subjected to prescribed harmonic thermal load: (i) the temperature of the top line is $u^{BC}(t) = 10 \exp[-\frac{i \pi t}{10}]$, while the temperature of the bottom line is zero; and (ii) all the lateral boundaries are adiabatic. The thermal properties are specified as $K^0 = 4 W / m \cdot K$, $K^1 = 8 W / m \cdot K$, $C_p^0 = 10 J / (m^3 \cdot K)$, and $C_p^1 = 5 J / (m^3 \cdot K)$. The DR-iBEM utilized $400$ boundary nodes and $200$ quadratic elements for the boundary mesh, and $64$ interior interpolation points are applied to improve the solution accuracy. The circular inhomogeneities do not require an interior mesh. 

The simulation results obtained in this section are compared with those predicted by the FEM software ANSYS 2026 R1. Because direct harmonic heat transfer analysis is not straightforward in the finite element software, verification is performed through a transient thermal analysis subjected to harmonic thermal loads. When the harmonic thermal load is applied for a sufficiently long time, such as $40$ s, the effects of the initial condition and transient responses decay, and the steady harmonic responses can be obtained. Subsequently, the magnitude and distribution of local thermal fields are compared with those from the DR-iBEM. Hence, the finite element analysis in this section demonstrates DR-iBEM's capability to compute boundary responses and reconstruct local thermal fields, which are crucial for evaluating averaged thermal fields. Because the FEM evaluation is only employed as a transient reference instead of a matched harmonic implementation, computational costs are not compared here. This section only lists computational usage of the DR-iBEM per case study. 

\subsubsection*{Case 1: A single circular inhomogeneity in a square sample}
\begin{figure}
    \centering
    \includegraphics[width=0.8\linewidth]{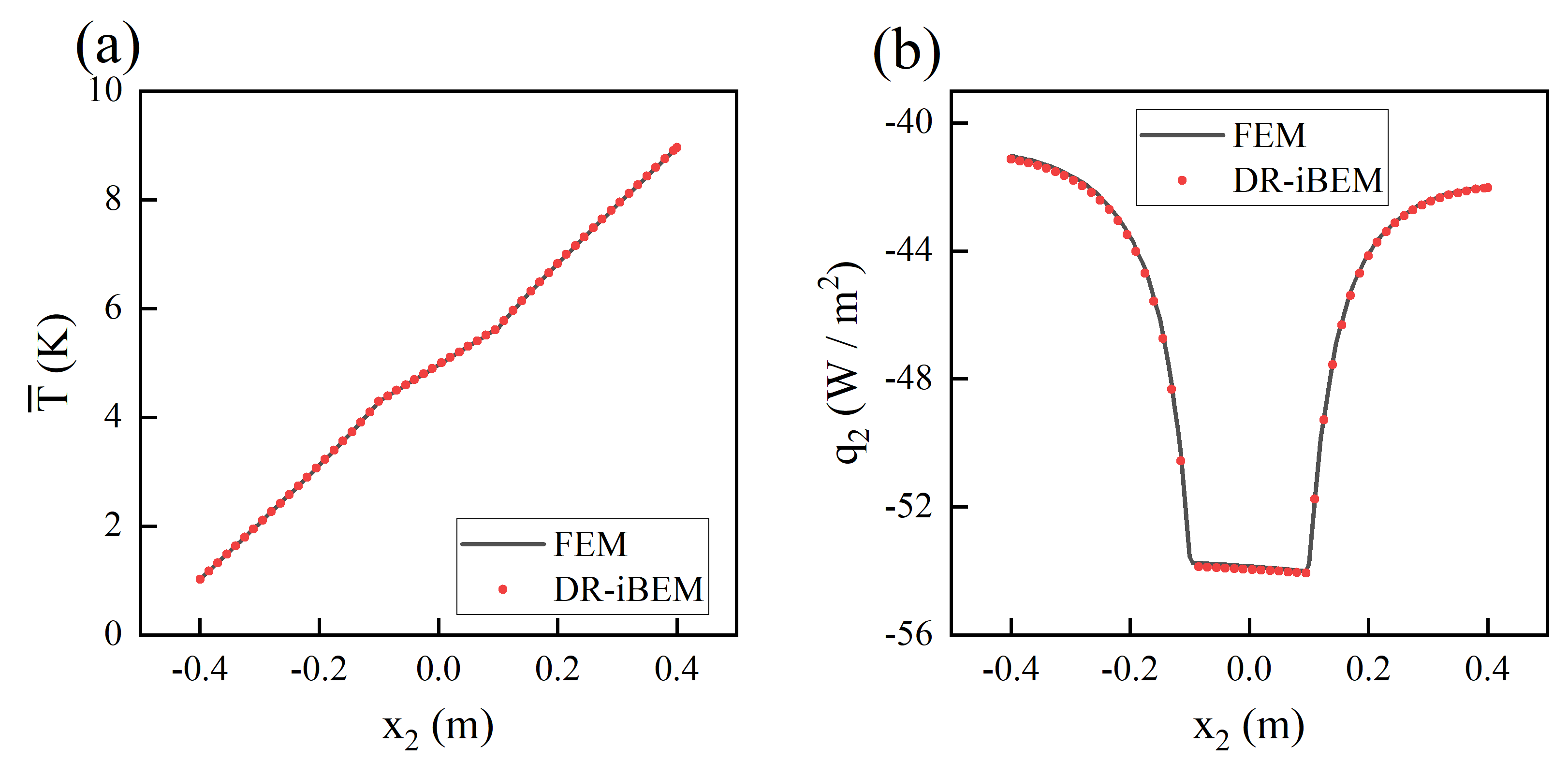}
    \caption{Comparison and variation of local thermal fields when the square sample contains a single circular inhomogeneity with radius $a = 0.1$ m. The (a) temperature and (b) heat flux are evaluated by the DR-iBEM and the FEM.}
    \label{fig:single_circle}
\end{figure}

Fig. \ref{fig:single_circle} plots the variation of local thermal fields along the vertical center line $x_2 \in [-0.4, 0.4]$ m. The DR-iBEM uses $0.045$ GB RAM and requires $1.489$ s for the computation. Since the top boundary is subjected to the harmonic thermal load, the temperature gradient in the vertical direction ($x_2$) is non-zero. When the excitation frequency is sufficiently low, the linear distribution of the unperturbed temperature can lead to a uniform ETG and an approximated linear EHS. However, unlike the steady-state problem, the temperature gradient is nonlinear when the excitation frequency $\omega \neq 0$, which results in non-uniform ETG and EHS in the inclusion $\Omega^1$. As shown in Fig. \ref{fig:single_circle} (a), the predictions by DR-iBEM with uniform eigen-fields approximation agree well with the finite element results, including the neighborhood of the interfaces. This agreement indicates that the uniform eigen-field approximation provides sufficient accuracy for the present single inhomogeneity configuration. 

\subsubsection*{Case 2: Two top-down circular inhomogeneities in a square sample}
\begin{figure}
    \centering
    \includegraphics[width=0.8\linewidth]{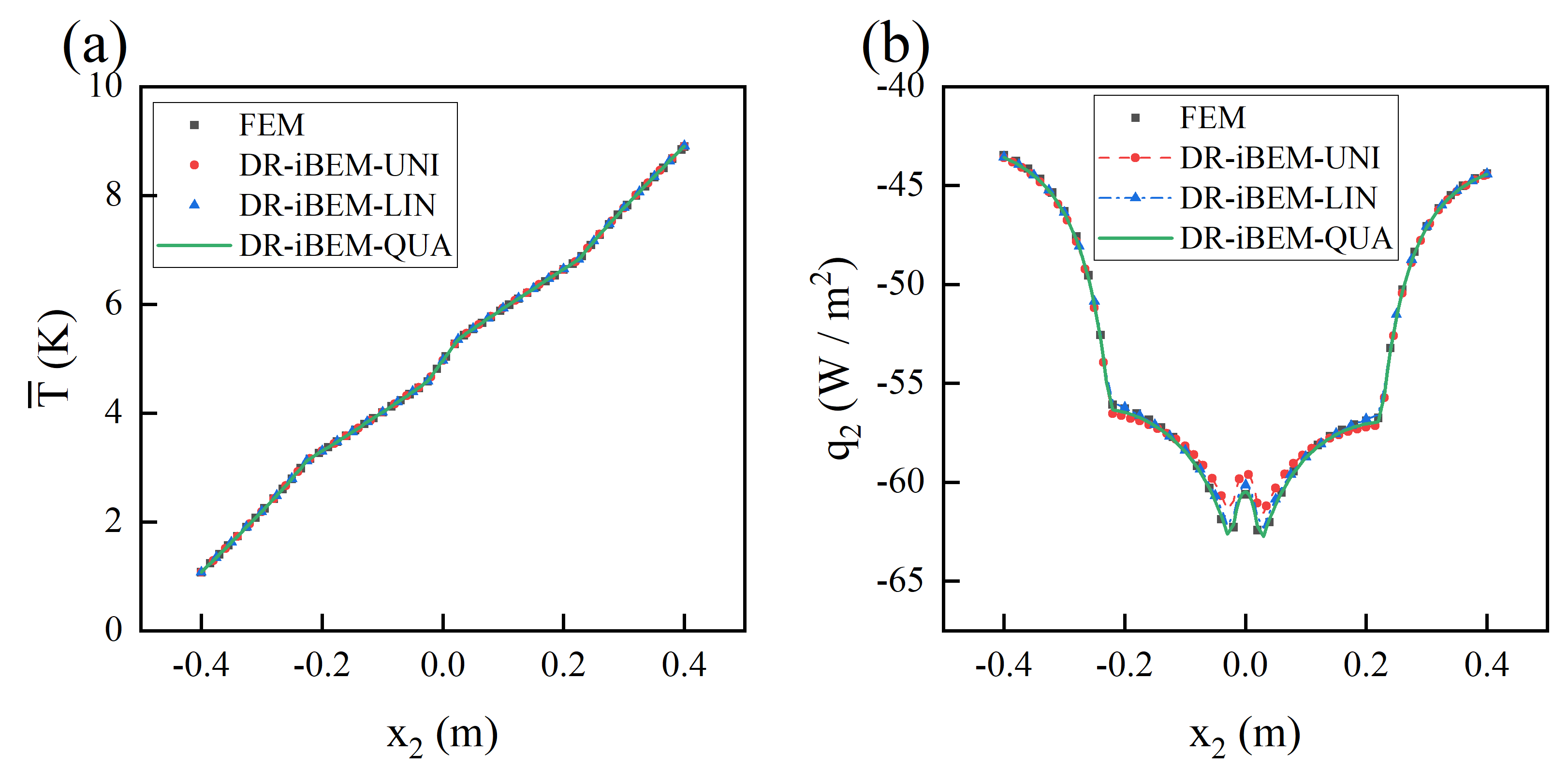}
    \caption{Comparison and variation of local thermal fields when the square sample contains two top-down circular inhomogeneities with radius $a = 0.1$ m, where the distance between two centers is $0.25$ m. The (a) temperature and (b) heat flux are evaluated by the DR-iBEM with three orders of polynomial-form eigen-fields. }
    \label{fig:two}
\end{figure}

Fig. \ref{fig:two} (a-b) plots the variation of local thermal fields when the square sample contains two top-down circular inhomogeneities, which are simulated through closed-form Eshelby's tensors. Due to the particle interactions, the assumption of uniform eigen-fields may be questionable. Following our previous work \cite{Wu2021,Wu2023,Wu2026}, we can extend the present formulation to polynomial eigen-fields and evaluate the accuracy and computational costs of the present assumption. Different orders of the eigen-fields are differentiated by UNI, LIN and QUA for uniform, linear, and quadratic eigen-fields, respectively. The DR-iBEM occupies $0.047,0.053,0.056$ GB RAM and $1.472,1.542,1.708$ s for uniform, linear, and quadratic eigen-fields, respectively. The distance between the two inhomogeneities' centers is $d = 0.25$ m, which results in intensive interactions between inhomogeneities. Note that even under steady-state heat conduction, such interactions can considerably disturb the local thermal fields (See Fig. 4(b) in \cite{Wu2023}), and the disturbance induced by particle interaction under harmonic heat transfer is even more significant. Based on the equivalent conditions in Eq. (\ref{eq:equiv_cond}), eigen-fields depend on the actual thermal fields multiplied by the difference in thermal properties, which implies that the non-uniform distribution of temperature and heat flux can lead to spatial variations of eigen-fields. Therefore, the uniform assumption of ETG and EHS can seldom provide accurate predictions on the local heat flux. 

As Fig. \ref{fig:two} (a) shows, the temperature predicted by DR-iBEM with three orders of polynomial-form eigen-fields agrees well with results by the FEM. Since the inhomogeneities exhibit greater thermal conductivity, the temperature gradients within them are smaller, and this feature is captured by all curves. In Eq. (\ref{eq:equiv_cond}), the pseudo-linear distribution of temperature and mismatch of heat capacity can lead to similar (non-uniform) variation of EHS, while it results in a pseudo-constant distribution of ETG within inhomogeneities. The minor discrepancies among the four curves indicate that the uniform ETG plays a significant role in the local disturbed temperature. However, Fig. \ref{fig:two} (b) clearly reveals that apparent discrepancies can be found between the ``DR-iBEM-UNI'' and ``FEM'' curves. Particularly, in the neighborhood of the interfaces, i.e., $x_2 = \pm 0.025$ m, the two curves exhibit the greatest differences. Because the equivalent conditions are constructed at the center of each subdomain, the curves ``DR-iBEM-UNI'' and ``FEM'' intersect at the centers, $x_2 = \pm 0.125$ m, as the equivalent conditions ensure consistency of the heat flux. As previously mentioned, the actual heat flux (temperature gradients) is no longer uniform; the uniform assumption of ETG cannot satisfy the equivalent heat flux condition at every interior point, which explains the greater discrepancies observed far from the center, particularly at interfaces. 

While the uniform eigen-fields approximation may not accurately predict the local thermal fields within the inhomogeneities, the discrepancies decay rapidly in the exterior regions. Such a phenomenon can be interpreted from both physical and mathematical aspects, (i) for harmonic heat transfer, the Green's function \cite{Wu2025a} is modified Helmholtz's potential functions, and the decaying behavior mixes the factor $r^{-1}$ and $\exp[-r]$ (including some other damping distance parameters), which exhibits faster decaying trend than the steady-state Green's function; (ii) the disturbances by eigen-fields are simulated by closed-form steady-state Eshelby's tensor, which exhibits $r^{-1}$, $r^{-2}$, and $r^{-3}$ for uniform, linear, and quadratic terms \cite{Mura1987}, respectively. Consequently, the influence of inaccurate eigen-fields predictions is restricted to a small region, for instance, within a distance comparable to several times the inhomogeneity radius from the center. The DR-iBEM with linear and quadratic eigen-fields improves the solution accuracy in the neighborhood of interfaces, particularly when the field point $x_2 \in [-0.025, 0.025]$ m, and minor discrepancies can be observed between DR-iBEM and FEM results. Based on two verification studies involving circular inhomogeneities and square-element approximations, the DR-iBEM algorithm can provide accurate and robust predictions for harmonic heat transfer, thereby supporting the subsequent work on Willis-type homogenization with averaged thermal fields.

\section{Homogenization scheme to extract effective properties under Bloch-form boundary conditions}
Based on the above-mentioned DR-iBEM, this section provides a detailed derivation of the homogenization scheme. In Eq. (\ref{eq:equiv_cond}), the heat flux and sensible heat storage can be expressed by the comparison medium and eigen-fields. Therefore, the averaged constitutive relations provide, 

\begin{equation}
    \langle F \rangle = \langle C_p u \rangle = C_p^c \langle u \rangle - \langle Q^* \rangle, \quad \langle q_i \rangle = \langle -K u_{,i} \rangle = -K^c \langle u_{,i} \rangle + K^c \langle u_{i}^* \rangle
    \label{eq:avg_cons}
\end{equation}
where the averaged sensible heat involves the averaged EHS, while the averaged heat flux relates to the averaged ETG. In steady-state thermal analysis, the effective thermal conductivity can be derived using the Fourier law, i.e., $\overline{K}_{ij} = -\langle q_i \rangle / \langle u_{,j} \rangle$. This treatment has a strong assumption that the averaged heat flux is merely related to the averaged temperature gradient, and similarly the averaged heat generation amount only relies on the averaged temperature change. However, in harmonic heat transfer, Eqs. (\ref{eq:equiv_cond}) and (\ref{eq:form_global}) clearly show that ETG and EHS are involved in the equivalent conditions, in which the two eigen-fields rely on both averaged temperature gradients and averaged temperature change. Therefore, the primary objective of this subsection is to express the averaged eigen-field in terms of the averaged temperature and temperature gradients, and thus the effective properties can be defined accordingly.

\subsection{Fluctuating thermal fields and their de-phased averages}
When the composite is subjected to the Bloch-form boundary conditions, the thermal fields can be written as, i.e., the temperature,

\begin{equation}
    u(\textbf{x}) = \hat{u}(\textbf{x}) \exp [i \bm{\zeta} \cdot \textbf{x}]
    \label{eq:thermal_}
\end{equation}
where $\bm{\zeta}$ refers to the 2D thermal wavenumber vector, and the symbol $\hat{(.)}$ denotes the de-phased thermal quantities, i.e., $\hat{u}$ is the de-phased temperature, which is a periodic function. When the weight function in Eq. (\ref{eq:def_avg}) is $f = \exp[-i \bm{\zeta} \cdot \textbf{x}]$, the de-phased average temperature becomes $\langle u \rangle = \langle \hat{u} \rangle^{f = 1}$. In the following, the de-phased thermal fields can be divided into two parts; the average and the fluctuation part, 

\begin{equation}
    \hat{u}(\textbf{x}) = \langle \hat{u} \rangle^{f=1} + \hat{u}^p(\textbf{x}) = \langle u \rangle + \hat{u}^p(\textbf{x}) 
\end{equation}
where $\langle \hat{u} \rangle^{f=1} = \langle u \rangle$ is applied. Here $\langle u \rangle$ is constant over the periodic composites; while $\hat{u}^p(\textbf{x})$ is a function of the position, which exhibits zero mean value over the composites. Moreover, the de-phased average temperature gradients and the average temperature are linearly dependent as, 

\begin{equation}
\begin{split}
    \langle u_{,i} \rangle & = \frac{1}{A_{\mathcal{D}}} \int_{\mathcal{D}} \exp[-i \zeta_m x_m] u_{,i}(\textbf{x}) \thinspace d\textbf{x} \\
    &= \frac{1}{A_\mathcal{D}} \int_{\mathcal{D}} \frac{\partial}{\partial x_i} \left( \exp[-i \zeta_m x_m] u(\textbf{x}) \right) \thinspace d\textbf{x} + i \zeta_i \frac{1}{A_{\mathcal{D}}} \int_{\mathcal{D}} \exp[-i \zeta_m x_m] u(\textbf{x}) \thinspace d\textbf{x} \\ 
    & = \underbrace{\frac{1}{A_{\mathcal{D}}} \int_{\partial \mathcal{D}} n_i(\textbf{x}) u(\textbf{x}) \exp[-i \zeta_m x_m] \thinspace d\textbf{x}}_{\text{Zero}} + i \zeta_i \frac{1}{A_{\mathcal{D}}} \int_{\mathcal{D}} \exp[-i \zeta_m x_m] u(\textbf{x}) \thinspace d\textbf{x} \\
    &= i \zeta_i \langle u \rangle
\end{split}
    \label{eq:relation_lin}
\end{equation}
where the Bloch-form boundary condition is used to zero out the boundary integral. Eq. (\ref{eq:relation_lin}) indicates that the macroscopic averages of temperature and temperature gradient are linearly dependent given the wavenumber vector. Hence, extracting effective properties directly from macroscopic thermal fields may lead to non-unique admissible constitutive representatives. This is the primary reason for introducing the residual fields, i.e., a residual temperature gradient $u_i^R$, so that $\langle u \rangle$ and $\langle u_{,i} - u_i^R \rangle$ are no longer linearly dependent. 

In general, the introduction of residual fields increases the computational burden that (i) the effects by residual fields are evaluated through domain integrals of Green's function, which generally require the entire domain calculation; (ii) the $n-$dimensional problem requires $n + 1$ independent loading cases, including one heat source-driven and $n$ residual-fields related problems, which requires evaluating averaged temperature, temperature gradients, sensible heat, and heat flux for $n + 1$ times. Such additional evaluations are comparatively manageable for 1D case or composites with a small number of inhomogeneities. Because the computational burdens significantly increase with the spatial dimension and number of inhomogeneities, it becomes cumbersome for 2D or 3D composites. Therefore, the present formulation does not determine the effective properties through such a retrieval procedure. Alternatively, we construct the operator relating the eigen-fields to generalized macroscopic inputs.

Based on the DR-iBEM post-process in Eq. (\ref{eq:post_temp}), the de-phased temperature at any interior field point $\textbf{x}$ can be written as, 

\begin{equation}
    \hat{u}(\textbf{x}) = \exp[-i \zeta_m x_m] \left( \textbf{H}(\textbf{x}) \textbf{U} + \textbf{S}(\textbf{x}) \textbf{Q} + \textbf{M}(\textbf{x}) \textbf{X} \right)
    \label{eq:temp_post}
\end{equation}
where the vectors $\textbf{U}, \textbf{Q}, \textbf{X}$ are invariants that do not change with the field point. Therefore, the de-phased average temperature can be obtained by averaging Eq. (\ref{eq:temp_post}) with $N_p$ sampling points, which can be obtained as, 

\begin{equation}
\begin{split}
    \langle u \rangle & = \frac{1}{A_{\mathcal{D}}} \sum_{p=1}^{N_p} \left\{ w(\textbf{x}^p) \exp[-i \zeta_m x^p_m] \left( \textbf{H}(\textbf{x}^p) \textbf{U} + \textbf{S}(\textbf{x}^p) \textbf{Q} + \textbf{M}(\textbf{x}^p) \textbf{X} \right) \right\}  
    = \overline{\textbf{H}} \textbf{U} + \overline{\textbf{S}} \textbf{Q} + \overline{\textbf{M}} \textbf{X}
\end{split}
    \label{eq:temp_post_avg}
\end{equation}
where $\textbf{x}^p$ represents the $p^\text{th}$ sampling point, and $w(\textbf{x}^p)$ is the weight function for $\textbf{x}^p$. For instance, $w(\textbf{x}^p) = \frac{A_{\mathcal{D}}}{N_p}$ for uniformly distributed sampling points. The de-phased averaged coefficient matrices are defined by the average of the sampling points. For example, 

\begin{equation}
\overline{\textbf{H}} = \frac{1}{A_{\mathcal{D}}} \sum_{p=1}^{N_p} w(\textbf{x}^p) \exp[-i \zeta_m x^p_m] \textbf{H}(\textbf{x}^p)
\label{eq:H_avg}
\end{equation}
while $\overline{\textbf{S}}$ and $ \overline{\textbf{M}}$ are defined in the same manner. 
As Eq. (\ref{eq:global_mat}) indicates, the vectors of $\textbf{U}, \textbf{Q}, \textbf{X}$ can be further assembled as a complete response vector, which can be expressed as, 

\begin{equation}
    \text{SVEC} = \begin{bmatrix} \bm{u}^B & \bm{u}^N & \bm{q}_n & \bm{Q}^* & \bm{u}^* \end{bmatrix}^T
    \label{eq:svec}
\end{equation}
where the superscripts of $\textbf{u}^B$ and $\textbf{u}^N$ refer to coefficients associated with boundary temperature and temperature at interior interpolation nodes, respectively; $\bm{q}_n$ is the outward heat flux; and $\bm{Q}^*$ and $\bm{u}^*$ are two eigen-fields. The vector $\textbf{X}$ collects the boundary and interior interpolation nodal temperatures, which is not an independent response vector. 

Subsequently, the local de-phased temperature in Eq. (\ref{eq:temp_post}) and its average in Eq. (\ref{eq:temp_post_avg}) can be further expressed with two larger coefficient matrix $\text{AMAT}(\textbf{x})$ and $\overline{\text{AMAT}}$ with the complete response vector $\text{SVEC}$,

\begin{equation}
\begin{aligned}
    \hat{u}(\textbf{x}) = \text{AMAT}(\textbf{x}) \thinspace \thinspace \text{SVEC}, \quad \langle u \rangle = \overline{\text{AMAT}} \thinspace \thinspace \text{SVEC}, \quad \hat{u}^p(\textbf{x}) = \underbrace{(\text{AMAT}(\textbf{x}) - \overline{\text{AMAT}})}_{\widetilde{\text{AMAT}}(\textbf{x})} \thinspace \thinspace \text{SVEC}
\end{aligned}
    \label{eq:temp_post_final}
\end{equation}
where 
\begin{equation}
\begin{aligned}
    \text{AMAT}(\textbf{x}) & = \begin{bmatrix} \textbf{H}^{B}(\textbf{x}) + \textbf{M}^{B}(\textbf{x}) & \textbf{M}^{N}(\textbf{x})& \textbf{S} (\textbf{x}) & \textbf{H}^{Q^*}(\textbf{x}) & \textbf{H}^{u^*} (\textbf{x})   \end{bmatrix} \\ 
    \overline{\text{AMAT}} & = \begin{bmatrix}
    \overline{\textbf{H}}^{B} + \overline{\textbf{M}}^{B} & \overline{\textbf{M}}^{N}& \overline{\textbf{S}} & \overline{\textbf{H}}^{Q^*} & \overline{\textbf{H}}^{u^*}   
    \end{bmatrix}
\end{aligned}
\end{equation}
where the superscripts $B$ and $N$ refer to coefficients associated with boundary temperature and temperature at interior interpolation nodes; while the superscripts $Q^*$ and $u^*$ stand for coefficients with the EHS and ETG, respectively. In particular, $\textbf{M}^B$ and $\textbf{M}^N$ are the partitions of the dual-reciprocity coefficient matrix $\textbf{M}$ related to $\textbf{u}^B$ and $\textbf{u}^N$, respectively. 

The periodic fluctuation part of the de-phased temperature is obtained by subtracting the average part from the local temperature field, which is shown as the third equation in Eq. (\ref{eq:temp_post_final}). The coefficient matrix $\text{AMAT}(\textbf{x})$ depends on the position of the field point, while the other one $\overline{\text{AMAT}}$ remains the same once the boundary conditions and microstructure are given. Consequently, the local de-phased temperature, its average, and its periodic fluctuation can be obtained through the same complete response vector. 

Although Eq. (\ref{eq:relation_lin}) has indicated that the de-phased average temperature gradients are correlated with the de-phased average temperature, this correlation only holds for de-phased average. For instance, when the conventional average without the de-phase factor, i.e., $w(\textbf{x}^p) \exp[-i \zeta_m x_m^p] = 1$ in Eq. (\ref{eq:H_avg}), the correlation is not applicable anymore. One example about the conventional average is illustrated in Section \ref{sec:conventional_average}. To obtain the general relationship between the temperature gradient and its average, we follow the same manner in Eq. (\ref{eq:temp_post_final}), and provide two coefficient matrices $\text{BMAT}(\textbf{x})$ and $\overline{\text{BMAT}}$ as follows: 

\begin{equation}
    \widehat{u_{,i}}(\textbf{x}) = \text{BMAT}(\textbf{x}) \thinspace \thinspace \text{SVEC}, \quad \langle u_{,i} \rangle = \overline{\text{BMAT}} \thinspace \thinspace \text{SVEC}, \quad \widehat{u_{,i}}^p(\textbf{x}) = \underbrace{(\text{BMAT}(\textbf{x}) - \overline{\text{BMAT}})}_{\widetilde{\text{BMAT}}(\textbf{x})} \thinspace \thinspace \text{SVEC}
    \label{eq:temp_grad_post_final}
\end{equation}
where 
\begin{equation}
    \begin{aligned}
    \text{BMAT}(\textbf{x}) & =\begin{bmatrix}
     \textbf{H}^{\prime B}(\textbf{x}) + \textbf{M}^{\prime B}(\textbf{x}) & \textbf{M}^{\prime N}(\textbf{x})& \textbf{S}^{\prime} (\textbf{x}) & \textbf{H}^{\prime Q^*}(\textbf{x}) & \textbf{H}^{\prime u^*} (\textbf{x})  
    \end{bmatrix} \\ 
    \overline{\text{BMAT}} & = \begin{bmatrix}
    \overline{\textbf{H}}^{\prime B} + \overline{\textbf{M}}^{\prime B} & \overline{\textbf{M}}^{\prime N} & \overline{\textbf{S}}^{\prime} & \overline{\textbf{H}}^{\prime Q^*} & \overline{\textbf{H}}^{\prime u^*}   
    \end{bmatrix}
    \end{aligned}
\end{equation}
where the superscript $(.)^{\prime}$ refers to the partial derivative with respect to field point $\textbf{x}$. 

\subsection{Relations between eigen-fields and averaged thermal fields}
As Eqs. (\ref{eq:temp_post_final}) and (\ref{eq:temp_grad_post_final}) indicate, the de-phased thermal fields and their averages can be expressed as linear operations of the solution vectors. This subsection aims to seek the relations between eigen-fields and averaged temperature and temperature gradients. Based on the local constitutive relations in Eq. (\ref{eq:equiv_cond}), the de-phased temperature and temperature gradients at the field point $\textbf{x}$ can be inversely expressed by eigen-fields, 

\begin{equation}
    \hat{u}(\textbf{x}) = \exp [-i\zeta_m x_m] \frac{Q^*(\textbf{x})}{C_p^c - C_p(\textbf{x})}, \quad \widehat{u_{,i}}(\textbf{x}) = \exp [-i\zeta_m x_m]\frac{K^c u_{i}^*(\textbf{x})}{K^c - K(\textbf{x})}
    \label{eq:thermal_by_eigen}
\end{equation}
which is satisfied at all field points when the comparison medium is different from the existing material phases. When the comparison medium is the matrix, Eq. (\ref{eq:thermal_by_eigen}) should only be applied within the inhomogeneities as the eigen-fields in the matrix are zero. However, the exact polarization method \cite{Willis1980} is applicable to an arbitrary selection of the comparison medium, which is confirmed in \ref{sec:verify_square} for local fields and \ref{sec:1D_case} for effective properties. Since the effective properties are independent of the comparison medium, we select the matrix as the comparison medium to  avoid constructing equivalent conditions in the matrix region, which reduces the total degrees of freedom. 

Without loss of generality, substituting Eq. (\ref{eq:thermal_by_eigen}) into Eq. (\ref{eq:temp_post_final}) yields the de-phased EHS at the $I^{\text{th}}$ subdomain: 

\begin{equation}
    \exp [-i\zeta_m x_m] \frac{Q^{I*}}{\Delta C_p^I} = \langle u \rangle + \hat{u}^p(\textbf{x}) \quad \text{or} \quad \exp [-i\zeta_m x_m] \frac{Q^{I*}}{\Delta C_p^I} - \hat{u}^p(\textbf{x}) = \langle u \rangle
    \label{eq:equiv_extend_1}
\end{equation}
Similarly, substituting Eq. (\ref{eq:thermal_by_eigen}) into Eq. (\ref{eq:temp_grad_post_final}) yields the de-phased ETG as: 

\begin{equation}
    \exp [-i\zeta_m x_m] \frac{K^c u_{i}^{I*}}{\Delta K^I} = \langle u_{,i} \rangle + \widehat{u_{,i}}^p(\textbf{x}) \quad \text{or} \quad \exp [-i\zeta_m x_m] \frac{K^c u_{i}^{I*}}{\Delta K^I} - \widehat{u_{,i}}^p(\textbf{x}) = \langle u_{,i} \rangle
    \label{eq:equiv_extend_2}
\end{equation}
Now substituting the detailed expression of fluctuation parts of the de-phased temperature in Eq. \eqref{eq:temp_post_final} and temperature gradients in Eq. \eqref{eq:temp_grad_post_final} into Eqs. (\ref{eq:equiv_extend_1}) and (\ref{eq:equiv_extend_2}) provides:

\begin{equation}
\begin{aligned}
    \langle u \rangle = \left( -\widetilde{\text{AMAT}}(\textbf{x}^{IC}) + \exp[-i \zeta_m x_m] \dfrac{1}{\Delta C_p^I} \textbf{R}_{Q^*}^{I} \right) \thinspace \thinspace \text{SVEC} = \mathcal{A}(\textbf{x}^{IC}) \thinspace \thinspace \text{SVEC} \\ 
    \langle u_{,i} \rangle = \left(-\widetilde{\text{BMAT}}(\textbf{x}^{IC}) + \exp[-i \zeta_m x_m] \dfrac{K^c}{\Delta K^I} \textbf{R}^{I}_{u^*} \right) \thinspace \thinspace \text{SVEC} =  \mathcal{B}_{i}(\textbf{x}^{IC}) \thinspace \thinspace \text{SVEC} \\
\end{aligned}
    \label{eq:detail_equiv}
\end{equation}
where the vectors $\textbf{R}_{Q^*}^I$ and $\textbf{R}_{u^*}^I$ denote the selection matrices that extract the EHS and ETG in the $\text{I}^{\text{th}}$ subdomain from the complete response vector, $Q^{I*} = \textbf{R}_{Q^*}^{I} \thinspace \text{SVEC}$ and $\textbf{u}^{I*} = \textbf{R}_{u^*}^I \thinspace \text{SVEC}$. Repeating such process over each subdomain, a global linear equation system can be reconstructed from Eq. \eqref{eq:detail_equiv} into the form of $\bm{\mathcal{C}} \thinspace \thinspace \text{SVEC} = \textbf{I} \thinspace \thinspace \hat{\textbf{U}}$ as follows: 

\begin{equation}
    \underbrace{\begin{bmatrix} \mathcal{A}(\textbf{x}^{1C}) \\ \mathcal{B}_{1}(\textbf{x}^{1C}) \\ \mathcal{B}_{2}(\textbf{x}^{1C}) \\ \vdots \\ \mathcal{A}(\textbf{x}^{N_I C}) \\ \mathcal{B}_{1}(\textbf{x}^{N_I C}) \\ \mathcal{B}_{2}(\textbf{x}^{N_I C})
    \end{bmatrix}}_{\mathcal{C}} \thinspace \thinspace \underbrace{\begin{bmatrix} \textbf{u}^B \\ \textbf{u}^N \\ \textbf{q}_n \\ \textbf{Q}^* \\ \textbf{u}^* \end{bmatrix}}_{\text{SVEC}} = 
    \begin{bmatrix} \langle u \rangle \\ \langle u_{,1} \rangle \\ \langle u_{,2} \rangle \\ \vdots \\ \langle u \rangle \\ \langle u_{,1} \rangle \\ \langle u_{,2} \rangle  \end{bmatrix}
    =
    \underbrace{\begin{bmatrix} \textbf{I}^{(3)} \\ \vdots \\ \textbf{I}^{(3)} \end{bmatrix}}_{\bm{\mathcal{I}}} \underbrace{\begin{bmatrix} \langle u \rangle \\ \langle u_{,1} \rangle \\ \langle u_{,2} \rangle \end{bmatrix}}_{\hat{\textbf{U}}}
    \label{eq:global_avg}
\end{equation} 
where $\textbf{I}^{(3)}$ refers to the $3 \times 3$ identity matrix. The eigen-fields of each subdomain are still related to the boundary responses and temperature at interior interpolation nodes. According to Eq. (\ref{eq:form_global}), the boundary responses, temperature at interior interpolation nodes can be expressed through the eigen-fields only under the Bloch-form boundary conditions. The left-hand side of Eq. (\ref{eq:global_avg}) can be equivalently expressed by a modified coefficient matrix multiplied by eigen-fields in each subdomain only. Consequently, the relations between eigen-fields in each subdomain and de-phased averages of the thermal quantities ($\hat{\textbf{U}}$) are obtained. The implementation of the Bloch-form boundary constraints and the corresponding elimination procedure are presented below.

\subsection{Rearrangement of the global matrix under the Bloch-form boundary condition}
Consider a rectangular composite sample centered at the origin, with horizontal and vertical lengths $L_1$ and $L_2$, respectively. The sample is subjected to the Bloch-form boundary condition with the thermal wavenumber vector $\bm{\zeta} = (\zeta_1, \zeta_2)$. Let $L, R$ and $U, D$ denote the thermal quantities on the left, right, upper, and lower boundaries, respectively. The Bloch-form boundary condition requires the temperature and outward heat flux to satisfy:

\begin{equation}
\begin{aligned}
    u^L \exp[i \zeta_1 L_1] = u^R, \quad q_n^L \exp[i \zeta_1 L_1] + q_n^R = 0 \\ 
    u^D \exp[i \zeta_2 L_2] = u^U, \quad q_n^D \exp[i \zeta_2 L_2] + q_n^U = 0
    \end{aligned}
    \label{eq:bloch-form}
\end{equation}
When the thermal wavenumber vector becomes zero, the Bloch-form boundary condition reduces to the periodic boundary condition in the steady-state homogenization.

The variables of the DR-iBEM global equation system in Eq. \eqref{eq:form_global} include: (i) boundary temperature, $NN$ variables; (ii) boundary outward heat flux, $NN$ variables (explained in the following); (iii) temperature at interior interpolation nodes, $NS$ variables; and (iv) eigen-fields within each subdomain, i.e., for the uniform case, $3 \times N_I$ variables. For boundary conditions such as prescribed temperature, heat flux, or convection, continuous boundary elements can be applied. However, under the Bloch-form boundary condition, both the temperature and outward heat flux on opposite boundaries are paired. Particularly, the corner nodes are shared with two sets of pairing conditions, which causes the collapse of two unknowns into a single unknown. Consequently, the outward heat flux (in two directions) of corner nodes becomes undefined. Following the treatment in \cite{Beer2008}, discontinuous corner elements are adopted in this work, so that one additional corner node is created to impose the Bloch-form boundary conditions in a consistent manner. 

With the pairing conditions in Eq. (\ref{eq:bloch-form}), the global matrix of the DR-iBEM can be expressed in terms of the one-sided boundary responses. Without the loss of any generality, this subsection adopts boundary responses on the left and lower boundaries. For compact notation, the boundary responses and coefficient components of the left/lower and right/upper are jointly denoted by superscripts $LL$ and $RU$, respectively. Eq. (\ref{eq:global_mat}) can be further divided into the following components,                              

\begin{equation}
\begin{aligned}
   \textbf{H} = \begin{bmatrix} \textbf{H}^{T, LL} & \textbf{H}^{T, RU} & \textbf{H}^{T, Q^*} & \textbf{H}^{T, u^*} \\ \textbf{H}^{E, LL} & \textbf{H}^{E, RU} & \textbf{H}^{E, Q^*} & \textbf{H}^{E,u^*} \end{bmatrix}, \quad \textbf{S} = \begin{bmatrix} \textbf{S}^{T, LL} & \textbf{S}^{T, RU} \\ S^{E, LL} & S^{E, RU} \end{bmatrix} \\
    \textbf{M} = \begin{bmatrix} \textbf{M}^{T, LL} & \textbf{M}^{T, RU} & \textbf{M}^{T, N} \\ \textbf{M}^{E, LL} & \textbf{M}^{E, RU} & \textbf{M}^{E, N} \end{bmatrix}
\end{aligned}
\end{equation}
where the superscript $T$ refers to integral equations constructed at boundary/interior nodes, and the superscript $E$ represents equivalent conditions established within each subdomain. For instance, $\textbf{H}^{T, LL}$ refers to matrix components of $\textbf{H}$ for integral equations at boundary/interior nodes with thermal quantities on the left/lower boundaries. Let $\bm{\lambda}_u = \text{diag}(\exp[i \zeta_1 L_1], \exp[i \zeta_2 L_2])$ and $\bm{\lambda}_{q} = - \bm{\lambda}_u$ denote Bloch-form phase factor matrices, 

\begin{small}
\begin{equation}
\begin{aligned}
    \bm{\mathcal{H}}^{LL} = \begin{bmatrix} \textbf{H}^{T, LL} + \textbf{H}^{T, RU} \bm{\lambda}_{u} + \textbf{M}^{T, LL} + \textbf{M}^{T, RU} \bm{\lambda}_{u} \quad \textbf{S}^{T, LL} + \textbf{S}^{T, RU} \bm{\lambda}_{q} \quad \textbf{M}^{T, N} \end{bmatrix} \\ \bm{\mathcal{S}}^{LL} = \begin{bmatrix} \textbf{u}^{LL} & \textbf{u}^N & \textbf{q}_n^{LL} \end{bmatrix}, \quad
    \begin{bmatrix} \bm{\mathcal{H}}^{LL} & \textbf{H}^{T, Q^*} & \textbf{H}^{T, u^*} \end{bmatrix} \begin{bmatrix} \bm{\mathcal{S}}^{LL} & \textbf{Q}^* & \textbf{u}^* \end{bmatrix}^T = \begin{bmatrix} \textbf{0} 
    \end{bmatrix}
\end{aligned}
    \label{eq:global_bloch}
\end{equation}
\end{small}
where $\bm{\mathcal{H}}^{ LL}$ and $\bm{\mathcal{S}}^{LL}$ collect contributions and responses from the left/lower boundaries and interior interpolation points, respectively. Subsequently, Eq. (\ref{eq:detail_equiv}) is partitioned based on the coefficients associated with boundary responses ($LL$) and eigen-fields ($\mathcal{E}$), 

\begin{equation}
    \text{SVEC} = \begin{bmatrix} \bm{\mathcal{S}}^{LL} & \bm{\mathcal{E}} \end{bmatrix}^T, \quad \bm{\mathcal{E}} = \begin{bmatrix} \textbf{Q}^* & \textbf{u}^* \end{bmatrix}, \quad \bm{\mathcal{C}} = \begin{bmatrix} \bm{\mathcal{C}^{LL}} & \bm{\mathcal{C}}^{\mathcal{E}} \end{bmatrix}
    \label{eq:partition_detail}
\end{equation}
where $\bm{\mathcal{E}}$ collects all eigen-fields in subdomains. Hence, the boundary responses and temperature at interior interpolation points can be expressed in terms of eigen-fields only, 

\begin{equation}
    \bm{\mathcal{S}}^{LL} = -\left( \bm{\mathcal{H}}^{LL} \right)^{-1} \left[ \textbf{H}^{T, Q^*} Q^* + \textbf{H}^{T, u^*} u_{i}^{*} \right] \quad \text{or} \quad \bm{\mathcal{S}}^{LL} = -\left( \bm{\mathcal{H}}^{LL} \right)^{-1} \bm{\mathcal{H}}^{\mathcal{E}} \bm{\mathcal{E}} 
    \label{eq:temp_eigen}
\end{equation}
where $\bm{\mathcal{H}}^{\mathcal{E}} = [\textbf{H}^{T, G} \thinspace \thinspace \textbf{H}^{T, H}]$. Substituting Eq. (\ref{eq:temp_eigen}) into Eq. (\ref{eq:global_avg}) provides, 

\begin{equation}
    \bm{\mathcal{C}}^{LL} \bm{\mathcal{S}}^{LL} + \bm{\mathcal{C}}^{\mathcal{E}} \bm{\mathcal{E}} = \bm{\mathcal{I}} \hat{\textbf{U}}
\end{equation}
Consequently, the eigen-fields in each subdomain can be related to the de-phased average temperature and temperature gradients, 

\begin{equation}
    \bm{\mathcal{E}} = \bm{\mathcal{R}} \thinspace \hat{\textbf{U}}, \qquad \bm{\mathcal{R}} = \left[ \bm{\mathcal{C}}^{\mathcal{E}} - \bm{\mathcal{C}}^{LL} \left( \bm{\mathcal{H}}^{LL}\right)^{-1} \bm{\mathcal{H}}^{\mathcal{E}} \right]^{-1} \thinspace \bm{\mathcal{I}}
\end{equation}
where matrix $\bm{\mathcal{R}}$ bridges the eigen-fields in each subdomain to the de-phased averaged (macroscopic) temperature/temperature gradients, which exhibits the dimension $(3 \times N_I) \times 3$. Although the de-phased temperature and its gradient satisfy the Bloch kinematic identity in Eq. (\ref{eq:relation_lin}), the three columns of $\bm{\mathcal{R}}$ are defined through independent generalized macroscopic inputs. The columns of $\bm{\mathcal{R}}$ should not be interpreted as independent physical Bloch solutions, but form the basis to identify the macroscopic constitutive operator. Once the constitutive operator is determined, the physical Bloch state automatically satisfies the kinematic relation in Eq. (\ref{eq:relation_lin}). Therefore, the complete constitutive matrix is determined at the operator level, which avoids introducing residual fields to construct several independent macroscopic solutions. Provided that the eigen-field operator, $\left[ \bm{\mathcal{C}}^{\mathcal{E}} - \bm{\mathcal{C}}^{LL} \left( \bm{\mathcal{H}}^{LL}\right)^{-1} \bm{\mathcal{H}}^{\mathcal{E}} \right]$, is not singular, the eigen-fields are uniquely determined for any generalized macroscopic fields. Because the averaged eigen-fields in Eq. (\ref{eq:avg_dephase_eigens}) are obtained through a linear projection of the eigen-fields, the effective properties in Eq. (\ref{eq:effec_properties}) are uniquely determined.

With the aid of  $\bm{\mathcal{R}}$, the averaged equivalent conditions in Eq. (\ref{eq:avg_cons}) can be converted into the effective constitutive relations. Specifically, the averaged eigen-fields can be written as, 

\begin{equation}
    \begin{bmatrix} \langle Q^* \rangle \\ \langle u_{1}^* \rangle \\ \langle u_{2}^* \rangle \end{bmatrix} = \begin{bmatrix} \mathcal{G}_{11} & \mathcal{G}_{12} & \mathcal{G}_{13} \\  \mathcal{G}_{21} & \mathcal{G}_{22} & \mathcal{G}_{23} \\ \mathcal{G}_{31} & \mathcal{G}_{32} & \mathcal{G}_{33} \end{bmatrix} \begin{bmatrix} \langle u \rangle \\ \langle u_{,1} \rangle \\ \langle u_{,2} \rangle \end{bmatrix}
    \label{eq:avg_dephase_eigens}
\end{equation}
where the matrix components $\mathcal{G}_{ij} = \sum_{I=1}^{N_I} \frac{A_{\Omega_{I}}}{A_{\mathcal{D}}} \mathcal{R}(3(I-1) + i,j)$, in which $\mathcal{R}(3(I-1) + i, j)$ refers to the component of the matrix at the $(3(I-1) + i)^\text{th}$ row and $j^\text{th}$ column;  $A_{\Omega_I}/A_{\mathcal{D}}$ refers to the weight (area fraction) of the $\text{I}^\text{th}$ subdomain. Substituting the above equation into Eq. (\ref{eq:avg_cons}) and collecting coefficients as Eq. (\ref{eq:cons_law}) provides, 

\begin{equation}
    \begin{aligned}
    \langle F \rangle = (C_p^c - \mathcal{G}_{11}) \langle u \rangle - \mathcal{G}_{12} \langle u_{,1} \rangle - \mathcal{G}_{13} \langle u_{,2} \rangle \\ 
    \langle q_{,1} \rangle = K^c \mathcal{G}_{21} \langle u \rangle - (K^c - K^c \mathcal{G}_{22}) \langle u_{,1} \rangle - K^c \mathcal{G}_{23} \langle u_{,2} \rangle \\
    \langle q_{,2} \rangle = K^c \mathcal{G}_{31} \langle u \rangle - K^c \mathcal{G}_{32} \langle u_{,1} \rangle - (K^c - K^c \mathcal{G}_{33}) \langle u_{,2} \rangle
    \end{aligned}
\end{equation}
where the effective properties in Eq. (\ref{eq:cons_law}) can be derived explicitly as, 

\begin{equation}
    \begin{bmatrix}
    \langle F \rangle \\ \langle q_{,1} \rangle \\ \langle q_{,2} \rangle 
    \end{bmatrix} = \begin{bmatrix} C_p^c - \mathcal{G}_{11} & -\mathcal{G}_{12} &  -\mathcal{G}_{13} \\ K^c \mathcal{G}_{21} & -K^c + K^c \mathcal{G}_{22} & -K^c \mathcal{G}_{23} \\ K^c \mathcal{G}_{31} & -K^c \mathcal{G}_{32} & -K^c + K^c \mathcal{G}_{33} \end{bmatrix} 
    \begin{bmatrix} \langle u \rangle \\ \langle u_{,1} \rangle \\ \langle u_{,2} \rangle \end{bmatrix}
    \label{eq:effec_properties}
\end{equation}
which provides the complete Willis-type thermal constitutive operator obtained from the eigen-field formulation. Instead of extracting thermal properties from several independent macroscopic solutions, no residual field is introduced. The constitutive relations are determined through the relations between the eigen-fields and generalized macroscopic fields. The current formulation is validated against Willis' exact homogenization method on a 1D laminate, which employs the residual temperature gradient field in \ref{sec:1D_case}.

\section{Fourier-series-based homogenization and symmetry relations}
\label{sec:fourier_main}
To verify the results of the DR-iBEM-based homogenization framework for the effective thermal properties of a 2D particulate composite, this section develops a Fourier-series-based homogenization method for periodic composites containing square inhomogeneities, in parallel with Srivastava and Nemat-Nasser's work on elastodynamics \cite{Srivastava2011}, which utilized the eigen-momentum and eigenstress to simulate mismatch in mass density and stiffness, respectively. Unlike the DR-iBEM framework using Green's function and Eshelby's tensors, the Fourier formulae handle the heterogeneous composite problem in the reciprocal space. Hence, the two methods can cross verify each other on the prediction of Willis-type effective thermal properties. In the following, the Fourier formulae are first derived by expanding the fluctuation thermal fields into spatial Fourier series, which quantifies fluctuation of thermal fields caused by two eigen-fields in subdomains. Subsequently, a global linear equation system is established to determine the relations between averaged eigen-fields and averaged temperature/temperature gradients, providing Willis-type effective thermal properties. Based on the properties of Fourier operators, we explicitly establish reciprocity relations connecting the effective properties at opposite wavenumbers. For a microstructure with a reflective symmetry, additional parity relations are derived that identify which effective properties are even or odd with respect to the corresponding wavenumber component and explain why the Willis coupling terms must vanish under symmetry-compatible wavenumbers. The complete proofs of these relations are provided in \ref{sec:proof}. Cross-verification against the DR-iBEM formulation is presented in Section 6, while the one-dimensional analytical solution benchmark is provided in \ref{sec:1D_case}.

\subsection{Fourier representation of fluctuating thermal fields}
Based on the de-phased definition in Eq. (\ref{eq:thermal_}), the constitutive relations with eigen-fields can be written as, 
\begin{equation}
    \hat{F}(\textbf{x}) = C_p^c \hat{u}(\textbf{x}) - \hat{Q}^{*}(\textbf{x}), \quad \hat{q}_i(\textbf{x}) = -K^c \left(i \zeta_i + \frac{\partial}{\partial x_i} \right) \hat{u}(\textbf{x})+ K^c \hat{u}_{i}^*(\textbf{x})
    \label{eq:eigen_dephase}
\end{equation}
Substituting Eq. (\ref{eq:eigen_dephase}) into the heat equation provides, 
\begin{equation}
    K^c \left( i \zeta_i + \frac{\partial}{\partial x_i} \right) \left( i \zeta_i + \frac{\partial}{\partial x_i} \right) \hat{u}(\textbf{x}) + i \omega C_p^c \hat{u}(\textbf{x}) = i \omega \hat{Q}^*(\textbf{x}) + K^c \left( i \zeta_i + \frac{\partial}{\partial x_i} \right) \hat{u}_{i}^{*}(\textbf{x}) 
    \label{eq:heat_fluc}
\end{equation}
where $\hat{Q}^*$ and $\hat{u}_{i}^*$ represent the de-phased EHS and ETG, respectively. Here, the operator $\left( i \zeta_i + \frac{\partial}{\partial x_i} \right)$ exists due to $\frac{\partial u(\textbf{x})}{\partial x_i} = i \zeta_i \hat{u}(\textbf{x}) \exp[i \bm{\zeta} \cdot \textbf{x}] + \hat{u}_{,i}(\textbf{x}) \exp[i \bm{\zeta} \cdot \textbf{x}]$, where $\hat{u}(\textbf{x})$ is the de-phased temperature. Because the sample is subjected to the Bloch-form boundary conditions, the thermal quantities can be expressed through Fourier series. For instance, the temperature can be written as, 
\begin{equation}
\begin{aligned}
    \hat{u}(\textbf{x}) = \langle \hat{u} \rangle^{f=1} + \hat{u}^p(\textbf{x}) = \breve{u}(\textbf{0}) + \sum_{\bm{\eta} \neq 0} \breve{u}(\bm{\eta}) \exp[i \bm{\eta} \cdot \textbf{x}] \\ 
    \breve{u}(\bm{\eta}) = \frac{1}{A_{\mathcal{D}}} \int_{\mathcal{D}} \exp[-i \bm{\eta} \cdot \textbf{x}] \hat{u}(\textbf{x}) \thinspace d\textbf{x}, \quad \bm{\eta} = \left( \frac{2 \pi m_1}{L_1}, \frac{2 \pi m_2}{L_2}\right), \quad m_1, m_2\in\mathbb{Z}
\end{aligned}
    \label{eq:fourier_series}
\end{equation}
when $\bm{\eta} = \textbf{0}$, the zero Fourier component reduces to the conventional spatial average of the de-phased field, $\breve{u}(\textbf{0}) = \langle \hat{u} \rangle^{f=1} = \langle u \rangle$; and the accent $\breve{(.)}$ refers to the Fourier coefficient. 




\subsection{Fluctuating thermal fields and eigen-fields}
Since the fluctuation part is expressed in terms of Fourier series, the left-hand side of Eq. (\ref{eq:heat_fluc}) can be further simplified as, 
\begin{equation}
    \begin{aligned}
     &K^c \left( i \zeta_i + \frac{\partial}{\partial x_i} \right) \left( i \zeta_i + \frac{\partial}{\partial x_i} \right) \hat{u}(\textbf{x}) + i \omega C_p^c \hat{u}(\textbf{x}) \\ 
     &= \left[ i \omega C_p^c -K^c \zeta_i \zeta_i \right] \breve{u}(\textbf{0}) +  \sum_{\bm{\eta} \neq 0} \breve{u}(\bm{\eta}) \left[ i\omega C_p^c - K^c \left(\zeta_i + \eta_i \right) \left(\zeta_i + \eta_i \right)  \right] \exp[i \bm{\eta} \cdot \textbf{x}] \\ 
     & = \sum_{\bm{\eta}} \left[ i\omega C_p^c - K^c \left(\zeta_i + \eta_i \right) \left(\zeta_i + \eta_i \right)  \right] \breve{u}(\bm{\eta}) \exp[i \bm{\eta} \cdot \textbf{x}]
    \end{aligned}
    \label{eq:heat_fluc_sim}
\end{equation}
Following the similar procedure, the right-hand side of Eq. (\ref{eq:heat_fluc}) can be rewritten as, 
\begin{equation}
    i \omega \hat{Q}^* + K^c \left( i \zeta_i + \frac{\partial}{\partial x_i} \right) \hat{u}_{i}^{*} = \sum_{\bm{\eta}} \left[ i \omega \breve{Q}^*(\bm{\eta})  + i K^c (\zeta_i + \eta_i) \breve{u}_{i}^*(\bm{\eta}) \right] \exp[i \bm{\eta} \cdot \textbf{x}]
    \label{eq:heat_fluc_sim_2}
\end{equation}
where the de-phased eigen-fields are expressed as $\hat{Q}^*(\textbf{x}) = \sum_{\bm{\eta}}\breve{Q}^*(\bm{\eta}) \exp[i\bm{\eta}\cdot\textbf{x}]$ and $\hat{u}_{i}^{*}(\textbf{x}) = \sum_{\bm{\eta}}\breve{u}_{i}^*(\bm{\eta}) \exp[i\bm{\eta}\cdot\textbf{x}]$. The equivalence of Eqs. (\ref{eq:heat_fluc_sim}) and (\ref{eq:heat_fluc_sim_2}) provides the following equation at each base wave: 
\begin{equation}
    \left[ i \omega C_p^c - K^c (\zeta_m + \eta_m) (\zeta_m + \eta_m) \right] \breve{u}(\bm{\eta}) = i \omega \breve{Q}^{*}(\bm{\eta}) + i K^c (\zeta_i + \eta_i) \breve{u}_{i}^{*}(\bm{\eta})
\end{equation}
Therefore, for non-zero Fourier components, the fluctuation coefficient for temperature and temperature gradients can be formally expressed as follows:
\begin{equation}
\begin{aligned}
    & \hat{u}^{p}(\textbf{x}) = \sum_{\bm{\eta} \neq 0} \left[ \mathcal{F}(\bm{\zeta}, \bm{\eta}) \breve{Q}^*(\bm{\eta}) + \mathcal{M}_{i}(\bm{\zeta}, \bm{\eta}) \breve{u}^*_{i}(\bm{\eta}) \right] \exp[i \bm{\eta} \cdot \textbf{x}] \\ 
    & \widehat{u_{,i}}^p(\textbf{x}) =  \sum_{\bm{\eta} \neq 0} \left[ \mathcal{N}_{i}(\bm{\zeta}, \bm{\eta}) \breve{Q}^*(\bm{\eta}) + \mathcal{T}_{ij}(\bm{\zeta}, \bm{\eta}) \breve{u}_{j}^{*}(\bm{\eta}) \right] \exp[i \bm{\eta} \cdot \textbf{x}]
\end{aligned}
    \label{eq:temp_fluc_series}
\end{equation}
where the temperature gradients related to temperature through the factor $i \zeta_i + i \eta_i$ for each corresponding Fourier components is used; and four tensors are defined as, 
\begin{equation}
\begin{aligned}
    \mathcal{F}(\bm{\zeta}, \bm{\eta}) = \frac{i \omega }{i \omega C_p^c - K^c (\zeta_m + \eta_m) (\zeta_m + \eta_m)}, \quad \mathcal{M}_{i}(\bm{\zeta}, \bm{\eta}) = \frac{i (\zeta_i + \eta_i) K^c}{i \omega C_p^c - K^c (\zeta_m + \eta_m) (\zeta_m + \eta_m)} \\ 
    \mathcal{N}_{i}(\bm{\zeta}, \bm{\eta}) = \frac{- \omega (\zeta_i + \eta_i) }{i \omega C_p^c - K^c (\zeta_m + \eta_m) (\zeta_m + \eta_m)}, \quad \mathcal{T}_{ij}(\bm{\zeta}, \bm{\eta}) = \frac{- (\zeta_i + \eta_i)  (\zeta_j + \eta_j) K^c}{i \omega C_p^c - K^c (\zeta_m + \eta_m) (\zeta_m + \eta_m)}
\end{aligned}
    \label{eq:four_tensors}
\end{equation}
which satisfy the following properties on reciprocal-lattice vectors: 
\begin{equation}
    \begin{aligned}
    \mathcal{F}(-\bm{\zeta}, -\bm{\eta}) = \mathcal{F}(\bm{\zeta}, \bm{\eta}), \quad -\mathcal{M}_i(-\bm{\zeta}, -\bm{\eta}) = \mathcal{M}_i(\bm{\zeta}, \bm{\eta}) \\ 
    -\mathcal{N}_{i}(-\bm{\zeta}, -\bm{\eta}) = \mathcal{N}_{i}(\bm{\zeta}, \bm{\eta}), \quad \mathcal{T}_{ij}(-\bm{\zeta}, -\bm{\eta}) = \mathcal{T}_{ij}(\bm{\zeta}, \bm{\eta})\\
    \mathcal{M}_{i}(\bm{\zeta},\bm{\eta})=\frac{iK^c}{\omega} \mathcal{N}_{i}(-\bm{\zeta},-\bm{\eta}).
    \end{aligned}
    \label{eq:fourier_kernel_property}
\end{equation}
The above properties will be applied to derive the opposite-wavenumber reciprocity relation. 

\subsection{The Effect of eigen-fields on effective thermal properties}
The de-phased temperature field can be decomposed into the average part and eigen-field induced fluctuation part, 
\begin{equation}
    \begin{aligned}
    & \hat{u}(\textbf{x}) = \langle u \rangle + \sum_{\bm{\eta} \neq 0} \left[ \mathcal{F}(\bm{\zeta}, \bm{\eta}) \breve{Q}^*(\bm{\eta}) + \mathcal{M}_{i}(\bm{\zeta}, \bm{\eta}) \breve{u}^*_{i}(\bm{\eta}) \right] \exp[i \bm{\eta} \cdot \textbf{x}] \\ 
    & \widehat{u_{,i}}(\textbf{x}) = \langle u_{,i} \rangle + \sum_{\bm{\eta} \neq 0} \left[ \mathcal{N}_{i}(\bm{\zeta}, \bm{\eta}) \breve{Q}^*(\bm{\eta}) + \mathcal{T}_{ij}(\bm{\zeta}, \bm{\eta}) \breve{u}_{j}^{*}(\bm{\eta}) \right] \exp[i \bm{\eta} \cdot \textbf{x}]
    \end{aligned}
    \label{eq:temp_series}
\end{equation}
where conventional spatial average of de-phased fields is replaced with de-phased average of actual fields. The purpose of the homogenization is to establish the relation between the de-phased eigen-fields and the de-phased averaged thermal quantities, such as temperature and temperature gradients. Specifically, we derive the relation by averaging the fluctuation fields over each subdomain, which is in analogy with the previous work \cite{Srivastava2011}. However, when the inhomogeneity is divided into adequately small elements, the de-phased eigen-fields can be assumed as uniform over each small subdomain. This approximation is consistent with the proposed DR-iBEM algorithm on square elements. Therefore, the spatial averages of fluctuation temperature and temperature gradients over each subdomain $\Omega^I$ are obtained as: 
\begin{equation}
\begin{aligned}
    \hat{u}^{p,I} = \frac{1}{A_{\Omega^I}} \int_{\Omega^{I}} \hat{u}^{p}(\textbf{x}) \thinspace d\textbf{x} = \sum_{\bm{\eta} \neq 0} g^I(\bm{\eta}) \left[ \mathcal{F}(\bm{\zeta}, \bm{\eta}) \breve{Q}^{*}(\bm{\eta}) + \mathcal{M}_{i}(\bm{\zeta}, \bm{\eta}) \breve{u}_{i}^{*}(\bm{\eta}) \right] \\ 
    \widehat{u_{,i}}^{p, I} = \frac{1}{A_{\Omega^I}} \int_{\Omega^I}  \widehat{u_{,i}}^p(\textbf{x}) \thinspace d\textbf{x} = \sum_{\bm{\eta} \neq 0} g^{I}(\bm{\eta}) \left[ \mathcal{N}_{i}(\bm{\zeta}, \bm{\eta}) \breve{Q}^{*}(\bm{\eta}) + \mathcal{T}_{ij}(\bm{\zeta}, \bm{\eta}) \breve{u}_{j}^{*}(\bm{\eta}) \right]
\end{aligned}
    \label{eq:fluctuation_part}
\end{equation}
where $A_{\Omega^I}$ refers to the area of the $I^{th}$ subdomain (rectangle); and $g^{I}(\bm{\eta}) = \dfrac{1}{A_{\Omega^I}} \int_{\Omega^I} \exp[i \bm{\eta} \cdot \textbf{x}] \thinspace d\textbf{x}$. The Fourier coefficients $\breve{Q}^{*}(\bm{\eta})$ and $\breve{u}_{i}^{*}(\bm{\eta})$ can be replaced with finite sum \cite{Srivastava2011}, 
\begin{equation}
\begin{aligned}
    \breve{Q}^*(\bm{\eta}) = \frac{1}{A_{\mathcal{D}}} \int_{\mathcal{D}} \hat{Q}^*(\textbf{x}) \exp[-i \bm{\eta} \cdot \textbf{x}] \thinspace d\textbf{x} = \sum_{I=0}^{N_I} \frac{A_{\Omega^I}}{A_{\mathcal{D}}} g^{I}(-\bm{\eta}) \thinspace \hat{Q}^{I*} \\ 
    \breve{u}^*_{i}(\bm{\eta}) = \frac{1}{A_{\mathcal{D}}} \int_{\mathcal{D}} \hat{u}_i^*(\textbf{x}) \exp[-i \bm{\eta} \cdot \textbf{x}] \thinspace d\textbf{x} = \sum_{I=0}^{N_I} \frac{A_{\Omega^I}}{A_{\mathcal{D}}} g^{I}(-\bm{\eta}) \thinspace \hat{u}_i^{I*}
\end{aligned}
    \label{eq:finite_sum}
\end{equation}
where the eigen-fields $\hat{Q}^{I*}$ and $\hat{u}^{I*}_i$ are uniform (de-phased) eigen-fields within the $I^{th}$ subdomain. Considering the matrix material as the comparison medium, we substitute Eqs. (\ref{eq:finite_sum}) into (\ref{eq:fluctuation_part}) and obtain: 
\begin{equation}
\begin{aligned}
    \hat{u}^{p, I} = \sum_{\bm{\eta} \neq 0} \left[ g^{I}(\bm{\eta}) \mathcal{F}(\bm{\zeta}, \bm{\eta}) \sum_{J=0}^{N_I} \left[ \frac{A_{\Omega^J}}{A_{\mathcal{D}}} g^{J}(-\bm{\eta}) \hat{Q}^{J*} \right] + g^{I}(\bm{\eta}) \mathcal{M}_{i}(\bm{\zeta}, \bm{\eta}) \sum_{J=0}^{N_I} \left[ \frac{A_{\Omega^J}}{A_{\mathcal{D}}} g^{J}(-\bm{\eta}) \hat{u}_{i}^{J*} \right] \right] \\
    \widehat{u_{,i}}^{p, I} = \sum_{\bm{\eta} \neq 0} \left[ g^{I}(\bm{\eta}) \mathcal{N}_{i}(\bm{\zeta}, \bm{\eta}) \sum_{J=0}^{N_I} \left[ \frac{A_{\Omega^J}}{A_{\mathcal{D}}}g^{J}(-\bm{\eta}) \hat{Q}^{J*} \right] + g^{I}(\bm{\eta}) \mathcal{T}_{ij}(\bm{\zeta}, \bm{\eta}) \sum_{J=0}^{N_I} \left[ \frac{A_{\Omega^J}}{A_{\mathcal{D}}}g^{J}(-\bm{\eta}) \hat{u}_{j}^{J*} \right] \right]
\end{aligned}
    \label{eq:fluctuate_final}
\end{equation}

Under the piecewise-constant presentation of eigen-fields in Eq. (\ref{eq:equiv_cond}), the local thermal fields are related to the eigen-fields in each subdomain, which can be written as, 
\begin{equation}
\begin{aligned}
    \hat{Q}^{I*} = (C_p^c - C_p^I) \left[ \langle u \rangle + \hat{u}^{p,I} \right], \quad
    K^c \hat{u}_{i}^{I*} = (K^c - K^I) \left[ \langle u_{,i} \rangle + \widehat{u_{,i}}^{p, I} \right]
\end{aligned}
    \label{eq:equiv_dephase}
\end{equation}
Substituting Eq. (\ref{eq:fluctuate_final}) into Eq. (\ref{eq:equiv_dephase}) provides, 
\begin{equation}
    \begin{aligned}
    \langle u \rangle = \frac{\hat{Q}^{I*}}{C_p^c - C_p^I} -\sum_{\bm{\eta} \neq 0} \left[ g^{I}(\bm{\eta}) \mathcal{F}(\bm{\zeta}, \bm{\eta}) \sum_{J=0}^{N_I} \left[ \frac{A_{\Omega^J}}{A_{\mathcal{D}}} g^{J}(-\bm{\eta}) \hat{Q}^{J*} \right] + g^{I}(\bm{\eta}) \mathcal{M}_{i}(\bm{\zeta}, \bm{\eta}) \sum_{J=0}^{N_I} \left[ \frac{A_{\Omega^J}}{A_{\mathcal{D}}} g^{J}(-\bm{\eta}) \hat{u}_{i}^{J*} \right] \right] \\ 
    \langle u_{,i} \rangle = \frac{K^c u_{i}^{I*}}{K^c - K^I} - \sum_{\bm{\eta} \neq 0} \left[ g^{I}(\bm{\eta}) \mathcal{N}_{i}(\bm{\zeta}, \bm{\eta}) \sum_{J=0}^{N_I} \left[ \frac{A_{\Omega^J}}{A_{\mathcal{D}}}g^{J}(-\bm{\eta}) \hat{Q}^{J*} \right] + g^{I}(\bm{\eta}) \mathcal{T}_{ij}(\bm{\zeta}, \bm{\eta}) \sum_{J=0}^{N_I} \left[ \frac{A_{\Omega^J}}{A_{\mathcal{D}}}g^{J}(-\bm{\eta}) \hat{u}_{j}^{J*} \right] \right]
    \end{aligned}
    \label{eq:avg_dephase_fourier}
\end{equation}

Note that Eq. (\ref{eq:avg_dephase_fourier}) should be satisfied at each subdomain, and repeating such a process over each subdomain provides a global system of linear equations relating the de-phased eigen-fields in each subdomain to de-phased averaged temperature/temperature gradients. Formally, the global system of linear equations is analogous to Eq. (\ref{eq:avg_dephase_eigens}), and the effective properties can be extracted following the same procedure described in Eq. (\ref{eq:effec_properties}). A complete block-matrix illustration is provided in \ref{sec:proof}. 

\subsection{Opposite-wavenumber reciprocity and reflection-symmetry relations}
Shmuel and Willis \cite{Gal2025a} first established the reciprocal relation between the two thermal Willis coupling operators within a source-driven homogenization framework. They showed that when the microscopic (local) thermal conductivity tensor is symmetric, i.e., $\bm{K}(\textbf{x}) = \bm{K}^T(\textbf{x})$, and the averaging weights are the same $w^{(1)}(\textbf{x}) = w^{(2)}(\textbf{x})$, the two nonlocal cross-coupling kernels form an adjoint pair. Note that the authors selected $s \theta = i \omega \theta$ as the second variable, in that representation, $\check{\bm{\xi}} = \hat{\bm{\xi}}^\dagger$, where the symbol $\dagger$ represents the formal adjoint with respect to the bilinear pairing (see Eqs. (4.15c, 4.15d) and Eqs. (5.1), (5.2) in \cite{Gal2025a}). In the present paper, the constitutive relation is based on the averaged temperature itself, the reciprocal relation is $\bm{\chi}^\text{eff} = i \omega \left( \bm{\xi}^\text{eff} \right)^\dagger$, which can be expressed as, 
\begin{equation}
    \bm{\chi}^\text{eff}(\textbf{x}, \textbf{x}', \omega) = i \omega \left[ \bm{\xi}^\text{eff}(\textbf{x}', \textbf{x}, \omega) \right]^T
    \label{eq:formal_adjoint}
\end{equation}
where the adjoint interchanges the source and field points. Since the effective kernels depend on the relative coordinate, $\textbf{r} = \textbf{x} - \textbf{x}'$, interchanging $\textbf{x}$ and $\textbf{x}'$ changes $\textbf{r}$ to $- \textbf{r}$. Hence, the corresponding Bloch-form representation becomes, 
\begin{equation}
\begin{aligned}
    C^\text{eff}(\bm{\zeta}, \omega) = C^\text{eff}(-\bm{\zeta}, \omega), \qquad K^\text{eff}(\bm{\zeta}, \omega) = \left[ K^\text{eff}(-\bm{\zeta}, \omega) \right]^T \\ 
    \bm{\chi}^\text{eff}(\bm{\zeta}, \omega) = i \omega \left[ \bm{\xi}^\text{eff}(-\bm{\zeta}, \omega) \right]^T, \qquad \bm{\chi}^\text{eff}(-\bm{\zeta}, \omega) = i \omega \left[ \bm{\xi}^\text{eff}(\bm{\zeta}, \omega) \right]^T
\end{aligned}
    \label{eq:reci_final}
\end{equation}
Therefore, reciprocity generally pairs the coupling terms evaluated at opposite Bloch wavenumbers, instead of enforcing a same-wavenumber relation. 

Additional parity relations arise when the microstructure exhibits mirror symmetry. Consider a microstructure symmetric about the $x_{1}$ axis. When the Bloch wavenumber changes from $\bm{\zeta} = (\zeta_{1}, \zeta_{2})$ to $\overline{\bm{\zeta}} = (\zeta_{1}, -\zeta_{2})$, the effective heat capacity, diagonal components of the thermal conductivity, and the $\xi_{1}^\text{eff}, \chi^\text{eff}_{1}$ are even functions of $\zeta_{2}$, 
\begin{equation}
\begin{aligned}
    C^\text{eff}(\bm{\zeta}, \omega) = C^\text{eff}(\overline{\bm{\zeta}}, \omega), \quad K_{11}^\text{eff}(\bm{\zeta}, \omega) = K_{11}^\text{eff}(\overline{\bm{\zeta}}, \omega), \quad K_{22}^\text{eff}(\bm{\zeta}, \omega) = K_{22}^\text{eff}(\overline{\bm{\zeta}}, \omega) \\ 
    \xi_{1}^\text{eff}(\bm{\zeta}, \omega) = \xi_{1}^\text{eff}(\overline{\bm{\zeta}}, \omega), \quad \chi_{1}^\text{eff}(\bm{\zeta}, \omega) = \chi_{1}^\text{eff}(\overline{\bm{\zeta}}, \omega)
\end{aligned}
\end{equation}
In contrast, the off-diagonal components of thermal conductivity and $\xi_{2}^\text{eff}, \chi^\text{eff}_{2}$ are odd functions of $\zeta_{2}$, 
\begin{equation}
\begin{aligned}
   K_{12}^\text{eff}(\bm{\zeta}, \omega) = -K_{12}^\text{eff}(\overline{\bm{\zeta}}, \omega), \quad K_{21}^\text{eff}(\bm{\zeta}, \omega) = -K_{21}^\text{eff}(\overline{\bm{\zeta}}, \omega) \\ 
    \xi_{2}^\text{eff}(\bm{\zeta}, \omega) = -\xi_{2}^\text{eff}(\overline{\bm{\zeta}}, \omega), \quad \chi_{2}^\text{eff}(\bm{\zeta}, \omega) = -\chi_{2}^\text{eff}(\overline{\bm{\zeta}}, \omega)
\end{aligned}
\end{equation}
The opposite-wavenumber reciprocity relations and reflection-symmetry parity relations are derived and proved in \ref{sec:proof} with the complete block analysis. The numerical examples are provided in Section \ref{sec:homo_verify} validating the DR-iBEM and Fourier-series-based methods, and the parity relations will be utilized to explain the vanishing and occurrence of cross-coupling terms. 

\subsection{Limitations of Fourier-series-based homogenization method}
Although the Fourier-series-based homogenization method and the homogenization scheme with the DR-iBEM exhibit good agreement in the following case studies, this method has two practical limitations. (i) Because the fluctuation parts of thermal fields are evaluated through Fourier series summation, the accuracy of the results highly depends on the number of reciprocal lattice vectors. For problems with greater thermal wavenumber vectors or higher frequency, the convergence of the Fourier series may become  slow. For instance, all numerical examples with the Fourier-series-based homogenization method utilize $301 \times 301 - 1$ reciprocal lattice to obtain convergent results. (ii) The Fourier-series-based homogenization method is applicable to the unit cell and inhomogeneities with simple geometries, such as rectangles. Particularly, for rectangular inhomogeneities, there exists an analytical solution for $g(\bm{\eta})$. Moreover, the computational burden to construct the global coefficient matrix significantly increases with the number of inhomogeneities, in addition to multiple evaluations with the reciprocal lattice. Hence, the DR-iBEM is more suitable for particulate composites with complex geometries and a large number of inhomogeneities.

\section{Validation and examples}

\label{sec:homo_verify}


Because neither the DR-iBEM nor the Fourier series method is an exact solution, an independent 1D benchmark against Willis' unique residual-field formulae (exact homogenization method) is provided in \ref{sec:1D_case}. This exact benchmark demonstrates that the present homogenization framework can recover the same constitutive relations without introducing residual fields. This section further cross-validates the effective thermal properties predicted by the DR-iBEM formulae in Section 4 and the Fourier formulae in Section 5. Subsequently, the influence of the averaging option is discussed and compared between the de-phased average and conventional spatial average. Finally, the frequency-wavenumber dependence of effective thermal properties is illustrated with circular particulate composites.

\subsection{Cross-verification by the case with square inhomogeneities}
\begin{figure}
    \centering
    \includegraphics[width=0.5\linewidth]{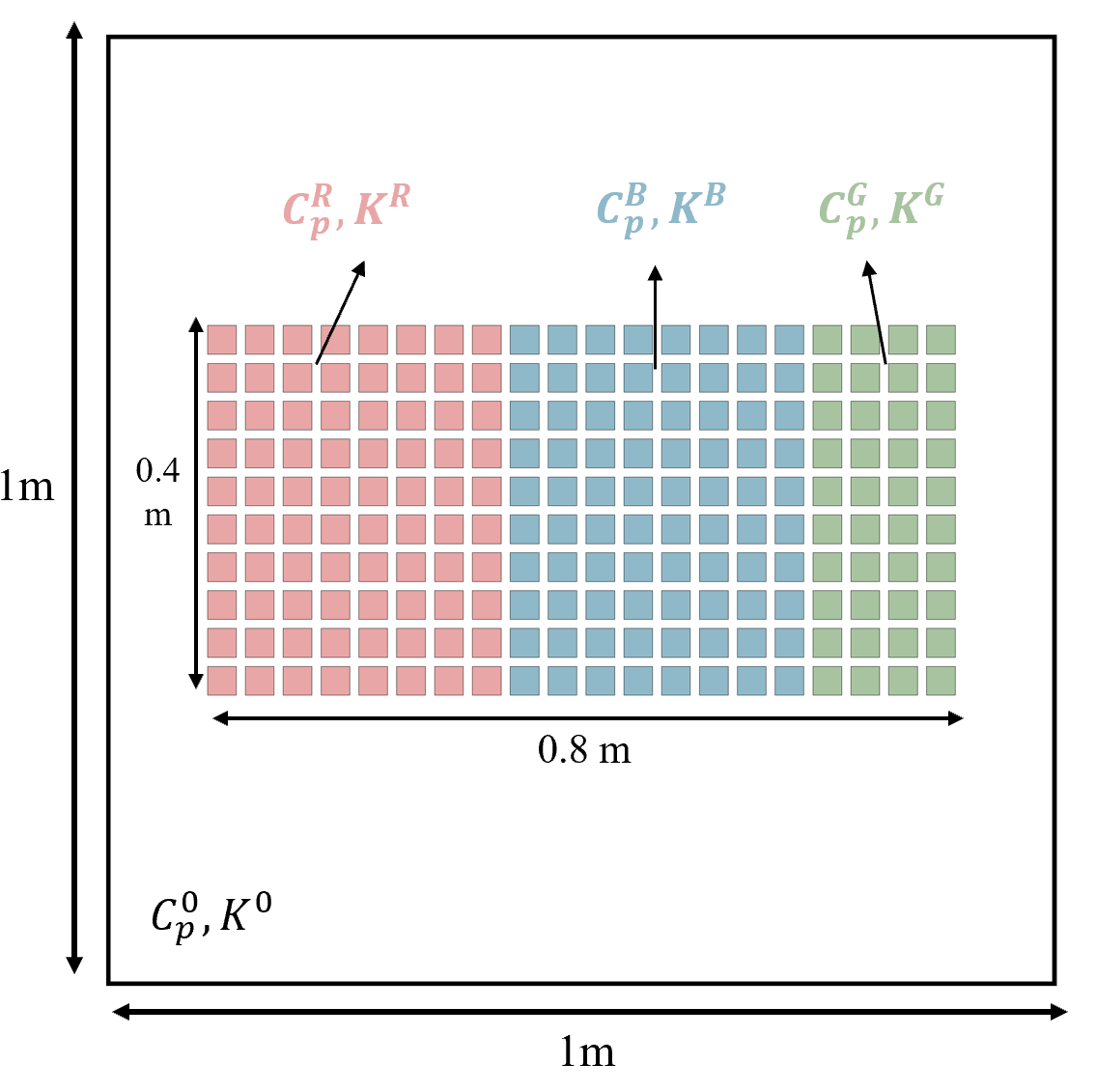}
    \caption{Schematic plot of a square composite containing uniformly distributed inhomogeneities within the region $0.8 \times 0.4$ m, which is subjected to Bloch-form boundary conditions with the thermal wavenumber vector $\bm{\zeta}$. The comparison medium is identical to the matrix material, and three types of materials with superscripts $R, G, B$ (for red, green, and blue regions) are involved. $200$ uniformly distributed square inhomogeneities with volume fraction $20\%$}
    \label{fig:homo_state}
\end{figure}

Fig. \ref{fig:homo_state} shows the schematic plot of a square composite ($1 \times 1$ m) containing $200$ uniformly distributed square inhomogeneities within the region of $0.8 \times 0.4$ m, which will be utilized for cross-verification between the DR-iBEM algorithm and the Fourier-series-based homogenization in Section 5. In Fig. \ref{fig:homo_state}, the volume fraction of inhomogeneities is 20\% ($\sum_{I=1}^{N_I} A_{\Omega^I} / A_{\mathcal{D}} = 0.2$), and thus the length of each square inhomogeneity is $0.0316$ m.  

The square composite is subjected to the Bloch-form boundary condition, and three thermal wavenumber vectors, $\bm{\zeta} = (0, 0), (10, 0)$ and $(0, 10)$ are considered. This section divides the region $0.8 \times 0.4$ m into three sub-regions, from left to right, $8$, $8$, and $4$ columns of inhomogeneities are assigned with different thermal properties. Specifically, the material properties are: (i) the matrix, $K^0 = 4 W / m \cdot K$, $C_p^0 = 10 J / m^3 \cdot K$ , which is selected as the comparison medium; (ii) the red region, $K^R = 200 W / m \cdot K$, $C_p^R = 5 J / m^3 \cdot K$; (iii) the blue region, $K^B = 10 W / m \cdot K$, $C_p^B = 17 J / m^3 \cdot K$; and (iv) the green region, $K^G = 2.4 W / m \cdot K$, $C_p^G = 2 J / m^3 \cdot K$. The circular frequency $\omega \in [0, 30]$ rad/s are considered. 

In the following numerical case studies, the DR-iBEM model utilizes: 
\begin{itemize}
    \item 404 boundary nodes, where four additional nodes are introduced to define the heat flux at corner nodes; and 200 quadratic boundary elements, including 8 elements adjacent to the corners are handled as discontinuous elements. 
    \item $20 \times 20$ uniformly distributed interior interpolation nodes; 
    \item $150 \times 150$ uniformly distributed sampling points to evaluate the averaged thermal fields. 
\end{itemize}
Since the comparison medium is selected to be identical to the matrix, the DR-iBEM does not require any interior discretization in the matrix phase. The case that the comparison medium is different from the matrix is provided for the 1D case study in \ref{sec:1D_case}.

This subsection provides the cross-verification between the DR-iBEM and the Fourier-series-based homogenization method. Note that the Fourier-series-based method constructed equivalent conditions on each subdomain in Fig. \ref{fig:homo_state}, and the number of Fourier series terms is selected as $301 \times 301 - 1$ to provide convergent results, in which $\bm{\eta} = (0, 0)$ does not require the evaluation as it provides the de-phased average quantity.

\subsection{Homogenization limit \texorpdfstring{$\bm{\zeta} = (0, 0)$}{}}
\label{sec:00}
\begin{figure}
    \centering
    \includegraphics[width=1\linewidth]{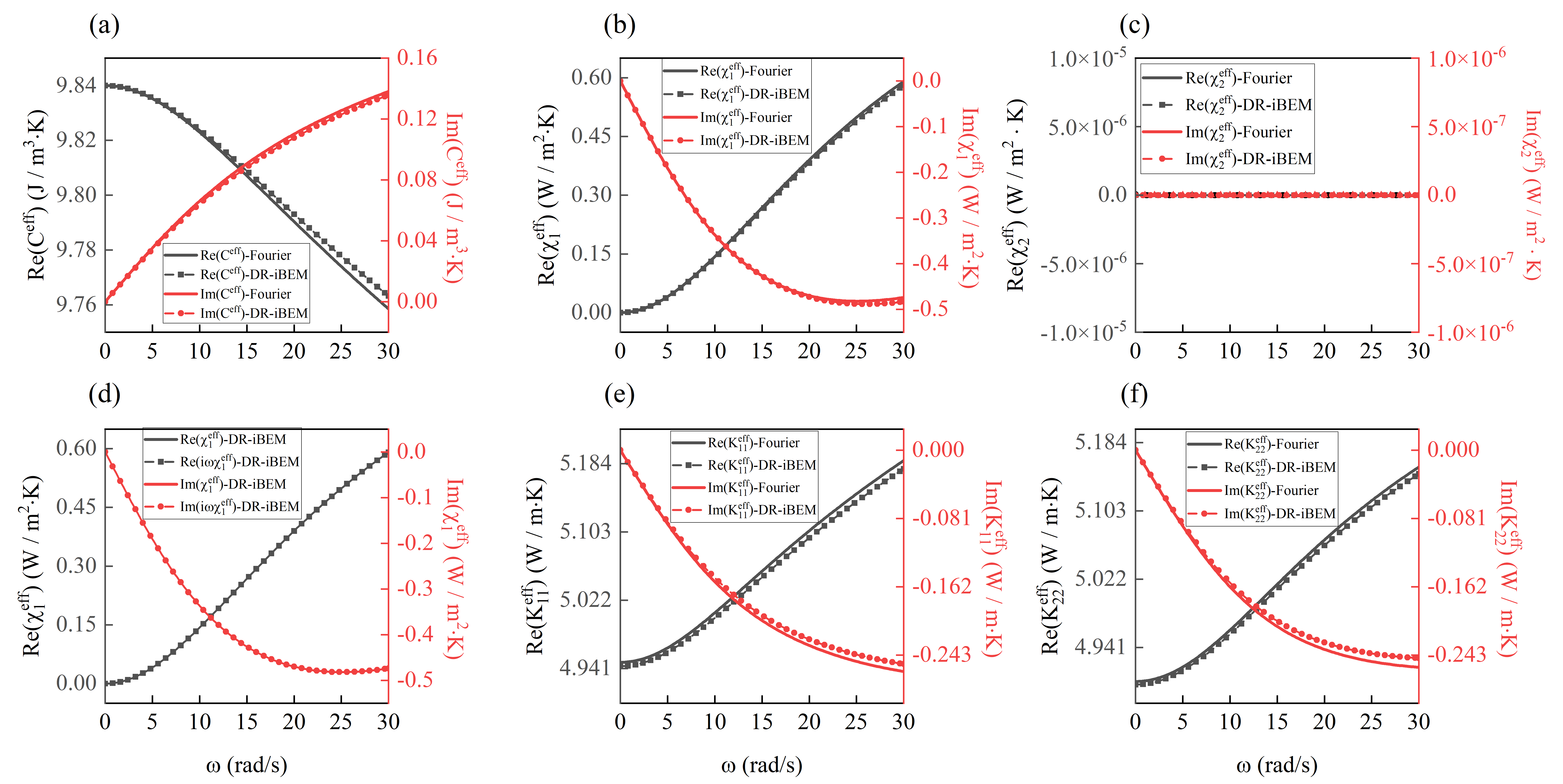}
    \caption{Comparison and variation of five effective properties between the DR-iBEM and Fourier-series-based method, (a) $C^{\text{eff}}$, (b) $\chi_{1}^{\text{eff}}$, (c) $\chi_{2}^{\text{eff}}$, (d) comparison between $\chi_{1}^{\text{eff}}$ and $i \omega \xi_{1}^\text{eff}$, (e) $K_{11}^{\text{eff}}$, and (f) $K_{22}^{\text{eff}}$. The circular frequency $\omega \in [0, 30]$ rad/s, and the thermal wavenumber vector $\bm{\zeta} = (0, 0)$, namely the homogenization limit.}
    \label{fig:limit_verify}
\end{figure}

Figs. \ref{fig:limit_verify} (a-f) show the comparison and variation of effective thermal properties between the DR-iBEM and Fourier-series-based method, when the circular frequency $\omega \in [0, 30]$ rad/s and the thermal wavenumber $\bm{\zeta} = (0, 0)$, namely the homogenization limit. The solid curves represent the predictions by the Fourier-series-based method, while the dashed curve with markers denotes the DR-iBEM results. Overall, good agreement is observed between the two independent numerical approaches for all effective properties in Figs. \ref{fig:limit_verify} (a-f). Fig. \ref{fig:limit_verify} (a) plots the effective heat capacity $C^{\text{eff}}$. Specifically, the real part of the effective heat capacity decreases gradually with frequency, while its imaginary part increases monotonically, indicating that the effective heat capacity is frequency-dependent. In previous homogenization models, the rule of mixtures has been widely used to estimate the effective heat capacity. For instance, it provides, 
\begin{equation}
    \overline{C}^\text{eff} = \phi^{R} C_p^R + \phi^B C_p^B + \phi^G C_p^G + \phi^0 C_p^0 = 9.84 J / m^3 \cdot K
    \label{eq:ROM}
\end{equation}
where $\phi^{R}, \phi^{B}, \phi^{G}$, and $\phi^{0}$ refer to the area fraction of the red, blue, green, and the matrix components, which are $0.08, 0.08, 0.04, 0.8$, respectively. When the circular frequency $\omega \to 0$, the real part of the effective heat capacity is identical to $9.84 J / m^3 \cdot K$, while its imaginary part is zero. However, when the circular frequency increases, Fig. \ref{fig:limit_verify} (a) shows that the effective heat capacity deviates from Eq. (\ref{eq:ROM}), which implies the conventional steady-state estimation through the rule of mixtures is not applicable for harmonic heat transfer. Moreover, Fig. \ref{fig:limit_verify} (a) shows that even if all material properties are real, the effective heat capacity can still be complex, and the increasing imaginary part is associated with phase lag and diffusive dynamic response of heat transfer in composites. 

Subsequently, Figs. \ref{fig:limit_verify} (b) and (c) plot the variation of the coupling terms $\chi_{1}^\text{eff}$ and $\chi_{2}^\text{eff}$, respectively. The maximum discrepancies exist at the frequency $30$ rad/s. Specifically, the Fourier-series-based method predicts $\chi^{\text{eff}}_{1} = 0.5908 -0.4738 i $, while the DR-iBEM provides $\chi^{\text{eff}}_{1} = 0.5808 - 0.4831 i$, and thus the errors are $1.7\%$ and $1.9\%$ for the real and imaginary parts, respectively. Since both DR-iBEM and Fourier-series-based methods are numerical methods, it is difficult to determine which is more accurate in these circumstances. However, the minor discrepancies between the two independent methods indicate that the extracted Willis-type coupling terms are reliable. As Fig. \ref{fig:limit_verify} (b) shows, the real part of $\chi^{\text{eff}}_{1}$ monotonically increases with the frequency, while the magnitude of its imaginary part decreases at higher frequencies. It should be emphasized that when the frequency $\omega \to 0$, the coupling term $\chi^{\text{eff}}_{1}$ reduces to zero, which reduces to the steady-state thermal homogenization. Specifically, the zero-coupling terms imply that the averaged thermal field is linked only to the averaged temperature gradients, analogous to the averaged stress and strain in elastostatics. However, the frequency-dependent coupling terms show that the averaged heat flux is related to both the averaged temperature and the temperature gradients, which is substantially different from the steady-state homogenization schemes. Consequently, the homogenization properties at the steady state are not applicable for harmonic heat transfer. 

Fig. \ref{fig:limit_verify} (c) shows that the coupling term $\chi_{2}^\text{eff}$ remains zero over the entire frequency range. For the Bloch wavenumber, $\bm{\zeta} = (0, 0)$, the reflection about the horizontal axis leaves the wavenumber unchanged. Therefore, the parity relation in Eq. (\ref{eq:effec_fanzhuan}) requires $\chi_{2}^\text{eff}(\textbf{0}) = -\chi_{2}^\text{eff}(\textbf{0})$, leading to vanishing of $\chi_{2}^\text{eff}$. The phenomenon can be interpreted as the vertical symmetry suppresses the coupling between the averaged temperature and the vertical component of the averaged heat flux, while the horizontal asymmetry allows for a nonzero $\chi^\text{eff}_{1}$. A similar phenomenon in elastodynamics has been shown in \cite{Srivastava2011}. However, there is an interesting yet fundamental question: does the vanishing of coupling terms require both the microstructure's symmetry and macroscopic boundary conditions? This question will be answered in comments associated with Fig. \ref{fig:x_verify} and Fig. \ref{fig:y_verify}. Fig. \ref{fig:limit_verify} (d) compares the relations between two coupling terms $\chi^{\text{eff}}_1$ and $\xi_{1}^{\text{eff}}$. It can be observed that the curves $\chi_{1}^{\text{eff}}$ and $i \omega \xi_{1}^\text{eff}$ have good agreement, and almost no discrepancies can be found. As shown in \ref{sec:proof}, the reciprocity relations between two coupling vectors hold for opposite Bloch wavenumber vectors. Since the wavenumber vector $\bm{\zeta} = (0, 0)$, the opposite wavenumber vectors coincide. The good agreement confirms the previous conclusion by Shmuel and Willis \cite{Gal2025a, Shmuel2025}, validating the homogenization scheme implemented with DR-iBEM. To the authors' best knowledge, this is the first work demonstrating such correlations in thermal particulate composites. 

Figs. \ref{fig:limit_verify} (e) and (f) plot the variation of effective thermal conductivity $K_{11}^\text{eff}$ and $K_{22}^\text{eff}$, respectively. Similar to the effective heat capacity in Fig. \ref{fig:limit_verify} (a), two effective thermal conductivities are purely real at the steady-state case, and become complex at non-zero frequencies. This phenomenon indicates that even though all material phases exhibit real material properties, the effective thermal conductivities can be complex. For micromechanical homogenization models, such as the Mori-Tanaka and self-consistent methods, these models generally yield effective properties that cannot explain the complex effective thermal conductivities. Particularly, the monotonically increasing imaginary part of the effective thermal conductivity characterizes the phase lag and diffusive dynamic responses of the composite. Therefore, conventional static/mixture-based models cannot reproduce these frequency-dependent complex nonlocal effective properties. 

\subsection{Horizontal wave: \texorpdfstring{$\bm{\zeta} = (10, 0)$}{}}
\begin{figure}
    \centering
    \includegraphics[width=1\linewidth]{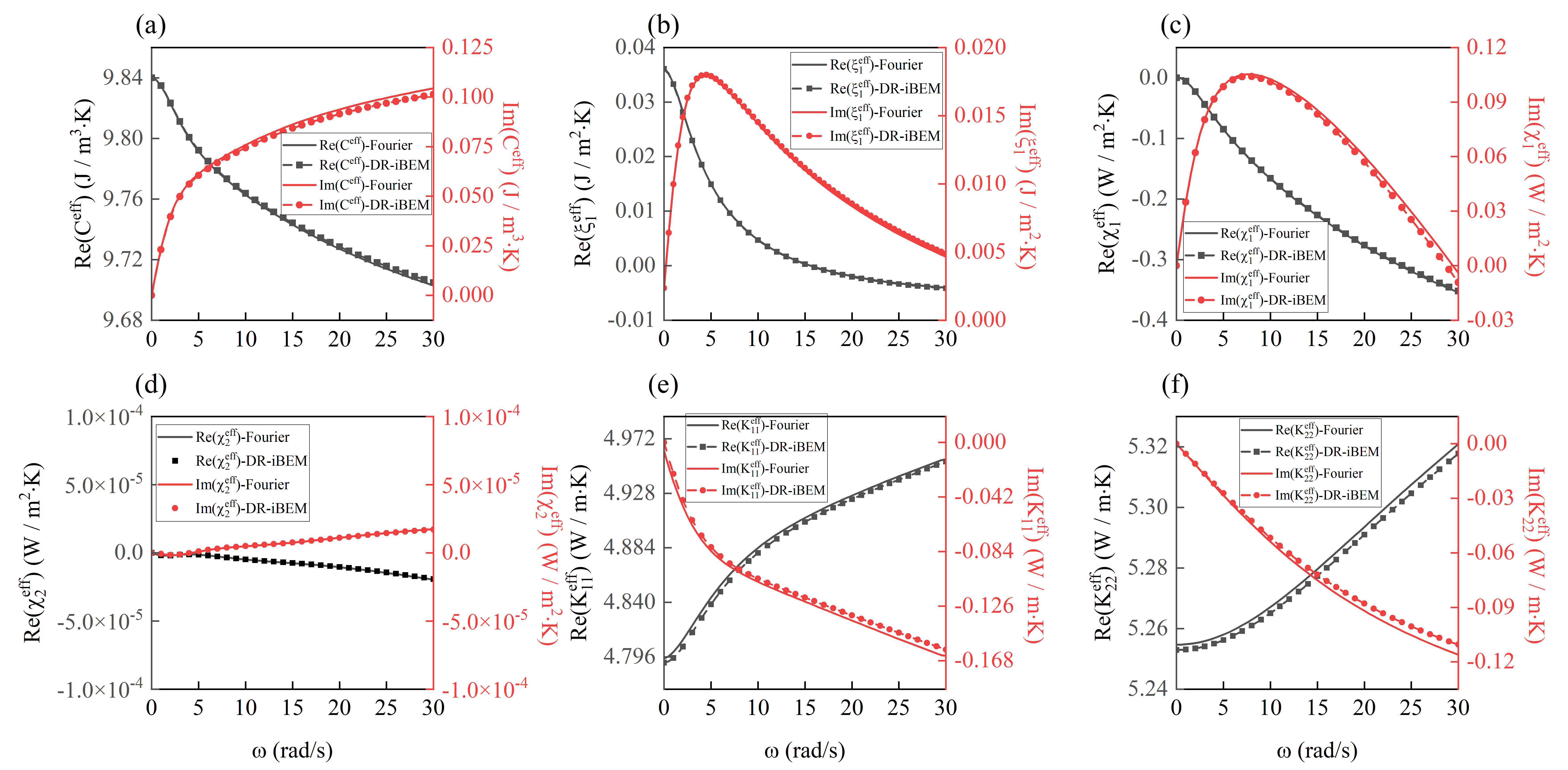}
    \caption{Comparison and variation of effective properties between the DR-iBEM and Fourier-series-based method, (a) $C^{\text{eff}}$, (b) $\xi_{1}^{\text{eff}}$, (c) $\chi_{1}^{\text{eff}}$, (d) $\chi_{2}^{\text{eff}}$, (e) $K_{11}^{\text{eff}}$, and (f) $K_{22}^{\text{eff}}$. The circular frequency $\omega \in [0, 30]$ rad/s, and the thermal wavenumber vector $\bm{\zeta} = (10, 0)$.}
    \label{fig:x_verify}
\end{figure}

Figs. \ref{fig:x_verify} (a-f) compare the effective properties by the DR-iBEM and Fourier-series-based homogenization methods, when the thermal wavenumber vector is $\bm{\zeta} = (10, 0)$. Before interpreting effective properties at higher frequencies, we emphasize an interesting phenomenon at $\omega \to 0$. Comparing Figs. \ref{fig:x_verify} with  \ref{fig:limit_verify}, when $\omega \to 0$, the effective heat capacity, thermal conductivities with $\bm{\zeta} = (10, 0)$ are different from the results with homogenization limit. Specifically, the effective heat capacity in Fig. \ref{fig:limit_verify} (a) recovers the steady-state homogenization result in Eq. (\ref{eq:ROM}); but Fig. \ref{fig:x_verify} (a) shows $C^\text{eff} = 9.816 J / m^3 \cdot K$. Moreover, the effective thermal conductivity $K_{11}^\text{eff} = 5.126$ and $4.791 W / m \cdot K$ in Figs. \ref{fig:limit_verify} (e) and \ref{fig:x_verify} (e), respectively. The discrepancies suggest that the effective thermal properties are not only dependent on frequency but also on the thermal wavenumber vector (boundary conditions), which reflects the nonlocal feature of the homogenized responses. Therefore, the steady-state homogenization results can only provide rational approximations when the frequency is low and the thermal wavenumber is sufficiently small, such as the homogenization limit, which further confirms that the conventional steady-state homogenization can rarely be applied to harmonic heat transfer for general cases.

Although Fig. \ref{fig:x_verify} (a) shows that the real and imaginary parts of the effective heat capacity are frequency-dependent, the slopes are different compared to Fig. \ref{fig:limit_verify} (a), which should be interpreted as the effect of the thermal wavenumber vector. The predictions by the DR-iBEM agree well with the Fourier-series-based homogenization method results, and the maximum discrepancy occurs at the frequency $\omega = 30$ rad/s, which are $9.703 + 0.104 i$ and $9.705 + 0.101 i$, respectively. Particularly, Figs. \ref{fig:x_verify} (b) and (c) indicate that the real part of the coupling terms $\xi_{1}^\text{eff}$ and $\chi_{1}^\text{eff}$ monotonically decreases, while their imaginary parts first increase and then decrease. It should be emphasized that the signs of both the real and imaginary parts of $\chi_{1}^\text{eff}$ are opposite to those observed in Fig. \ref{fig:limit_verify} (b). The opposite signs of $\chi_{1}^\text{eff}$ should not be interpreted as an inconsistency. As previously discussed, when the thermal wavenumber vector changes, the imposed macroscopic field no longer corresponds to the classic steady-state case (long thermal wavelength). Therefore, the extracted coupling term is a finite-wavenumber effective kernel \cite{Gal2025a}, which can differ from that in the homogenization limit. The good agreement in Fig. \ref{fig:x_verify} (c) indicates that the opposite sign of coupling terms is a finite-wavenumber effect. 

Subsequently, Fig. \ref{fig:x_verify} (d) plots the variation of the coupling term $\chi_{2}^\text{eff}$. For the Bloch wavenumber, $\bm{\zeta} = (10, 0)$, the reflection about the horizontal axis leaves the wavenumber unchanged, as $(\zeta_1, \zeta_{2}) = (10, 0)$, $(\zeta_{1}, -\zeta_{2}) = (10, 0)$. Based on the parity relation derived in Eq. (\ref{eq:effec_fanzhuan}), $\chi^\text{eff}_{2}(10, 0) = -\chi^\text{eff}_2(10, 0)$, which requires $\chi^\text{eff}_{2} = 0$ at any circular frequency. The magnitude of $\chi_{2}^\text{eff}$ is less than $2 \times 10^{-5}$, and it is approximately three orders of magnitude smaller than $\chi_{1}^\text{eff}$, which can be regarded as numerical residuals. Note that the numerical residuals are expected in both the DR-iBEM and the Fourier-series-based homogenization methods: (i) the DR-iBEM evaluates the local thermal fields through boundary integrals, dual-reciprocity approximated source density, polynomial-form eigen-fields, and a finite number of sampling points; and (ii) the Fourier-series-based homogenization method relies on truncated reciprocal lattice summation. Therefore, a similar observation can be found as the zero coupling term in Fig. \ref{fig:limit_verify} (c). Their close agreement further validates the consistency of the effective properties extracted by the two homogenization methods. 

Figs. \ref{fig:x_verify} (e) and (f) compare the effective thermal conductivity components $K_{11}^\text{eff}$ and $K_{22}^{\text{eff}}$, respectively. For $K_{11}^\text{eff}$, the real part increases monotonically with the circular frequency, while its imaginary part decreases from zero to negative values. For $K_{22}^\text{eff}$, a different trend can be observed, where the real part first decreases slightly in low frequencies and then increases with the frequency, while its imaginary part monotonically decreases with the frequency. As shown in Fig. \ref{fig:homo_state}, the microstructure is arranged such that it maintains symmetry in the vertical direction while being asymmetric in the horizontal direction. The different variations of effective thermal conductivities should be interpreted as both the effect of the finite-wavenumber and the anisotropic nature of the microstructure, which will be further confirmed in Fig. \ref{fig:y_verify} with the thermal wavenumber vector $\bm{\zeta} = (0, 10)$ subsequently. In addition, the effective thermal conductivities are complex-valued when the circular frequency is non-zero, while all material phases exhibit the real-valued thermal properties. This observation further confirms the conclusion drawn in the previous subsection that the conventional steady-state homogenization models cannot explain such frequency-dependent and complex-valued thermal properties. Overall, the predictions by DR-iBEM and the Fourier-series-based homogenization method are close, although some minor discrepancies can be observed. Specifically, when $\omega = 30$ rad/s, the DR-iBEM provides $K^{\text{eff}}_{11} = 4.954 - 0.159 i$ and $K^{\text{eff}}_{22} = 5.318 - 0.111 i$, while the Fourier-series-based homogenization method predicts $K^{\text{eff}}_{11} = 4.956 - 0.164 i$ and $K^{\text{eff}}_{22} = 5.328 - 0.116 i$. The relative errors of the real parts are only about $0.04\%$ and $0.18\%$ for $K^{\text{eff}}_{11}$ and $K^{\text{eff}}_{22}$, respectively. Note that the relative discrepancies of the imaginary parts are slightly larger because their magnitudes are much smaller than their real parts, but the absolute differences remain minor.

\begin{figure}
    \centering
    \includegraphics[width=1\linewidth]{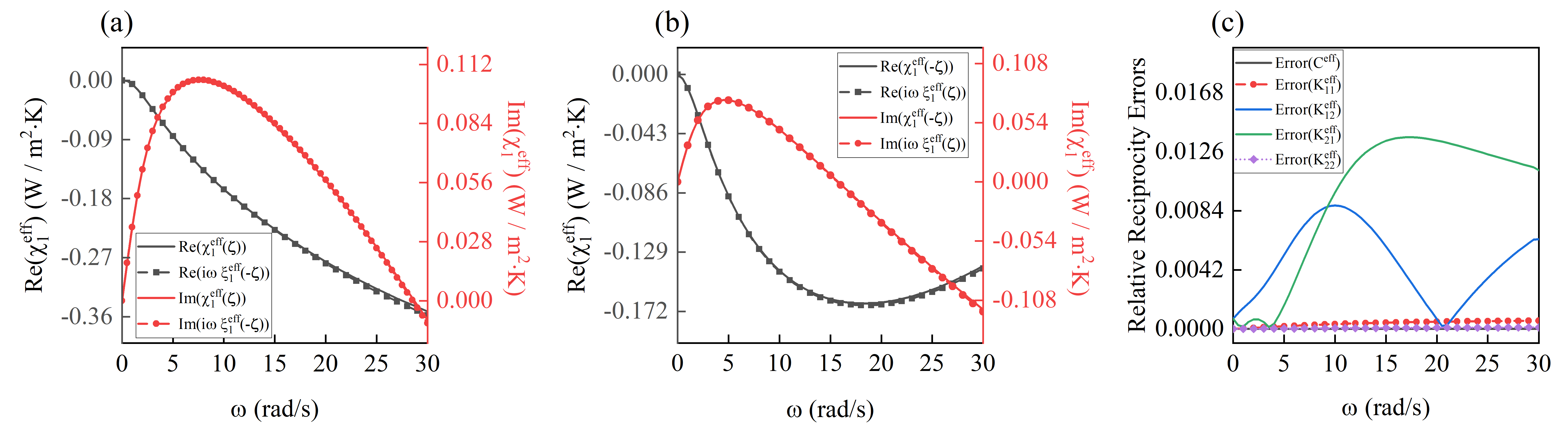}
    \caption{Comparison and variation of effective thermal properties between the opposite-wavenumber vectors by DR-iBEM, (a) $\chi^\text{eff}_{1}(\bm{\zeta})$ and $i \omega \xi_{1}^\text{eff}(-\bm{\zeta})$; (b) $\chi^\text{eff}_{1}(-\bm{\zeta})$ and $i \omega \xi_{1}^\text{eff}(\bm{\zeta})$; and (c) relative errors between the effective heat capacity $C^\text{eff}(\bm{\zeta}), C^\text{eff}(-\bm{\zeta})$, and the effective thermal conductivity matrix components $\bm{K}^\text{eff}(\bm{\zeta})$ and $(\bm{K}^\text{eff}(-\bm{\zeta}))^T$. The circular frequency $\omega \in [0, 30]$ rad/s, and the thermal wavenumber vector $\bm{\zeta} = (10, 0)$.}
    \label{fig:pos_neg_check}
\end{figure}

Subsequently, Figs. \ref{fig:pos_neg_check} (a-c) examine the reciprocity relations of effective properties between two Bloch opposite wavenumber vectors. As proved in \ref{sec:proof}, two cross-coupling vectors should satisfy the reciprocity relation as Eq. (\ref{eq:reci_final}). Since the Fourier-series-based homogenization method naturally satisfies the relation, it is important to check whether DR-iBEM can reproduce the theoretically required behavior. Fig. \ref{fig:pos_neg_check} (a) compares the real and imaginary parts of $\chi^\text{eff}_{1}(\bm{\zeta})$ and $i \omega \xi_{1}^\text{eff}(-\bm{\zeta})$, while Fig. \ref{fig:pos_neg_check} (b) presents $\chi^\text{eff}_{1}(-\bm{\zeta})$ and $i \omega \xi_{1}^\text{eff}(\bm{\zeta})$. The slight differences between the two figures confirm that the cross-coupling terms extracted by the DR-iBEM scheme are consistent with Willis' homogenization theory \cite{Gal2025a}. Because $\chi^\text{eff}_{2}$ remains numerically negligible, as shown in Fig. \ref{fig:x_verify} (d), the second-component reciprocity relation is not plotted. 

Moreover, the reciprocity relation also imposes constraints on the effective heat capacity and effective thermal conductivity tensor. Fig. \ref{fig:pos_neg_check} (c) plots the relative reciprocity errors defined in Eq. (\ref{eq:rec_errors}). As Fig. \ref{fig:pos_neg_check} (c) indicates, the relative errors of $C^\text{eff}$, $K_{11}^\text{eff}$, and $K_{22}^\text{eff}$ are slight, while the relative errors of $K_{12}^\text{eff}$ and $K_{21}^\text{eff}$ are approximately $1.4\%$. These comparatively larger relative errors arise from the extremely small magnitudes of the off-diagonal conductivity components. For instance, when $\omega = 30$ rad/s, the magnitude, $|K_{12}^\text{eff}(\bm{\zeta})| \approx 3.43 \times 10^{-5} $, while $|K_{22}^\text{eff}(\bm{\zeta})| \approx 5.33$. Hence, the off-diagonal component is five orders of magnitude smaller than the diagonal components, and its relative error does not indicate the violation of reciprocity. 

\begin{equation}
    \text{Error}(C^\text{eff}) = \frac{|C^\text{eff}(\bm{\zeta}) - C^\text{eff}(-\bm{\zeta})|}{|C^\text{eff}(\bm{\zeta})|}, \quad \text{Error}(K_{ij}^\text{eff}) = \frac{|K^\text{eff}_{ij}(\bm{\zeta}) - K_{ji}^\text{eff}(-\bm{\zeta})|}{|K^\text{eff}_{ij}(\bm{\zeta})|}
    \label{eq:rec_errors}
\end{equation}

\subsection{Vertical wave: \texorpdfstring{$\bm{\zeta} = (0, 10)$}{}}
\begin{figure}
    \centering
    \includegraphics[width=1\linewidth]{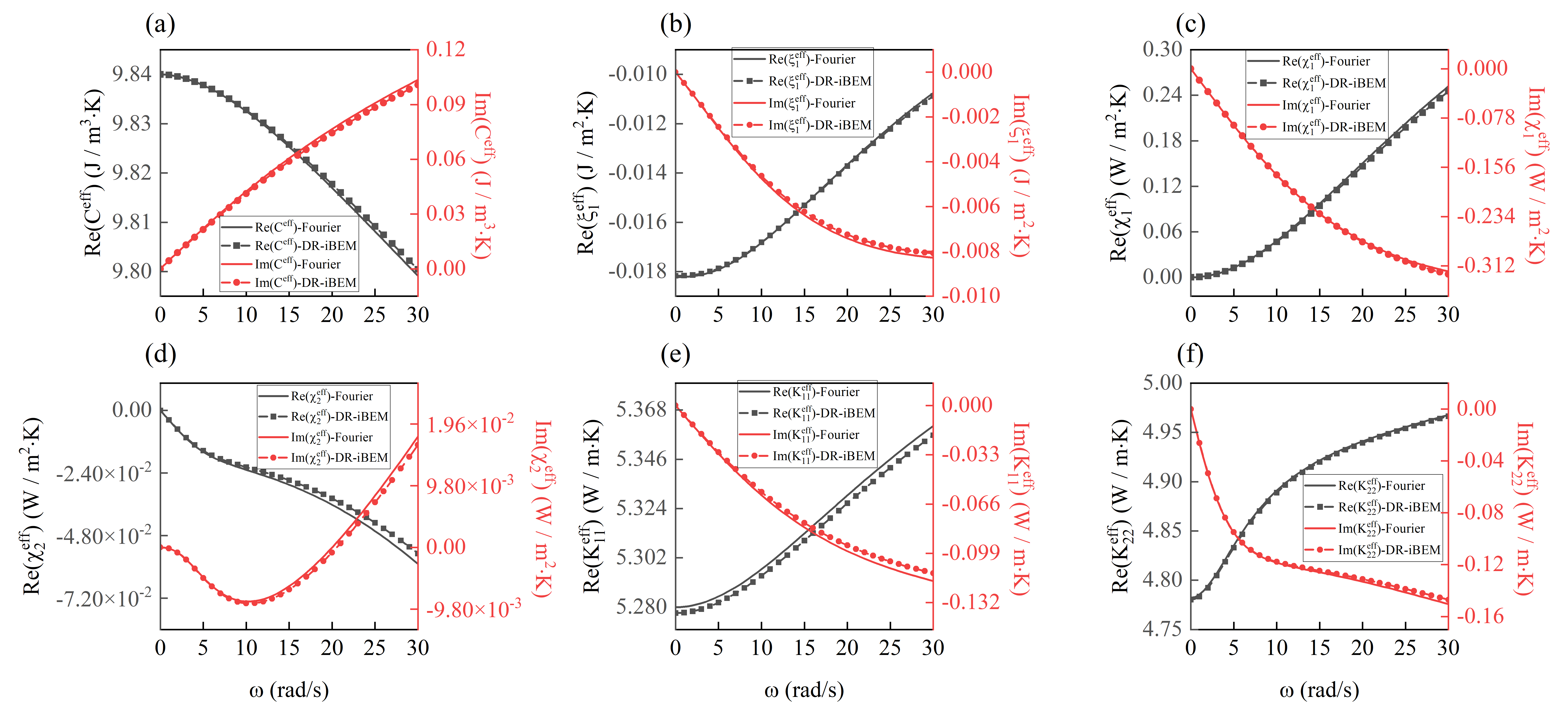}
    \caption{Comparison and variation of effective properties between the DR-iBEM and Fourier-series-based method, (a) $C^{\text{eff}}$, (b) $\xi_{1}^{\text{eff}}$, (c) $\chi_{1}^{\text{eff}}$, (d) $\chi_{2}^{\text{eff}}$, (e) $K_{11}^{\text{eff}}$, and (f) $K_{22}^{\text{eff}}$. The circular frequency $\omega \in [0, 30]$ rad/s, and the thermal wavenumber vector $\bm{\zeta} = (0, 10)$.}
    \label{fig:y_verify}
\end{figure}

Fig. \ref{fig:y_verify} (a-f) compares the effective properties by the DR-iBEM and Fourier-series-based homogenization methods, when the thermal wavenumber vector is $\bm{\zeta} = (0, 10)$. Comparing Fig. \ref{fig:y_verify} with Fig. \ref{fig:x_verify}, even when the frequency $\omega \to 0$, the effective heat capacity and thermal conductivities differ, which implies that the effective thermal properties depend on both the frequency and the thermal wavenumber vectors. 

The maximum discrepancies at the frequency $\omega = 30$ rad/s between two methods are listed: (i) for $C^\text{eff}$ in Fig. \ref{fig:y_verify} (a), the DR-iBEM provides $9.800 + 0.100 i$, while the Fourier-series-based homogenization method predicts $9.799 + 0.104 i$; (ii) for $\chi_{1}^\text{eff}$ in Fig. \ref{fig:y_verify} (c), the DR-iBEM provides $0.245 - 0.325 i$, while the Fourier-series-based homogenization method predicts $0.252 - 0.321 i$; and (iii) for $K_{11}^\text{eff}$ and $K_{22}^\text{eff}$ in Figs. \ref{fig:y_verify} (e) and (f), the DR-iBEM provides $5.356 - 0.112 i$ and $4.966 - 0.147 i$, while the Fourier-series-based homogenization method predicts $5.360 - 0.117 i$ and $4.967 - 0.150 i$, respectively. Based on the above data, the discrepancies in the real parts of the effective properties are minor. However, the relatively larger differences in the imaginary parts are due to the small magnitudes, while the absolute differences remain minor. Therefore, the comparison further confirms the reliability of the DR-iBEM in extracting complex-valued effective thermal properties.

Since the general frequency-dependent behaviors, complex-valued properties, and the agreement between the DR-iBEM and Fourier-series-based homogenization method have been discussed above, this case study primarily addresses the role of microstructural symmetry in the vanishing of Willis-type coupling terms. Fig. \ref{fig:y_verify} (d) shows that the coupling term $\chi^\text{eff}_{2}$ is nonzero over the considered frequency range, which appears to contrast with results in Figs. \ref{fig:limit_verify} (c) and \ref{fig:x_verify} (d). As Eq. (\ref{eq:effec_fanzhuan}) shows, for the Bloch wavenumber vector $\bm{\zeta} = (0, 10)$ and its mirrored case $\overline{\bm{\zeta}} = (0, -10)$, the second component of the cross-coupling terms $\xi_{2}^\text{eff}$ and $\chi_{2}^\text{eff}$ reverse the sign. However, this does not require $\chi_{2}^\text{eff}(0, 10) = -\chi^{\text{eff}}_{2} (0, 10)$, which allows for non-zero cross-coupling terms. Therefore, when the Bloch thermal wave propagates perpendicular to the horizontal symmetric axis, the transverse components of cross-coupling terms can be nonzero, which will not vanish due to the microstructural symmetry. This resolves the question raised in Section \ref{sec:00}: (i) microstructural symmetry alone does not necessarily lead to the vanishing of coupling terms; (ii) the vanishing of coupling terms occurs when the Bloch wavenumber is invariant under the corresponding mirror operation. 

\begin{figure}
    \centering
    \includegraphics[width=1\linewidth]{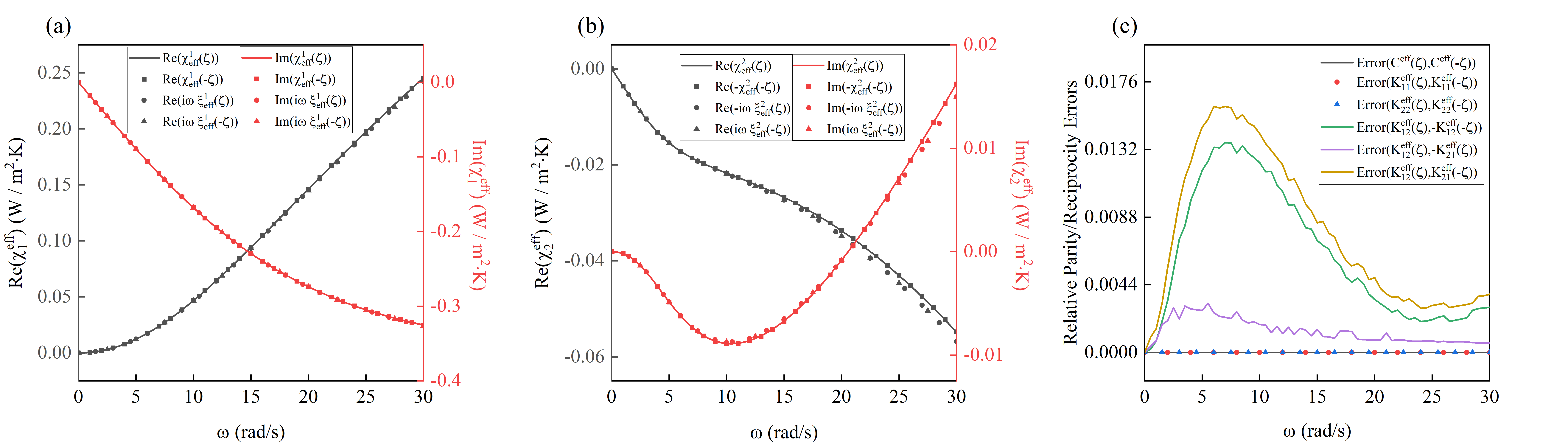}
    \caption{Comparison and variation of effective thermal properties between the opposite wavenumber vectors by DR-iBEM, (a) $\chi^\text{eff}_{1}(\bm{\zeta})$, $\chi^\text{eff}_{1}(-\bm{\zeta})$, $i \omega \xi_{1}^\text{eff}(\bm{\zeta})$, and $i \omega \xi_{1}^\text{eff}(-\bm{\zeta})$; (b) $\chi^\text{eff}_{2}(\bm{\zeta})$, $-\chi^\text{eff}_{2}(-\bm{\zeta})$, $-i \omega \xi_{2}^\text{eff}(\bm{\zeta})$, and $i \omega \xi_{2}^\text{eff}(-\bm{\zeta})$; and (c) relative errors between the effective heat capacity $C^\text{eff}(\bm{\zeta}), C^\text{eff}(-\bm{\zeta})$, and the effective thermal conductivity matrix components $\bm{K}^\text{eff}(\bm{\zeta})$ and $(\bm{K}^\text{eff}(-\bm{\zeta}))^T$. The circular frequency $\omega \in [0, 30]$ rad/s, and the thermal wavenumber vector $\bm{\zeta} = (0, 10)$.}
    \label{fig:pos_neg_check_y}
\end{figure}

Figs. \ref{fig:pos_neg_check_y} (a-c) examine the reciprocity relations and parity relations for the wavenumber vector $\bm{\zeta} = (0, 10)$ and its opposite (reflected) one $-\bm{\zeta} = (0, -10)$. Because the microstructure is symmetric about the horizontal axis, Eqs. (\ref{eq:reci_final}) and (\ref{eq:effec_fanzhuan}) together provide: (i) $\xi_{1}^\text{eff}(\bm{\zeta}) = \xi_{1}^\text{eff}(\bm{-\zeta})$ and $\chi_{1}^\text{eff}(\bm{\zeta}) = \chi_{1}^\text{eff}(\bm{-\zeta})$; and (ii) $\xi_{2}^\text{eff}(\bm{\zeta}) = -\xi_{2}^\text{eff}(\bm{-\zeta})$ and $\chi_{2}^\text{eff}(\bm{\zeta}) = -\chi_{2}^\text{eff}(\bm{-\zeta})$; and (iii) $K_{12}^\text{eff}(\bm{\zeta}) = -K_{21}^\text{eff}(\bm{\zeta}) = -K_{12}^\text{eff}(-\bm{\zeta}) = K_{21}^\text{eff}(\bm{-\zeta})$. The close agreement among quantities in Figs. \ref{fig:pos_neg_check_y} (a) and \ref{fig:pos_neg_check_y} (b) demonstrates that the opposite-wavenumber reciprocity and the parity relations are satisfied simultaneously. Fig. \ref{fig:pos_neg_check_y} (c) plots the relative errors of the effective heat capacity, diagonal components of the thermal conductivity. The errors associated with $C^\text{eff}, K_{11}^\text{eff}, K_{22}^\text{eff}$ are close to zero, which confirms that they remain unchanged for two opposite wavenumbers. Moreover, Fig. \ref{fig:pos_neg_check_y} (c) further examines three relations among $K_{12}^\text{eff}(\bm{\zeta}), -K_{21}^\text{eff}(\bm{\zeta}), -K_{12}^\text{eff}(-\bm{\zeta})$, and $K_{21}^\text{eff}(\bm{-\zeta})$. The maximum discrepancy is approximately $1.6\%$. Because the magnitude of the off-diagonal components is about $0.02$, the corresponding absolute errors are less than $4 \times 10^{-4}$, which is negligible compared with the diagonal components, and its relative error does not indicate the violation of reciprocity.

\subsection{Discussion on the de-phased average and conventional average}
\label{sec:conventional_average}
\begin{figure}
    \centering
    \includegraphics[width=1\linewidth]{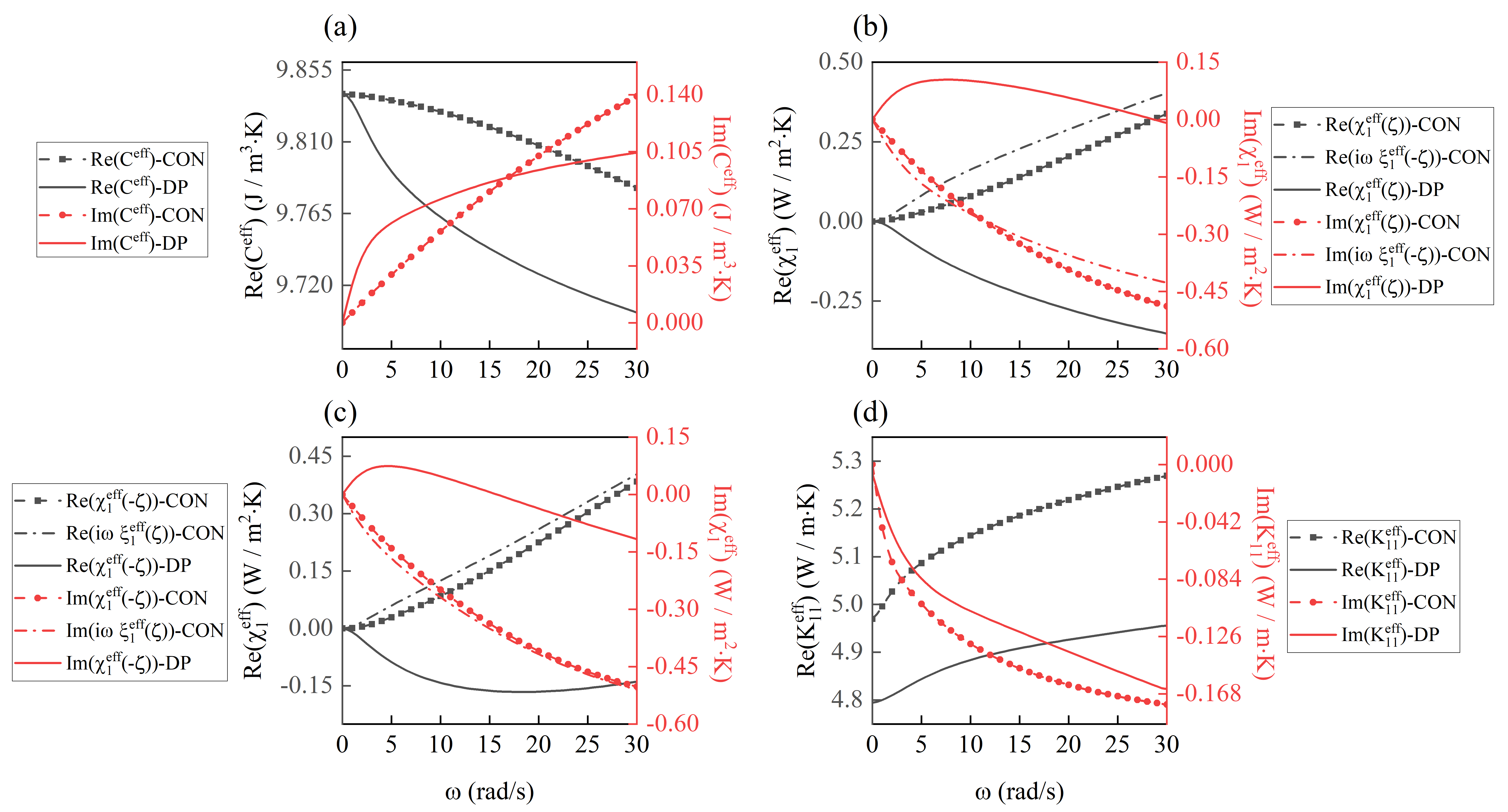}
    \caption{Comparison and variation of three effective properties evaluated by the DR-iBEM, (a) $C^{\text{eff}}$, (b) $\chi_{1}^\text{eff}(\bm{\zeta})$ and $i \omega \xi_{1}^\text{eff}(-\bm{\zeta})$; (c) $\chi_{1}^\text{eff}(-\bm{\zeta})$ and $i \omega \xi_{1}^\text{eff}(\bm{\zeta})$; and (d) $K_{11}^\text{eff}$. The composite contains uniformly distributed square inhomogeneities, and the volume fraction is $20\%$. The circular frequency $\omega \in [0, 30]$ rad/s, and the thermal wavenumber vector $\bm{\zeta} = (10, 0)$. The legends ``CON'' (uniform weight) and ``DP'' (de-phased weight) refer to the conventional and de-phased average, respectively.}
    \label{fig:avg_verify}
\end{figure}

Fig. \ref{fig:avg_verify} compares the effective properties extracted using the conventional spatial average and de-phased average, which are denoted as ``CON'' and ``DP'', respectively. The definition of the de-phased and conventional average is given in Eq. (\ref{eq:def_avg}), when the weight function $f(x) \equiv \exp[-i \bm{\zeta} \cdot \textbf{x}]$ and $1$. When $\omega = 0$, the two averaging schemes give the same prediction at the static limit. With increasing frequency, however, clear discrepancies can be found in both $C^\text{eff}$ and $K_{11}^\text{eff}$, as shown in Figs. \ref{fig:avg_verify} (a) and \ref{fig:avg_verify} (d). The difference primarily arises because the conventional average keeps the Bloch phase variation in the averaging process, while the de-phased average removes it. 

The influence of the averaging scheme is more evident from two coupling terms in Figs. \ref{fig:avg_verify} (b) and (c). The de-phased average and conventional average options have contradictory predictions on trends of the real and imaginary parts of the coupling term $\chi^{\text{eff}}_{1}$. Specifically, the de-phased average shows that the real part is a monotonically decreasing function, and the imaginary part increases at lower frequencies and then decreases. In contrast, the conventional average option shows a monotonically increasing real part and a monotonically decreasing imaginary part. Moreover, the opposite-wavenumber reciprocity relation is not satisfied using the conventional spatial average, which implies that the extracted effective thermal properties with the conventional average option are not physically consistent under the Bloch-form boundary conditions. Therefore, the difference in Fig. \ref{fig:avg_verify}(d) is expected. 

The discrepancies between the two averaging options can be interpreted from Eq. (\ref{eq:thermal_}), which indicates that under the Bloch-form boundary conditions, the temperature (or other thermal fields) contains a macroscopic thermal wavenumber vector (phase factor) and a periodic de-phased part. The conventional spatial option averages the physical field directly, which mixes contributions from both the phase factor and the de-phased part. Consequently, the extracted effective thermal properties are affected by the phase factor, which explains that such an averaging option cannot clearly represent the amplitude of the periodic de-phased parts. When the thermal wavenumber vector is sufficiently small, the effective thermal properties may be approximated using the conventional spatial average option. However, with greater thermal wavenumbers, such as $\bm{\zeta} = (10, 0)$ in Fig. \ref{fig:avg_verify}, the phase factor may cause cancellation or amplification, which leads to inconsistent physical representation of the effective thermal properties. In contrast, the de-phased average options remove the phase factor before averaging. Therefore, the averaged quantities only represent the amplitudes of the periodic parts. It should be emphasized that the extracted coupling terms by the de-phased average satisfy the reciprocal relation, while the conventional spatial average does not, which confirms that the de-phased averaging is necessary to extract effective thermal properties under the Bloch-form boundary conditions.

\subsection{Frequency-wavenumber dependent effective properties of circular particulate composites}
\begin{figure}
    \centering
    \includegraphics[width=1\linewidth]{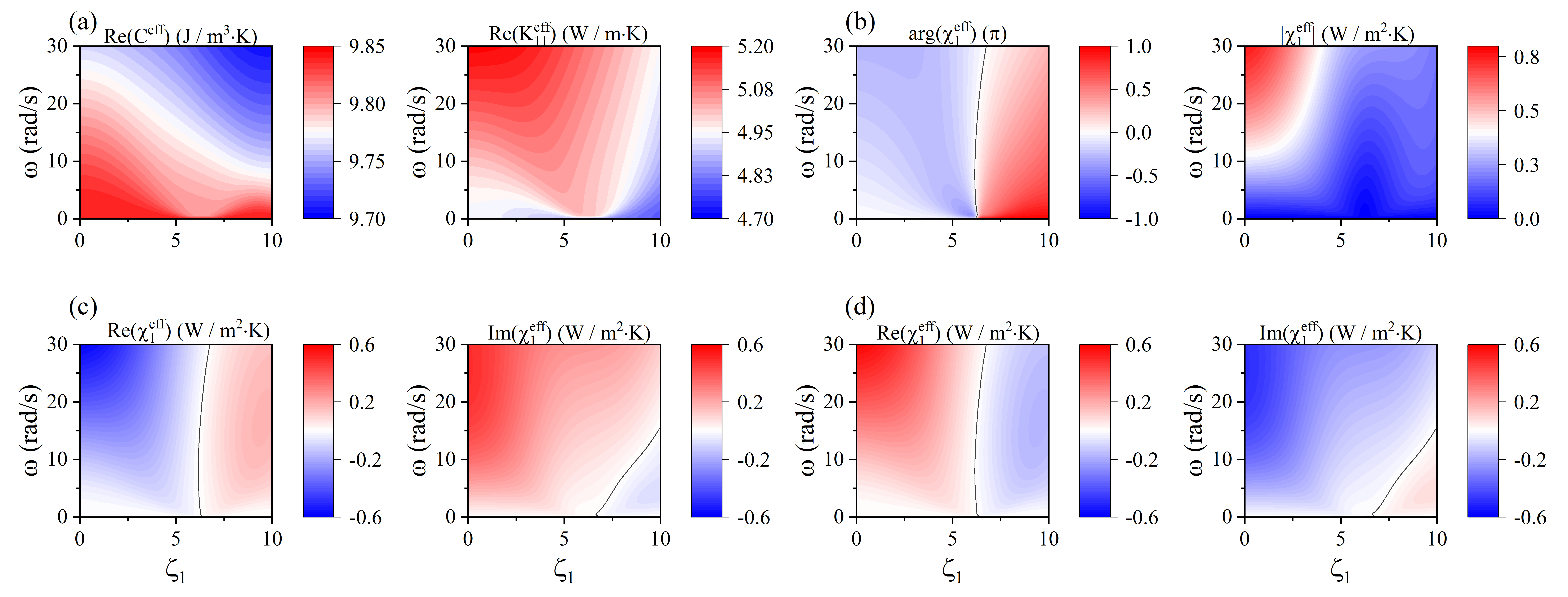}
    \caption{Frequency-wavenumber dependence of the effective thermal properties of a composite containing $200$ uniformly distributed circular inhomogeneities within the $0.8 \times 0.4 m^2$ region. (a) real parts of the heat capacity $C^\text{eff}$ and longitudinal thermal conductivity $K_{11}^\text{eff}$; (b) phase angle and magnitude of the coupling component $\chi^\text{eff}_{1}$; (c) real and imaginary parts of the coupling component $\chi_{1}^\text{eff}$ with the original microstructure and wavenumber $\bm{\zeta} = (\zeta_{1}, 0)$; and (d) real and imaginary parts of the coupling component $\chi_{1}^\text{eff}$ with the spatially reversed microstructure and wavenumber $\bm{\zeta} = (-\zeta_{1}, 0)$. The black curves in (b-d) refer to zero contours. $\omega \in [0, 30]$ rad/s, $\zeta_{1} \in [0, 10]$. }
    \label{fig:x_direction}
\end{figure}

The previous subsection has employed square inhomogeneities because their simple geometry allows for cross-verification by the Fourier-series-based method. To further demonstrate the applicability of the DR-iBEM algorithm to other shapes of inhomogeneities, this subsection replaces the square inhomogeneities in Fig. \ref{fig:homo_state} with circular inhomogeneities, while the geometric features and material properties remain, and the volume fraction is retained. Note that the microstructure remains asymmetric along the $x_{1}$ direction and symmetric in the $x_{2}$ direction about the horizontal axis. Fig. \ref{fig:x_direction} presents the effective properties over the frequency-wavenumber space, $\omega \in [0, 30]$ rad/s and $\zeta_{1} \in [0, 10]$. The smooth contour distributions indicate continuous variations of the effective thermal properties with frequency and wavenumber. Moreover, when the wavenumber is fixed, the effective thermal properties vary with frequency. Hence, a single set of frequency or wavenumber alone is inadequate to characterize the performance under Bloch-form settings. 

Fig. \ref{fig:x_direction} (a) shows the real parts of $C^\text{eff}$ and $K_{11}^\text{eff}$ for horizontal waves, $\bm{\zeta} = (\zeta_{1}, 0), \zeta_{1} \in [0, 10]$. The effective heat capacity $C^\text{eff}$ varies within a comparatively narrow range; however, Fig. \ref{fig:x_direction} (a) demonstrates that the effective heat capacity is both frequency- and wavenumber-dependent. For instance, at lower wavenumbers, the real part of the heat capacity decreases with increasing frequency. In contrast, the longitudinal thermal conductivity $Re(K_{11}^\text{eff})$ exhibits a stronger dependence on wavenumber. Specifically, at lower frequencies, $Re(K_{11}^\text{eff})$ exhibits a maximum around $\zeta_{1} \approx 6$. These observations confirm the dependence on boundary conditions; therefore, it is inappropriate to apply homogenized properties obtained using homogenization limits.

Fig. \ref{fig:x_direction} (b) presents the phase angle and magnitude of $\chi_{1}^\text{eff}$ for the original microstructure subjected to $\bm{\zeta} = (\zeta_{1}, 0)$. The coupling magnitude approaches zero in the low-frequency regions and increases with frequency, indicating that Willis-type homogenization theory is general, as the coupling terms vanish around the static limits. The magnitude is not a monotonic function of the wavenumber, and a region with comparatively small values can be observed close to $\zeta_{1} \approx 6$. The phase distribution is consistent with the real and imaginary parts in Fig. \ref{fig:x_direction} (c). Note that the phase angle of $\chi^\text{eff}_{1}$ represents the phase difference between the averaged heat flux contribution and the averaged temperature. As Fig. \ref{fig:x_direction} (b) indicates, the phase angle is both frequency- and wavenumber-dependent. In general, a negative phase angle denotes a lead of the coupling-induced heat flux relative to the averaged temperature, whereas a positive phase angle indicates a lag. With the wavenumber increase, the phase angle gradually shifted from negative to positive values. 

The real and imaginary parts in Fig. \ref{fig:x_direction} (c) provide a clear description of the variation of the phase angle. The real part of $\chi^\text{eff}_{1}$ changes from negative to positive as $\zeta_{1}$ crosses an intermediate value, while its imaginary part remains positive except in the higher-wavenumber and lower-frequency zone. In addition to the original microstructure, Fig. \ref{fig:x_direction} (d) plots the coupling component when the material arrangement spatially reverses, and the wave propagates from the opposite direction. Comparing Figs. \ref{fig:x_direction} (c) and (d), both the real and imaginary parts of the coupling component are opposite, while the zero contours remain unchanged. The results suggest that $-\chi_{1}^\text{eff, rev}(-\zeta_{1}, 0) = \chi_{1}^\text{eff, ori}(\zeta_{1}, 0)$, where the superscripts ``$\text{rev}$'' and ``$\text{ori}$'' refer to the spatially reversed and original microstructures, respectively. The coupling components exhibit direction-dependent behavior related to both the wave propagation and the microstructure. Reversing both the wave propagation direction and the asymmetric microstructure preserves the magnitude but reverses their signs.

\section{Conclusions}
This paper establishes a Willis-type thermal homogenization framework for two-dimensional particulate composites under harmonic heat transfer, which extracts the unique effective constitutive properties. While residual fields were introduced to generate additional independent macroscopic states in the literature, the present work constructs the complete constitutive operator by relating eigen-fields and macroscopic thermal inputs. Comparison with the one-dimensional exact residual-field homogenization formulation confirms that the present model can recover the unique Willis constitutive relations without introducing any residual fields. The framework is implemented using the DR-iBEM, in which mismatches in thermal conductivity and heat capacity are simulated by eigen-temperature-gradient (ETG) and an eigen-heat-source (EHS), respectively. The harmonic heat transfer problem is recast using the steady-state Green's function and classic Eshelby's tensors. The local thermal fields were verified against the finite element analysis, while the effective properties were cross-validated using an extended Fourier-series-based homogenization method for square inhomogeneities, which demonstrates consistent predictions of the effective heat capacity, thermal conductivity, and two Willis thermal coupling tensors. 

The Fourier formulae establish the opposite-wavenumber reciprocity relations between two coupling tensors, which are related by the factor $i \omega$. The effective heat capacity remains the same for opposite wavenumbers, while the effective thermal conductivity satisfies the corresponding transpose relation. Additional parity relations arise from the reflection symmetry of the microstructure. In particular, a coupling component vanishes only when both the microstructure and the prescribed wavenumber are invariant under the corresponding reflection. Therefore, reflection symmetry alone does not eliminate coupling terms, which is illustrated by the numerical examples for horizontal and vertical propagation. In general, the effective properties are frequency- and wavenumber-dependent, reflecting temporal dispersion and spatial non-locality. The two-dimensional thermal homogenization framework can be extended to general three-dimensional and multiphysical problems.

\section*{CRediT Author Statement}
\textbf{Chunlin Wu}: Conceptualization, Methodology, Data Curation, Validation, Visualization, Supervision, Writing-original draft, Writing Review \& Editing, Funding Acquisition; \textbf{Ning Ding}: Methodology, Data Curation, Validation, Writing-original draft; \textbf{Huiming Yin}: Methodology, Writing Review \& Editing. 

\section*{Acknowledgment}
CW's work is supported by National Natural Science Foundation of China Grant No. 12302086. HY's work was sponsored by the US Department of Agriculture NIFA \#2021-67021-34201 and the National Science Foundation (NSF) (IIP \#1738802 and IIP \#1941244). These supports are gratefully acknowledged.

\appendix 

\section{Numerical verification of square inhomogeneities} \label{sec:verify_square}
\begin{figure}
    \centering
    \includegraphics[width=0.7\linewidth]{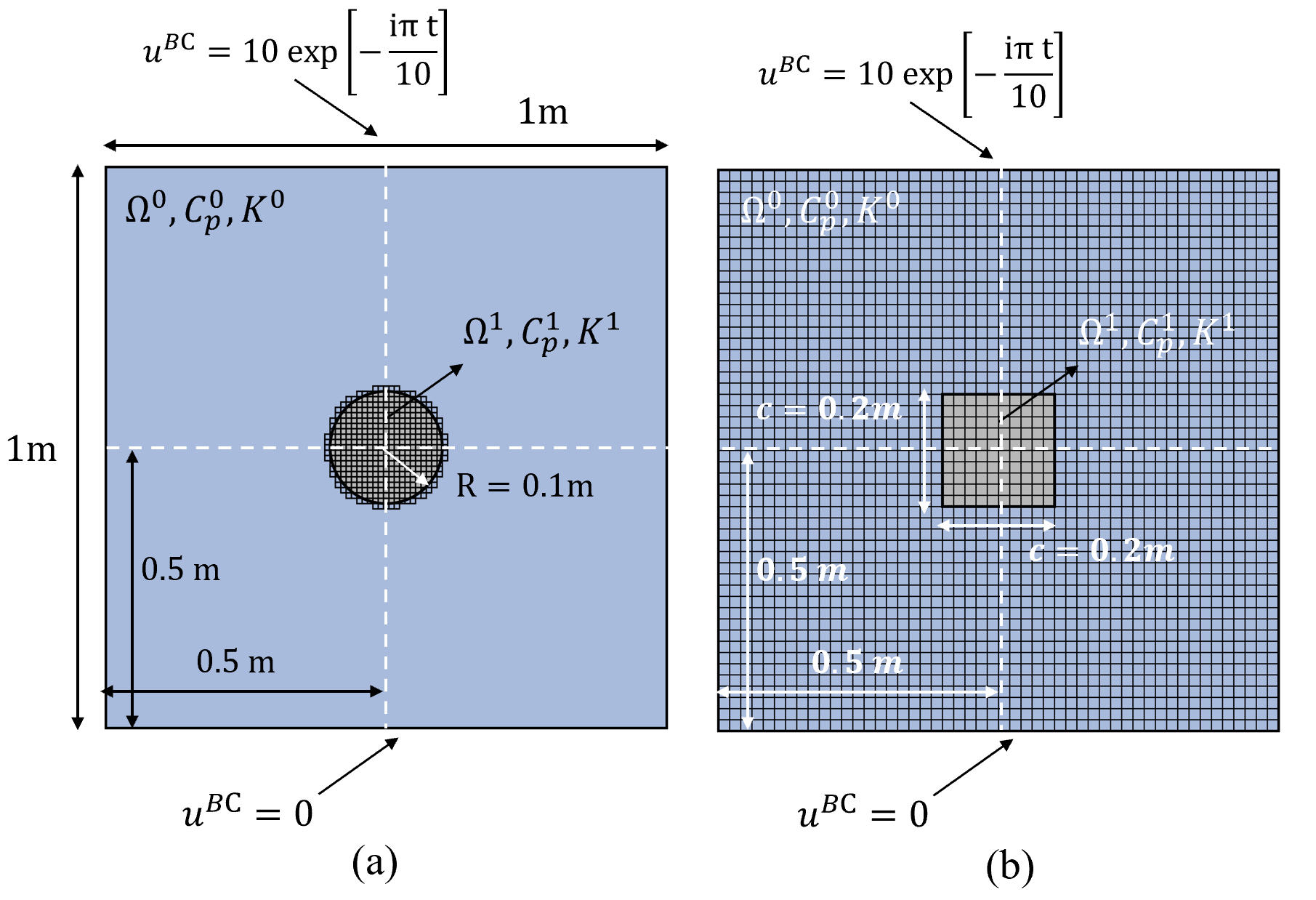}
    \caption{Geometry and boundary settings for numerical verification of the DR-iBEM algorithm on harmonic heat transfer. A square sample with length $1$ m contains (a) a single circular inhomogeneity placed at the center with radius $a = 0.1 m$, which is approximated by several square element sizes; (b) a square inhomogeneity with length $c = 0.2$ m placed at the center, where the entire domain is simulated by the comparison medium different from the matrix and the inhomogeneity. The top line is subjected to harmonic thermal load $u^{BC}(t) = 10 \exp[-\frac{i \pi t}{10}]$, while the temperature of the bottom line is zero.}
    \label{fig:case_square}
\end{figure}

This appendix provides verification of local thermal fields by the DR-iBEM algorithm against predictions by FEM. As Fig. \ref{fig:case_square} indicates, the boundary conditions and material properties are retained from Section \ref{sec:verify_circular}. In case (i), the matrix is selected as the comparison medium, while in case (ii) $K^c = 50 W / m \cdot K$, $C_p^c = 1 J / m^3 \cdot K$ as the comparison medium, which is different from the inhomogeneity and the matrix. For each case, the sizes of the square elements are provided. 

\subsection*{(i) One circular inhomogeneity approximated by multiple square elements}
\begin{figure}
    \centering
    \includegraphics[width=0.8\linewidth]{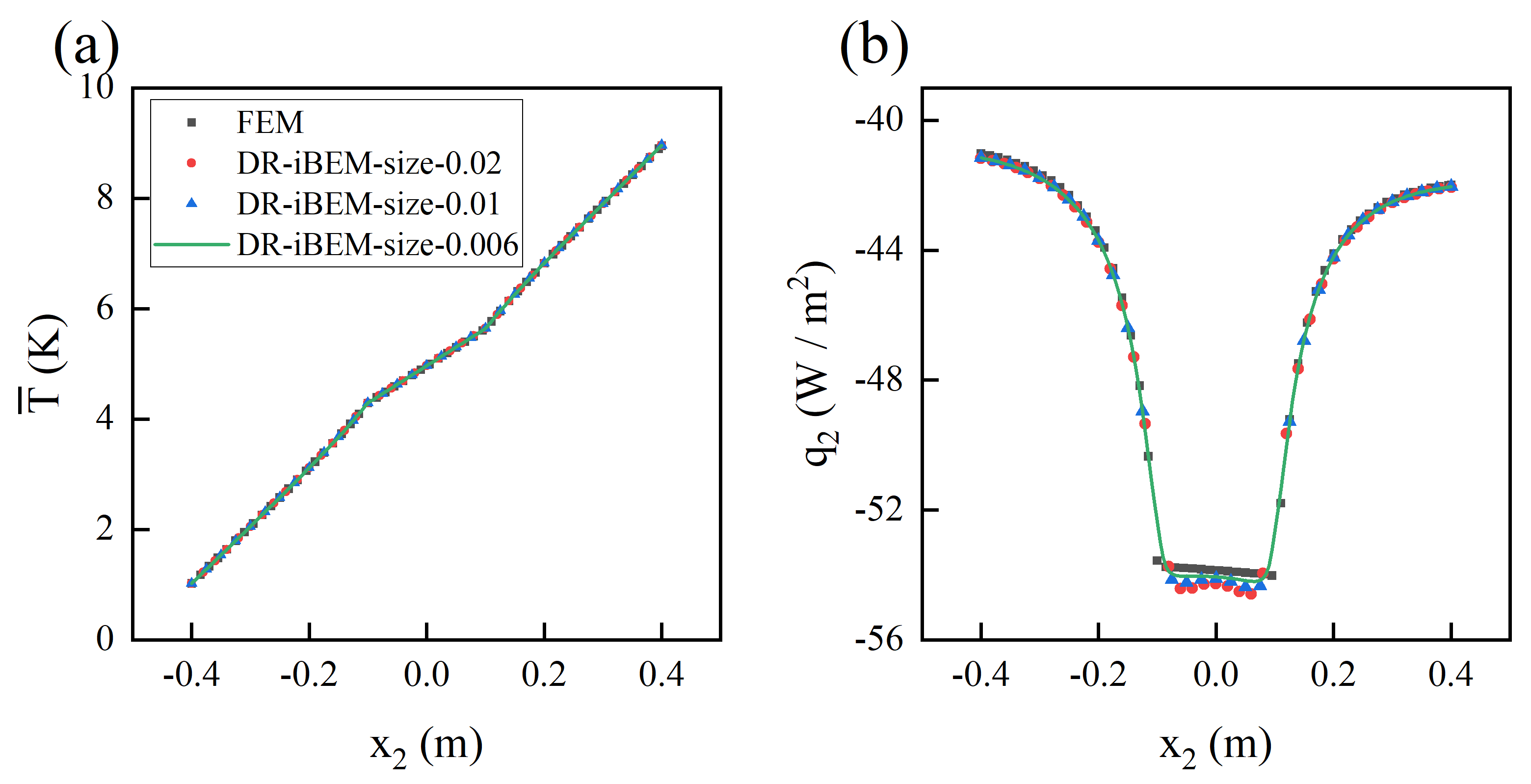}
    \caption{Comparison and variation of local thermal fields when the square sample contains a single circular inhomogeneity with radius $a = 0.1$ m. The (a) temperature and (b) heat flux evaluated by the DR-iBEM with square-elements (size $0.02, 0.01, 0.006$ m).}
    \label{fig:single_square}
\end{figure}

The primary reason for presenting this case study is that this discretization strategy has been widely applied in FFT-based methods, suggesting that a similar FFT-based EIM can be seamlessly integrated into the DR-iBEM framework. The DR-iBEM occupies $0.085,0.231,1$ GB RAM and $1.749,5.928,25.446$ s for element sizes $0.02, 0.01, 0.006$ m, respectively. Note that, unlike closed-form Eshelby's tensor for circular inclusions, the square elements can only approximate the geometric features. For instance, as shown in Fig. \ref{fig:case_square} (a), the square elements are kept if part of them lies in the circular subdomain. Hence, smaller square elements can better describe the actual shape of inhomogeneities. As Fig. \ref{fig:single_square} (a) indicates, predictions by square elements with sizes $0.02, 0.01, 0.006$ m agree well with results generated by the finite element analysis. However, the results of heat flux in Fig. \ref{fig:single_square} (b) exhibit apparent differences among the three sizes and the FEM. Based on the above discussion, the curve ``DR-iBEM-size-0.02'' exhibits a non-linear distribution of heat flux within the subdomain, and such discrepancies should be interpreted as inconsistent geometric features. In addition, when the element size decreases, curves ``DR-iBEM-size-0.01'' and ``DR-iBEM-size-0.006'' provide better predictions, though discrepancies can still be observed in the neighborhood of the inhomogeneities, such as $x_2 = \pm 0.1$ m. Therefore, the discrepancies caused by inconsistent geometric features will be eliminated for square inhomogeneities in the subsequent case (ii). 

\subsection*{(ii) One square inhomogeneity with a different comparison medium}
This subsection answers the fundamental question of whether the choice of comparison medium affects the predicted local thermal fields. This question is important since the comparison medium is usually introduced as a reference/auxiliary medium in Willis' polarization methods. As previously mentioned in Eq. (\ref{eq:equiv_cond}), if eigen-fields satisfy equivalent conditions with adequate accuracy, the eventual local thermal fields should be independent of the choice of the comparison medium. Therefore, eigen-fields and the comparison media are paired, so that changing the comparison medium alters the eigen-fields, but the change in the comparison medium should not alter the actual thermal fields in the original heterogeneous system. 

To answer this question, a square sample containing a single square inhomogeneity with length $c = 0.2$ m is considered. Unlike the two previous subsections, the thermal properties of the comparison medium $K^c$ and $C_p^c$ are different from those of any material phases. Therefore, eigen-fields are non-zero in both the matrix and inhomogeneity regions. The square is discretized into square elements with three sizes $0.02, 0.01$, and $0.005$ m, respectively. In such a case, this subsection validates: (i) whether the DR-iBEM algorithm is still applicable when the comparison medium is different from the matrix and inhomogeneities; and (ii) whether the solution converges when more square elements are applied. 

\begin{figure}
    \centering
    \includegraphics[width=0.8\linewidth]{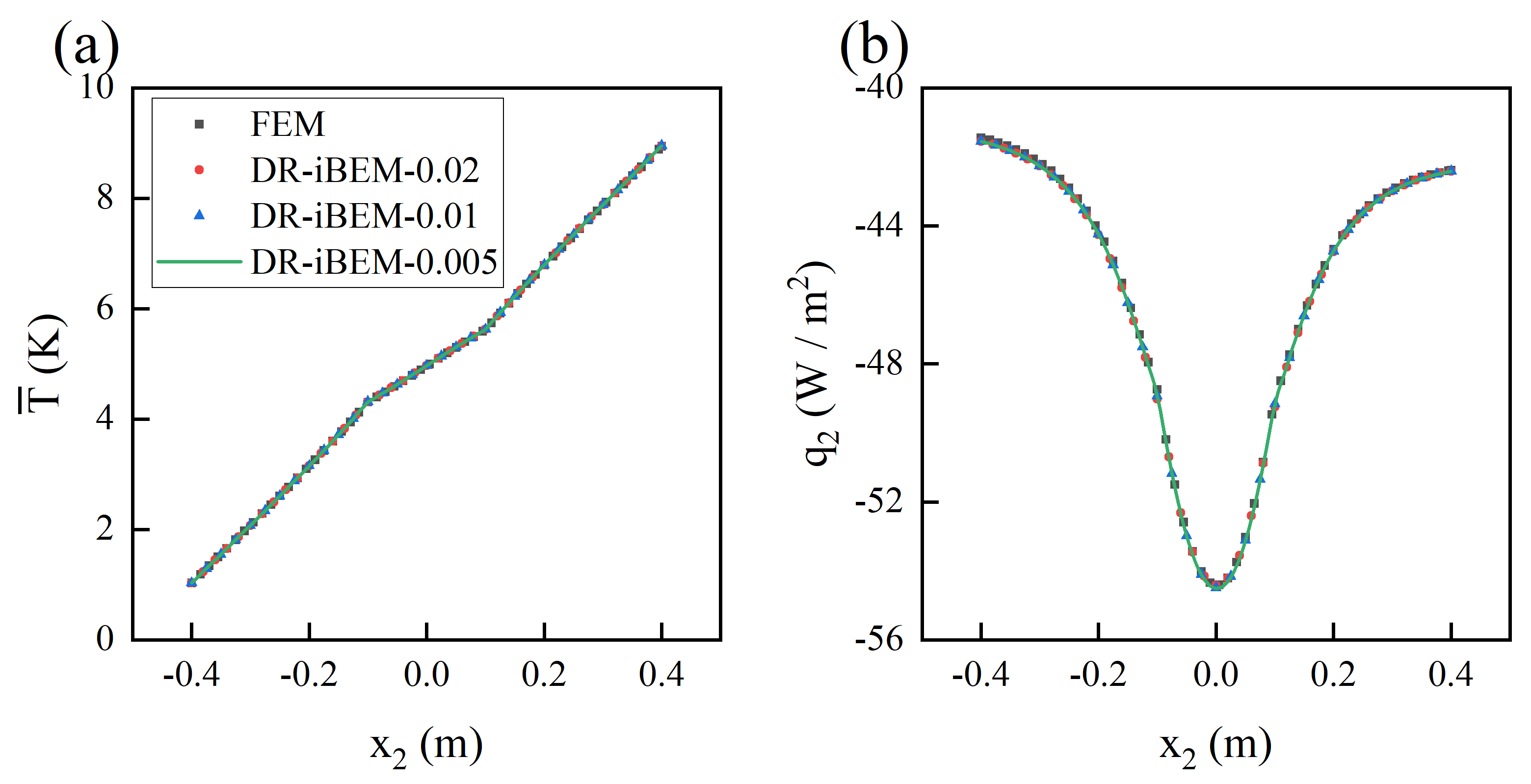}
    \caption{Comparison and variation of local thermal fields when the square sample contains a single square inhomogeneity with length $c = 0.2$ m. The DR-iBEM employs the comparison medium different from the matrix, and compares (a) temperature and (b) heat flux evaluated with element size $0.02, 0.01, 0.005$ m.}
    \label{fig:cmp}
\end{figure}

Fig. \ref{fig:cmp} (a) plots the temperature variations predicted by the DR-iBEM with three element sizes and the FEM. It can be observed that all curves from the DR-iBEM agree well with the FEM, and the temperature slope change is accurately captured in all curves. Subsequently, Fig. \ref{fig:cmp} (b) plots the heat flux variations. As numerical cases in Fig. \ref{fig:two} (b) indicate, the heat flux is a higher-order field than the temperature, which is supposed to be more sensitive to numerical errors. However, all curves by the DR-iBEM agree well with the FEM results. Particularly, although the temperature distribution in Fig. \ref{fig:cmp} (a) seems to be linear within the inhomogeneity, the spatial variation of heat flux in Fig. \ref{fig:cmp} (b) denies the assumption. Comparing Fig. \ref{fig:cmp} (b) with Fig. \ref{fig:single_square} (b), the spatial variation of heat flux is significantly influenced by the geometric shape of inhomogeneities. Specifically, since the square inhomogeneity can be exactly reproduced by several sub-elements, the discrepancies in Fig. \ref{fig:single_square} (b) do not exist in Fig. \ref{fig:cmp} (b), and therefore, it can be clearly attributed to geometric approximation errors. Moreover, as the element size decreases from $0.02$ to $0.005$ m, the numerical solutions remain stable and converge to the FEM results. 

However, it should be emphasized that the square inhomogeneity contains angular vertices. Our recent work \cite{Yang2026} rigorously proved that the ETG field and heat flux field must share the same singular exponent. The case study only aims to verify the robustness of the DR-iBEM algorithm when the comparison medium is selected differently from the matrix. Since the singular heat flux is highly localized and the plotted path is away from the corners, the comparison between the DR-iBEM and FEM results primarily reflects the regular part of the local thermal fields. Hence, the good agreement between the DR-iBEM and FEM results is sufficiently meaningful for homogenization analysis, where the effective properties are extracted from averaged thermal fields over thousands of sampling points. The above results confirm that the DR-iBEM algorithm is robust, as the choice of the comparison medium does not alter the predictions of local thermal fields once the eigen-fields are consistently solved. Together with two previous subsections, this case study confirms the accuracy and ability of the DR-iBEM algorithm and subsequent application in Willis-type homogenization. Moreover, since local solutions are independent of the comparison medium, the homogenization procedure can be conducted by setting the matrix as the comparison medium, which significantly reduces the total degrees of freedom, as no interior discretization in the matrix phase is required.

\section{Fourier proof of the opposite-wavenumber reciprocity relation} \label{sec:proof}

This appendix provides a proof of the opposite-wavenumber reciprocity relation based on the Fourier-series-based homogenization formulae in Section 5.1. Note that the following derivation considers $\omega \neq 0$ for general unsteady cases. When $\omega = 0$, the static case can be recovered from the zero-frequency limit. 

\subsection*{Inhomogeneity subdomain interaction coefficient matrices}
The sampling domain $A_{\mathcal{D}}$ contains $N_I$ subdomains. Since the matrix is selected as the comparison medium, nonzero eigen-fields only exist in inhomogeneity subdomains. The area fraction of the $I^\text{th}$ subdomain and geometric Fourier coefficient are defined as, $f^I = A_{\Omega^I} / A_{\mathcal{D}}$ and $g^I(\bm{\eta}) = 1 / A_{\Omega^I} \int_{\Omega^I} \exp[i \bm{\eta} \cdot \textbf{x}] \thinspace d\textbf{x}$, respectively. 

Based on Eq. (\ref{eq:avg_dephase_fourier}), the subdomain interaction coefficients can be assembled as, 

\begin{equation}
\begin{aligned}
    \overline{\mathcal{F}}^{IJ}(\bm{\zeta}) = f^{J} \sum_{\bm{\eta} \neq \textbf{0}} g^I(\bm{\eta}) \mathcal{F}(\bm{\zeta}, \bm{\eta}) g^J(-\bm{\eta}), \quad \overline{\mathcal{M}}_{j}^{IJ}(\bm{\zeta}) = f^{J} \sum_{\bm{\eta} \neq \textbf{0}} g^I(\bm{\eta}) \mathcal{M}_{j}(\bm{\zeta}, \bm{\eta}) g^J(-\bm{\eta}) \\ 
    \overline{\mathcal{N}}_{j}^{IJ}(\bm{\zeta}) = f^{J} \sum_{\bm{\eta} \neq \textbf{0}} g^I(\bm{\eta}) \mathcal{N}_{j}(\bm{\zeta}, \bm{\eta}) g^J(-\bm{\eta}), \quad \overline{\mathcal{T}}_{ij}^{IJ}(\bm{\zeta}) = f^{J} \sum_{\bm{\eta} \neq \textbf{0}} g^I(\bm{\eta}) \mathcal{T}_{ij}(\bm{\zeta}, \bm{\eta}) g^J(-\bm{\eta}) \\ 
\end{aligned}
    \label{eq:coef_asem}
\end{equation}
where the superscripts $I$ and $J$ represent the influence at the $I^\text{th}$ subdomain caused by the $J^\text{th}$ subdomain. Note that the reciprocal-lattice vectors $\pm \bm{\eta}$ are included in the centrosymmetric Fourier truncation. For instance, the case study in Section 6 utilizes $(2 M + 1) \times (2 M + 1) - 1$ truncation, where $M = 150$. Replacing $\bm{\eta}$ by $-\bm{\eta}$ in Eq. (\ref{eq:coef_asem}), interchanging the subdomain indices $I$ and $J$, and using Eq. (\ref{eq:fourier_kernel_property}), the following properties can be obtained, 

\begin{equation}
\begin{aligned}
    f^I \thinspace \overline{\mathcal{F}}^{IJ}(\bm{\zeta}) = f^J \thinspace \overline{\mathcal{F}}^{JI}(-\bm{\zeta}), \quad f^I \thinspace \overline{\mathcal{M}}_{j}^{IJ}(\bm{\zeta}) = \frac{i K^c}{\omega} f^J \overline{\mathcal{N}}_{j}^{JI}(-\bm{\zeta}) \\
    \quad \frac{i K^c}{\omega} f^I \thinspace \overline{\mathcal{N}}_{j}^{IJ}(\bm{\zeta}) = f^J \overline{\mathcal{M}}_{j}^{JI}(-\bm{\zeta}) , \quad
    f^I \thinspace \overline{\mathcal{T}}^{IJ}_{ij}(\bm{\zeta}) = f^J \thinspace \overline{\mathcal{T}}_{ji}^{JI}(-\bm{\zeta})
\end{aligned}
    \label{eq:coeff_property}
\end{equation}
Let $\textbf{W} = \text{diag} \begin{bmatrix} f^1 & f^2 & \cdots & f^{N_I} \end{bmatrix}$ and $\textbf{W}_{d} = \text{diag} \begin{bmatrix} f^1 \textbf{I}^{(2)} & f^2 \textbf{I}^{(2)} & \cdots & f^{N_I} \textbf{I}^{(2)} \end{bmatrix}$, where $\textbf{I}^{(2)}$ is the $2 \times 2$ identity matrix. In matrix notation, Eq. (\ref{eq:coeff_property}) can be expressed as, 

\begin{equation}
    \begin{aligned}
    \textbf{W} \bm{\mathcal{F}}(\bm{\zeta}) = \bm{\mathcal{F}}^T(-\bm{\zeta}) \textbf{W}, \quad \textbf{W} \bm{\mathcal{M}}(\bm{\zeta}) = \frac{i K^c}{\omega} \bm{\mathcal{N}}^T(-\bm{\zeta}) \textbf{W}_d \\ \frac{i K^c}{\omega} \textbf{W}_{d} \bm{\mathcal{N}}(\bm{\zeta}) = \bm{\mathcal{M}}^T(-\bm{\zeta}) \textbf{W}, \quad \textbf{W}_{d} \bm{\mathcal{T}}(\bm{\zeta}) = \bm{\mathcal{T}}^T(-\bm{\zeta}) \textbf{W}_{d}
    \end{aligned}
    \label{eq:coeff_property_matrix}
\end{equation}
where $\bm{\mathcal{F}},\bm{\mathcal{M}}, \bm{\mathcal{N}}, \bm{\mathcal{T}}$ collect coefficients presented in Eq. (\ref{eq:coef_asem}). 

\subsection*{Transpose relation of the global coefficient matrix}
The global coefficient matrix of the Fourier-series method can be written as, 

\begin{equation}
    \underbrace{\begin{bmatrix} \bm{\Delta C}_p - \bm{\mathcal{F}}(\bm{\zeta}) & - \bm{\mathcal{M}}(\bm{\zeta}) \\ - \bm{\mathcal{N}}(\bm{\zeta}) &  \bm{\Delta K} - \bm{\mathcal{T}}(\bm\zeta) \end{bmatrix}}_{\bm{\mathcal{J}}_\text{eig}(\bm{\zeta})} \underbrace{\begin{bmatrix} \hat{\bm{Q}}^* \\ \hat{\bm{U}}^* \end{bmatrix}}_{\hat{\bm{\mathcal{X}}}_\text{eig}} = \underbrace{\begin{bmatrix}
        \mathbf 1^{(N_I)} & \mathbf 0^{(N_I\times 2)}\\
        \mathbf 0^{(2N_I\times 1)} & \begin{bmatrix} \mathbf I^{(2)}\\ \vdots\\ \mathbf I^{(2)}
        \end{bmatrix}
    \end{bmatrix}}_{\bm{\mathcal{I}}_\text{eig}} \underbrace{\begin{bmatrix} \langle u \rangle \\ \langle u_{,1} \rangle \\ \langle u_{,2} \rangle \end{bmatrix}}_{\hat{\textbf{U}}}
    \label{eq:global_fourier}
\end{equation}
%
where $\bm{\mathcal{J}}_{\text{eig}}$ is the global coefficient matrix involving equivalent conditions; $\hat{\bm{\mathcal{X}}}_{\text{eig}}$ refers to the de-phased eigen-fields within each subdomain. Note that if one of material differences become zero, i.e., selecting the matrix material as the comparison medium, the corresponding eigen-field degree of freedom and equivalent condition equations are omitted. Therefore, Eq. (\ref{eq:global_fourier}) applies to a reduced eigen-field system. Specifically, $\hat{\bm{Q}}^* = \begin{bmatrix} \hat{Q}^{1*} & \cdots & \hat{Q}^{N_I*} \end{bmatrix}^T$ and $\hat{\bm{U}}^* = \begin{bmatrix} \hat{u}_{1}^{1*} & \hat{u}_{2}^{1*} & \cdots & \hat{u}_{1}^{N_I*} & \hat{u}_{2}^{N_I*} \end{bmatrix}^T$. $\bm{\Delta C}_p = \text{diag} \begin{bmatrix} \frac{1}{\Delta C_p^1} & \cdots & \frac{1}{\Delta C_p^{N_I}} \end{bmatrix}$ and $\bm{\Delta K} = \text{diag} \begin{bmatrix} \frac{K^c}{\Delta K^1} \textbf{I}^{(2)} & \cdots & \frac{K^c}{\Delta K^{N_I}}\textbf{I}^{(2)} \end{bmatrix}$, which are two diagonal matrices. The matrix $\bm{\mathcal{I}}_\text{eig}$ maps the macroscopic input vector $\hat{\bm{U}}$ to the subdomain equations. 

Consider a block weighting matrix, $\bm{\mathcal{S}}_{\text{eig}} = \begin{bmatrix} \mathbf W & \mathbf 0 \\ \mathbf 0 & \frac{iK^{c}}{\omega}\mathbf W_{d} \end{bmatrix}$. Since the local contrast matrices, $\bm{\Delta C}_p$ and $\bm{\Delta K}$, are diagonal, they should satisfy, $\mathbf{W} \bm{\Delta C}_p = \bm{\Delta C}_p^T \mathbf{W}$ and $\mathbf{W}_{d} \bm{\Delta K} = \bm{\Delta K}^T \mathbf{W}_{d}$. Based on Eq. (\ref{eq:coeff_property_matrix}), one can write, 


\begin{equation}
\begin{aligned}
    \bm{\mathcal{S}}_{\text{eig}} \bm{\mathcal{J}}_{\text{eig}}(\bm{\zeta}) = \begin{bmatrix} \textbf{W} \left[ \bm{\Delta C}_p - \bm{\mathcal{F}}(\bm{\zeta}) \right] & - \textbf{W} \bm{\mathcal{M}}(\bm{\zeta}) \\ -\frac{i K^c}{\omega} \textbf{W}_d \bm{\mathcal{N}}(\bm{\zeta}) & \frac{i K^c}{\omega} \left[ \bm{\Delta K - \bm{\mathcal{T}}(\bm{\zeta})} \right] \end{bmatrix} \\
    \bm{\mathcal{J}}_{\text{eig}}^T(-\bm{\zeta}) \bm{\mathcal{S}}_{\text{eig}} = \begin{bmatrix} \left[ \bm{\Delta C}_p - \bm{\mathcal{F}}(-\bm{\zeta}) \right]^T \textbf{W}  & -\frac{i K^c}{\omega} \bm{\mathcal{N}}^T(-\bm{\zeta}) \textbf{W}_d \\ -\bm{\mathcal{M}}^T(-\bm{\zeta}) \textbf{W} & \frac{i K^c}{\omega} \left[ \bm{\Delta K - \bm{\mathcal{T}}(-\bm{\zeta})} \right]^T \textbf{W}_d  \end{bmatrix}
\end{aligned}
\end{equation}
Substituting the relations in Eq. (\ref{eq:coeff_property_matrix}), the two above equations are equivalent, 

\begin{equation}
    \bm{\mathcal{S}}_{\text{eig}} \bm{\mathcal{J}}_{\text{eig}}(\bm{\zeta}) = \bm{\mathcal{J}}_{\text{eig}}^T(-\bm{\zeta}) \bm{\mathcal{S}}_{\text{eig}}
\end{equation}
where the above equation shows a weighted transpose relation between the global coefficient matrix related to the opposite Bloch wavenumbers. Specifically, 

\begin{equation}
    \left(\bm{\mathcal{J}}_{\text{eig}}(\bm{\zeta})\right)^{-1} = \bm{\mathcal{S}}_{\text{eig}}^{-1}
    \left( \bm{\mathcal{J}}_{\text{eig}}^{T}(-\bm{\zeta}) \right)^{-1}
    \bm{\mathcal{S}}_{\text{eig}}
    \label{eq:J_inverse_relation}
\end{equation}

Combining Eq. (\ref{eq:global_fourier}) and Eq. (\ref{eq:J_inverse_relation}), the de-phased eigen-fields in all subdomains are related to the macroscopic inputs as, 

\begin{equation}
 \hat{\bm{\mathcal{X}}}_{\text{eig}} = \left( \bm{\mathcal{J}}_{\text{eig}}(\bm{\zeta}) \right)^{-1} \bm{\mathcal{I}}_\text{eig} \hat{\textbf{U}}
    \label{eq:eigenfield_solution}
\end{equation}
and the de-phased averaged eigen-fields can be rewritten as, 

\begin{equation}
    \begin{bmatrix} \langle Q^*\rangle\\ \frac{i K^c}{\omega} \langle u_1^*\rangle\\ \frac{i  K^c}{\omega} \langle u_2^*\rangle \end{bmatrix} = \bm{\mathcal I}_\text{eig}^{T} \bm{\mathcal S}_{\text{eig}} \hat{\bm{\mathcal{X}}}_{\text{eig}} = \bm{\mathcal I}_\text{eig}^{T} \bm{\mathcal S}_{\text{eig}}
    \left(\bm{\mathcal J}_{\mathrm{eig}}(\bm{\zeta})\right)^{-1} 
    \bm{\mathcal I}_\text{eig} \hat{\textbf U}
    \label{eq:avg_eigen_macro_1}
\end{equation}

Consequently, using Eq. (\ref{eq:J_inverse_relation}), the de-phased averaged eigen-fields can be written in terms of the global coefficients associated with the opposite wavenumber, 

\begin{equation}
    \begin{aligned}
    \bm{\mathcal I}_\text{eig}^{T} \bm{\mathcal S}_{\text{eig}}
    \left(\bm{\mathcal J}_{\mathrm{eig}}(\bm{\zeta})\right)^{-1} 
    \bm{\mathcal I}_\text{eig} \hat{\textbf U} = \bm{\mathcal{I}}^T_\text{eig} \left(\bm{\mathcal J}^T_{\mathrm{eig}}(-\bm{\zeta})\right)^{-1} \bm{\mathcal S}_{\text{eig}} \bm{\mathcal{I}}_\text{eig} \hat{\textbf{U}} = \left[ \bm{\mathcal{I}}^T_\text{eig} \bm{\mathcal{S}}_\text{eig} \left( \bm{\mathcal{J}}_\text{eig}(-\bm{\zeta}) \right)^{-1} \bm{\mathcal{I}}_\text{eig}  \right]^T \hat{\textbf{U}}
    \end{aligned}
    \label{eq:avg_eigen_macro}
\end{equation}
where Eq. (\ref{eq:J_inverse_relation}) and the symmetric relation $\bm{\mathcal{S}}_{\text{eig}} = \bm{\mathcal{S}}^T_{\text{eig}}$ are applied. The right-hand side of Eq. (\ref{eq:avg_eigen_macro}) can be rewritten as $3 \times 3$ coefficient matrix multiplied by the macroscopic inputs. Specifically, for opposite wavenumber vectors, it shows, 

\begin{equation}
    \begin{bmatrix}
        \langle Q^*\rangle(\bm{\zeta}) \\ \frac{i K^c}{\omega} \langle u_1^*\rangle (\bm{\zeta})\\ \frac{i K^c}{\omega} \langle u_2^*\rangle (\bm{\zeta})
    \end{bmatrix} = \underbrace{\begin{bmatrix}
    \mathcal{G}^+_{11} & \mathcal{G}^+_{12} & \mathcal{G}^+_{13} \\
    \mathcal{G}^+_{21} & \mathcal{G}^+_{22} & \mathcal{G}^+_{23} \\
    \mathcal{G}^+_{31} & \mathcal{G}^+_{32} & \mathcal{G}^+_{33} \\
    \end{bmatrix}}_{\bm{\mathcal{G}}^+} \begin{bmatrix} \langle u \rangle \\ \langle u_{,1} \rangle \\ \langle u_{,2} \rangle\end{bmatrix}, \quad     \begin{bmatrix}
        \langle Q^*\rangle(-\bm{\zeta}) \\ \frac{i K^c}{\omega} \langle u_1^*\rangle (-\bm{\zeta})\\ \frac{i K^c}{\omega} \langle u_2^*\rangle (-\bm{\zeta})
    \end{bmatrix} = \underbrace{\begin{bmatrix}
    \mathcal{G}^-_{11} & \mathcal{G}^-_{12} & \mathcal{G}^-_{13} \\
    \mathcal{G}^-_{21} & \mathcal{G}^-_{22} & \mathcal{G}^-_{23} \\
    \mathcal{G}^-_{31} & \mathcal{G}^-_{32} & \mathcal{G}^-_{33} \\
    \end{bmatrix}}_{\bm{\mathcal{G}}^-}  \begin{bmatrix} \langle u \rangle \\ \langle u_{,1} \rangle \\ \langle u_{,2} \rangle\end{bmatrix}
    \label{eq:g_zeta}
\end{equation}
where two coefficient matrices $\bm{\mathcal{G}}^+ = \left(\bm{\mathcal{G}}^- \right)^T$. Therefore, the effective properties can be obtained as,  

\begin{equation}
    \begin{aligned}
        C^\text{eff}(\bm{\zeta})&=C_p^c-\mathcal{G}_{11}^{+}, & \xi_1^{\text{eff}} (\bm{\zeta}) &= -\mathcal{G}_{12}^{+},&\xi_2^{\text{eff}}(\bm{\zeta})&= -\mathcal{G}_{13}^{+} \\
        \chi_1^{\text{eff}}(\bm{\zeta}) & = -i\omega\mathcal{G}_{21}^{+}, & K_{11}^{\text{eff}}(\bm{\zeta})&=K^c + i \omega \mathcal{G}_{22}^{+},&  K_{12}^{\text{eff}}(\bm{\zeta})&=i\omega\mathcal{G}_{23}^{+}\\
        \chi_2^{\text{eff}}(\bm{\zeta})&=-i\omega\mathcal{G}_{31}^{+} , &K_{21}^{\text{eff}}(\bm{\zeta})&=i\omega\mathcal{G}_{32}^{+},&K_{22}^{\text{eff}}(\bm{\zeta})&=K^c+i\omega\mathcal{G}_{33}^{+} 
    \end{aligned}
    \label{eq:effec_fourier_pos}
\end{equation}
and
\begin{equation}
    \begin{aligned}
        C^\text{eff}(-\bm{\zeta})&=C_p^c-\mathcal{G}_{11}^{-}, & \xi_1^{\text{eff}} (-\bm{\zeta}) &= -\mathcal{G}_{12}^{-},&\xi_2^{\text{eff}}(-\bm{\zeta})&= -\mathcal{G}_{13}^{-} \\
        \chi_1^{\text{eff}}(-\bm{\zeta}) & = -i\omega\mathcal{G}_{21}^{-}, & K_{11}^{\text{eff}}(-\bm{\zeta})&=K^c + i \omega \mathcal{G}_{22}^{-},&  K_{12}^{\text{eff}}(-\bm{\zeta})&=i\omega\mathcal{G}_{23}^{-}\\
        \chi_2^{\text{eff}}(-\bm{\zeta})&=-i\omega\mathcal{G}_{31}^{-} , &K_{21}^{\text{eff}}(-\bm{\zeta})&=i\omega\mathcal{G}_{32}^{-},&K_{22}^{\text{eff}}(-\bm{\zeta})&=K^c+i\omega\mathcal{G}_{33}^{-} 
    \end{aligned}
    \label{eq:effec_fourier_neg}
\end{equation}
Because $\bm{\mathcal{G}}^+ = \left(\bm{\mathcal{G}}^- \right)^T$, the relations of effective properties of two opposite wavenumbers are, 

\begin{equation}
    \begin{aligned}
    C^\text{eff}(\bm{\zeta}) &= C^\text{eff}(-\bm{\zeta}) \quad \text{and} \quad \bm{K}^\text{eff}(\bm{\zeta}) = \left(\bm{K}^\text{eff}(-\bm{\zeta})\right)^T
    \end{aligned}
    \label{eq:final_relation}
\end{equation}
and the reciprocity relation is, 

\begin{equation}
    \bm{\chi}^\text{eff}(\bm{\zeta}) = i \omega \left( \bm{\xi}^\text{eff}(-\bm{\zeta}) \right)^T \quad \text{and} \quad \bm{\chi}^\text{eff}(-\bm{\zeta}) = i \omega \left( \bm{\xi}^\text{eff}(\bm{\zeta}) \right)^T
\end{equation}

\subsection*{Discussion of the parity relation for a reflection-symmetric microstructure}
This subsection provides discussion on the parity relation of effective properties when the microstructure is symmetric. Without loss of any generality, consider a microstructure symmetric about the horizontal axis $x_2 = 0$ (the coordinate origin is located on the mirror axis), as indicated in Fig. \ref{fig:homo_state}. For each subdomain $\Omega^I$, there exists a mirror subdomain $\Omega^{I'}$ about the horizontal axis, which exhibits the same thermal properties and geometry characteristics, such as area fraction, heat capacity, and thermal conductivity. Hence, the geometric Fourier coefficients of two subdomains are related as, 
\begin{equation}
    g^{I'}(\eta_{1}, \eta_{2}) = \frac{1}{A_{\Omega^{I'}}} \int_{\Omega^{I'}} \exp[i (\eta_{1} x_{1}' + \eta_{2} x_{2}')] \thinspace d\textbf{x}' = \frac{1}{A_{\Omega^{I}}} \int_{\Omega^I} \exp[i(\eta_{1} x_{1} - \eta_{2} x_{2})] \thinspace d\textbf{x} = g^{I}(\eta_{1}, -\eta_{2})
    \label{eq:g_reflection}
\end{equation}
and Fourier tensors in Eq. (\ref{eq:four_tensors}) with opposite $-\zeta_{2}, -\eta_{2}$ can be expressed as, ($\overline{\bm{\zeta}} = (\zeta_{1}, -\zeta_{2})$ and $\overline{\bm{\eta}} = (\eta_{1}, -\eta_{2})$), 

\begin{equation}
    \begin{aligned}
    \mathcal{F} (\overline{\bm{\zeta}}, \overline{\bm{\eta}}) & = \mathcal{F}(\bm{\zeta}, \bm{\eta}), & \mathcal{M}_{1} (\overline{\bm{\zeta}}, \overline{\bm{\eta}}) & = \mathcal{M}_{1}(\bm{\zeta}, \bm{\eta}), & \mathcal{M}_{2} (\overline{\bm{\zeta}}, \overline{\bm{\eta}}) & = -\mathcal{M}_{2}(\bm{\zeta}, \bm{\eta}) \\
    \mathcal{N}_{1} (\overline{\bm{\zeta}}, \overline{\bm{\eta}}) & = \mathcal{N}_{1}(\bm{\zeta}, \bm{\eta}), & \mathcal{N}_{2} (\overline{\bm{\zeta}}, \overline{\bm{\eta}}) & = -\mathcal{N}_{2}(\bm{\zeta}, \bm{\eta}), & \mathcal{T}_{11} (\overline{\bm{\zeta}}, \overline{\bm{\eta}}) & = \mathcal{T}_{11} (\bm{\zeta}, \bm{\eta}) \\
    \mathcal{T}_{12} (\overline{\bm{\zeta}}, \overline{\bm{\eta}}) & = -\mathcal{T}_{12} (\bm{\zeta}, \bm{\eta}), &  \mathcal{T}_{21} (\overline{\bm{\zeta}}, \overline{\bm{\eta}}) & = -\mathcal{T}_{21} (\bm{\zeta}, \bm{\eta}), &  \mathcal{T}_{22} (\overline{\bm{\zeta}}, \overline{\bm{\eta}}) & = \mathcal{T}_{22} (\bm{\zeta}, \bm{\eta})
    \end{aligned}
    \label{eq:kernel_reflection}
\end{equation}

For the system associated with $\overline{\bm{\zeta}}$, the mirror subdomains are relabeled that the subdomain $\Omega^{I'}$ occupies the position originally assigned to $\Omega^I$. Substituting Eqs. (\ref{eq:g_reflection}) and Eq. (\ref{eq:kernel_reflection}) into Eq. (\ref{eq:coef_asem}) provides, 

\begin{equation}
    \begin{aligned}
    \overline{\mathcal{F}}^{I'J'}(\overline{\bm{\zeta}}) & = \overline{\mathcal{F}}^{IJ}(\bm{\zeta}), & \overline{\mathcal{M}}_{1}^{I'J'}(\overline{\bm{\zeta}}) & = \overline{\mathcal{M}}_{1}^{IJ}(\bm{\zeta}), & \overline{\mathcal{M}}_{2}^{I'J'}(\overline{\bm{\zeta}}) & = -\overline{\mathcal{M}}_{2}^{IJ}(\bm{\zeta}) \\
    \overline{\mathcal{N}}_{1}^{I'J'}(\overline{\bm{\zeta}}) & = \overline{\mathcal{N}}_{1}^{IJ}(\bm{\zeta}), & \overline{\mathcal{N}}_{2}^{I'J'}(\overline{\bm{\zeta}}) & = -\overline{\mathcal{N}}_{2}^{IJ}(\bm{\zeta}), & \overline{\mathcal{T}}_{11}^{I'J'}(\overline{\bm{\zeta}}) & = \overline{\mathcal{T}}_{11}^{IJ}(\bm{\zeta}) \\ 
    \overline{\mathcal{T}}_{12}^{I'J'}(\overline{\bm{\zeta}}) & = -\overline{\mathcal{T}}_{12}^{IJ}(\bm{\zeta}), &\overline{\mathcal{T}}_{21}^{I'J'}(\overline{\bm{\zeta}}) & = -\overline{\mathcal{T}}_{21}^{IJ}(\bm{\zeta}), & \overline{\mathcal{T}}_{22}^{I'J'}(\overline{\bm{\zeta}}) & = \overline{\mathcal{T}}_{22}^{IJ}(\bm{\zeta}) 
    \end{aligned}
    \label{eq:coefficient_reflection}
\end{equation}

Let $\bm{D}^R = \text{diag} \begin{bmatrix} 1 & -1 & 1 & -1 & \cdots & 1 & -1 \end{bmatrix}$ denote a diagonal matrix with dimension $2 N_I \times 2 N_I$. Based on Eq. (\ref{eq:coefficient_reflection}), components of the global coefficient matrix for $\overline{\bm{\zeta}}$ can be expressed as, 
\begin{equation}
    \bm{\mathcal{F}}(\overline{\bm{\zeta}}) = \bm{\mathcal{F}}(\bm{\zeta}), \quad \bm{\mathcal{M}}(\overline{\bm{\zeta}}) = \bm{\mathcal{M}}(\bm{\zeta}) \bm{D}^{R}, \quad \bm{\mathcal{N}}(\overline{\bm{\zeta}}) = \bm{D}^R \bm{\mathcal{N}}(\bm{\zeta}), \quad \bm{\mathcal{T}}(\overline{\bm{\zeta}}) = \bm{D}^R \bm{\mathcal{T}}(\bm{\zeta}) \bm{D}^R
\end{equation}
Therefore, the global coefficient matrix in Eq. (\ref{eq:global_fourier}) for the wavenumber vector $\overline{\bm{\zeta}}$ can be written as, 

\begin{equation}
    \mathcal{J}_{\text{eig}}(\overline{\bm{\zeta}})= \begin{bmatrix} \bm{\Delta C}_p - \bm{\mathcal{F}}(\bm{\zeta}) & - \bm{\mathcal{M}}(\bm{\zeta}) \bm{D}^R \\ - \bm{D}^R \bm{\mathcal{N}}(\bm{\zeta}) &  \bm{\Delta K} - \bm{D}^R\bm{\mathcal{T}}(\bm\zeta)\bm{D}^R \end{bmatrix} = \begin{bmatrix} \textbf{I}^{(N_I)} & \textbf{0} \\ \textbf{0} & \bm{D}^R \end{bmatrix} \bm{\mathcal{J}}_\text{eig}(\bm{\zeta}) \begin{bmatrix} \textbf{I}^{(N_I)} & \textbf{0} \\ \textbf{0} & \bm{D}^R \end{bmatrix} 
    \label{eq:global_fourier_trans}
\end{equation}
where $\bm{D}^R \bm{\Delta K} \bm{D}^R = \bm{\Delta K}$ is applied. Substituting Eq. (\ref{eq:global_fourier_trans}) into Eq. (\ref{eq:avg_eigen_macro_1}) provides, 
\begin{equation}
    \begin{bmatrix} \langle Q^* \rangle (\overline{\bm{\zeta}}) \\ \frac{i K^c}{\omega} \langle u_{1}^* \rangle (\overline{\bm{\zeta}}) \\ \frac{i K^c}{\omega}\langle u_{2}^* \rangle (\overline{\bm{\zeta}}) \end{bmatrix} = \bm{\mathcal{I}}^T_\text{eig} \bm{\mathcal{S}}_\text{eig} \left( \bm{\mathcal{J}}_{\text{eig}}(\overline{\bm{\zeta}}) \right)^{-1} \bm{\mathcal{I}}_\text{eig} \hat{\bm{U}} = \underbrace{\bm{\mathcal{I}}^T_\text{eig} \bm{\mathcal{S}}_\text{eig} \begin{bmatrix} \textbf{I}^{(N_I)} & \textbf{0} \\ \textbf{0} & \bm{D}^R \end{bmatrix} \left( \bm{\mathcal{J}}_{\text{eig}}(\bm{\zeta}) \right)^{-1} \begin{bmatrix} \textbf{I}^{(N_I)} & \textbf{0} \\ \textbf{0} & \bm{D}^R \end{bmatrix}\bm{\mathcal{I}}_\text{eig} }_{\bm{\mathcal{G}}(\overline{\bm{\zeta}})} \hat{\bm{U}}
    \label{eq:avg_eigen_opposite}
\end{equation}
Because the microstructure is symmetric about the horizontal axis, the mirror pairs of particles share the same area fraction, which provides: 
\begin{equation}
    \begin{bmatrix} \textbf{I}^{(N_I)} & \textbf{0} \\ \textbf{0} & \bm{D}^R \end{bmatrix} \bm{\mathcal{S}}_\text{eig} = \bm{\mathcal{S}}_\text{eig} \begin{bmatrix} \textbf{I}^{(N_I)} & \textbf{0} \\ \textbf{0} & \bm{D}^R \end{bmatrix} \quad \text{and} \quad \begin{bmatrix} \textbf{I}^{(N_I)} & \textbf{0} \\ \textbf{0} & \bm{D}^R \end{bmatrix} \bm{\mathcal{I}}_\text{eig} = \bm{\mathcal{I}}_{\text{eig}} \begin{bmatrix} 1 & 0 & 0 \\ 0 & 1 & 0 \\ 0 & 0 & -1 \end{bmatrix}
\end{equation}
Therefore, Eq. (\ref{eq:avg_eigen_opposite}) can be further simplified as, 

\begin{equation}
    \bm{\mathcal{G}}(\overline{\bm{\zeta}}) = \begin{bmatrix} 1 & 0 & 0 \\ 0 & 1 & 0 \\ 0 & 0 & -1 \end{bmatrix} \bm{\mathcal{G}}(\bm{\zeta}) \begin{bmatrix} 1 & 0 & 0 \\ 0 & 1 & 0 \\ 0 & 0 & -1 \end{bmatrix} = \begin{bmatrix}
        \mathcal{G}_{11} & \mathcal{G}_{12} & -\mathcal{G}_{13} \\ 
        \mathcal{G}_{21} & \mathcal{G}_{22} & -\mathcal{G}_{23} \\
       - \mathcal{G}_{31} & -\mathcal{G}_{32} & \mathcal{G}_{33}
    \end{bmatrix}
    \label{eq:g_zeta_over}
\end{equation}

Comparing Eq. (\ref{eq:g_zeta_over}) with Eq. (\ref{eq:g_zeta}), the effective properties for $\overline{\bm{\zeta}}$ are, 

\begin{equation}
\begin{aligned}
    C^\text{eff}(\bm{\zeta}) & = C^\text{eff}(\overline{\bm{\zeta}}), & \xi^\text{eff}_{1}(\bm{\zeta}) & = \xi^\text{eff}_{1}(\overline{\bm{\zeta}}), & \xi^\text{eff}_{2}(\bm{\zeta}) & = -\xi^\text{eff}_{2}(\overline{\bm{\zeta}}) \\
    \chi^\text{eff}_{1}(\bm{\zeta}) & = \chi^\text{eff}_{1}(\overline{\bm{\zeta}}), & K_{11}^\text{eff}(\bm{\zeta}) & = K_{11}^\text{eff}(\overline{\bm{\zeta}}), & K^\text{eff}_{12}(\bm{\zeta}) & = -K^\text{eff}_{12}(\overline{\bm{\zeta}}) \\
    \chi^\text{eff}_{2}(\bm{\zeta}) & = -\chi^\text{eff}_{2}(\overline{\bm{\zeta}}), & K_{21}^\text{eff}(\bm{\zeta}) & = -K^{\text{eff}}_{21}(\overline{\bm{\zeta}}), & K_{22}^\text{eff}(\bm{\zeta}) & = K_{22}^\text{eff}(\overline{\bm{\zeta}})
\end{aligned}
\label{eq:effec_fanzhuan}
\end{equation}
The parity relations will be used in Sections 6.2 and 6.4 to characterize the wavenumber dependence of the coupling vectors and to explain why their second components have opposite signs for opposite vertical Bloch wavenumbers. 

\section{Unique constitutive relation for 1D thermal laminate and validation} \label{sec:1D_case}
\begin{figure}
    \centering
    \includegraphics[width=0.3\linewidth]{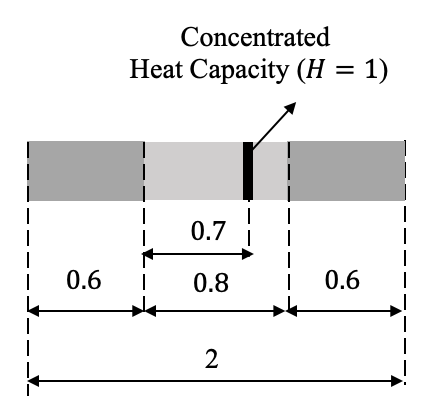}
    \caption{Schematic plot of a 1D laminate with normalized length $2$. The bimaterial unit cell consists of two material phases, the light and dark gray occupy $0.8$ and $1.2$, respectively; a concentrated heat capacity $H = 1$ (normalized) is placed to break microstructure's symmetry. }
    \label{fig:1d}
\end{figure}

This appendix compares the proposed homogenization scheme (with DR-iBEM) and the 1D analytical solution of the bimaterial laminates. Fig. \ref{fig:1d} shows the bimaterial unit cell with normalized length $2$ consists of two material phases, where the light and dark gray occupy $0.8$ and $1.2$, respectively. An additional concentrated heat capacity $H = 1$ is placed to break the microstructure's symmetry. Normalized thermal properties are employed as follows: (i) the heat capacities of the light gray and dark gray phases are $C_p^1 = 0.5$ and $C_p^2 = 0.5$, respectively; and (ii) the thermal conductivities of the light gray and dark gray phases are $K^1 = 0.05$ and $K^2 = 1$, respectively. Normalized frequency $\overline{\omega} = \frac{\omega C_p^2}{4 K^2} \in [0, 0.625]$ is considered with 1D thermal wavenumber $\zeta = 1$.

In this case study, the comparison medium is selected as $K^c = 0.4$ and $C_p^c = 0.7$, which differ from the properties of any of the involved phases. The entire composite is discretized into $200$ line intervals. Since each interval contains two eigen-field unknowns, namely the eigen-temperature-gradient and eigen-heat-source, and the concentrated heat capacity only introduces one additional EHS unknown. Therefore, the total number of unknowns for the eigen-fields is $200 \times 2 + 1 = 401$. The averaged thermal fields are evaluated using $2,000$ uniformly distributed sampling points.

Willis \cite{Willis2011} proposed that residual fields, such as inelastic strain in elastodynamics, can be introduced to resolve the nonuniqueness of the constitutive relation. Subsequently, Willis \cite{Willis2012} first derived the unique elastic effective constitutive relations of the one-dimensional laminate. Extending Willis' formulation (Eq. (24) in \cite{Willis2012}) to thermal homogenization, the unique effective thermal properties can be expressed as, 

\begin{equation}
\begin{aligned}
    \begin{bmatrix} C^\text{eff} & \xi^\text{eff} \\ \chi^\text{eff} & - K^\text{eff} \end{bmatrix} & = \begin{bmatrix} \overline{C} & \overline{\xi} \\ \overline{\chi} & - \overline{K} \end{bmatrix} + \frac{\overline{K} \thinspace \overline{C} + \overline{\chi} \overline{\xi}}{\zeta^2 \overline{K} - i \omega \overline{C}} \begin{bmatrix} -\zeta^2 & -i \zeta \\ -\zeta \omega & -i\omega \end{bmatrix} + \begin{bmatrix} \overline{C}^d & \overline{\xi}^d \\ \overline{\chi}^d & \overline{K}^d \end{bmatrix}
\end{aligned}
\end{equation}
Here, $\overline{C}, \overline{\xi}$, $\overline{\chi}$, and $\overline{K}$ given by Eq. (51) in \cite{Wu2026_rspa}, is an admissible representative of the effective constitutive relation obtained without introducing residual fields. The additional $\overline{C}^d$, $\overline{\xi}^d$, $\overline{\chi}^d$, and $\overline{K}^d$, following the residual-field construction of Willis \cite{Willis2012}, remove the nonuniqueness and provide the unique effective kernels.

\begin{equation}
    \begin{aligned}
    \overline{C}^d &= \langle C \rangle + 2 i \omega D \sum_{m=-\infty}^{\infty}  \frac{\left(\mu - \frac{i m \pi}{L}\right)c_m(\mu) c_{-m}(-\mu)}{\left(\mu - \frac{i m \pi}{L}\right)^2 + \zeta^2} \\
    \overline{\xi}^d & = -D \sum_{m = -\infty}^{\infty} \left[ \frac{c_m(\mu) b_{-m}(-\mu)}{\mu - \frac{i m \pi}{L} + i \zeta} + \frac{b_{m}(\mu) c_{-m}(-\mu)}{\mu - \frac{i m \pi}{L} - i \zeta} \right] \\
    \overline{\chi}^d & = -i D \omega \sum_{m = -\infty}^{\infty} \left[ \frac{b_m(\mu) c_{-m}(-\mu)}{\mu - \frac{i m \pi}{L} + i \zeta} + \frac{c_{m}(\mu) b_{-m}(-\mu)}{\mu - \frac{i m \pi}{L} - i \zeta} \right] \\ 
    \overline{K}^d & = -2 D \sum_{m = -\infty}^{\infty} \frac{\left(\mu - \frac{i m \pi}{L}\right) b_m(\mu) b_{-m}(-\mu)}{\left( \mu - \frac{i m \pi}{L} \right)^2 + \zeta^2}
    \end{aligned}
\end{equation}
where $\langle C \rangle = 1$ is the effective heat capacity at the static limit; $\mu$ is the Floquet exponent of the laminate, and $D$ can be seen in Eq. (40) in \cite{Wu2026_rspa}; $a_{m}, b_{m}, c_{m}$ are microstructure-specific Green's function-related Fourier series, which can be found in Eqs. (59-60) in \cite{Wu2026_rspa}.

\begin{figure}
    \centering
    \includegraphics[width=1\linewidth]{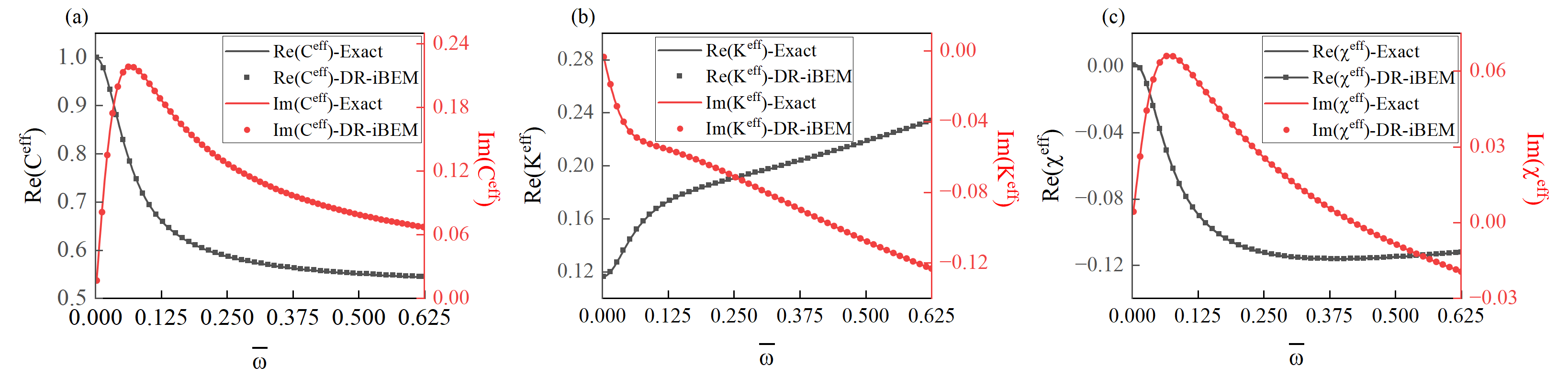}
    \caption{Comparison and variation of the effective thermal properties of the bimaterial laminate containing a concentrated heat capacity, (a) $C^\text{eff}$, (b) $\chi^\text{eff}$, and (c) $K^\text{eff}$, when the normalized frequency $\overline{\omega} \in [0, 0.625]$ and the thermal wavenumber $\zeta = 1$.}
    \label{fig:1d_cmp}
\end{figure}

Fig. \ref{fig:1d_cmp} (a-c) plots variations of $C^\text{eff}$, $K^\text{eff}$, and $\chi^\text{eff}$ predicted by the analytical solution and the proposed homogenization scheme, respectively. Fig. \ref{fig:1d_cmp} shows that (i) the proposed homogenization scheme is reliable and reproduces the  results of the analytical solution; and (ii) the variation of the comparison medium does not alter local thermal fields and extracted effective properties, once eigen-fields are determined accurately with the equivalent conditions. In addition, this result confirms that the comparison medium is only an auxiliary reference in Willis' polarization method. Moreover, the agreement demonstrates that the proposed homogenization scheme recovers the unique Willis constitutive relation rather than just one admissible constitutive representative, and it does not introduce any residual fields.

\end{document}